
\input harvmac
\newcount\figno
\figno=0
\def\fig#1#2#3{
\par\begingroup\parindent=0pt\leftskip=1cm\rightskip=1cm\parindent=0pt
\baselineskip=11pt
\global\advance\figno by 1
\midinsert
\epsfxsize=#3
\centerline{\epsfbox{#2}}
\vskip 12pt
{\bf Fig. \the\figno:} #1\par
\endinsert\endgroup\par
}
\def\figlabel#1{\xdef#1{\the\figno}}
\def\encadremath#1{\vbox{\hrule\hbox{\vrule\kern8pt\vbox{\kern8pt
\hbox{$\displaystyle #1$}\kern8pt}
\kern8pt\vrule}\hrule}}

\overfullrule=0pt

%
\def\tilde{\widetilde}
\def\bar{\overline}

\font\zfont = cmss10 

\def\bigone{\hbox{1\kern -.23em {\rm l}}}
\def\ZZ{\hbox{\zfont Z\kern-.4emZ}}

\Title{hep-th/9408074, HUTP-94/A017, IASSNS-HEP-94-54}
{\vbox{\centerline{A STRONG COUPLING TEST OF $S$-DUALITY}}}
\smallskip
\centerline{Cumrun Vafa}
\smallskip
\centerline{\it Lyman Laboratory of Physics, Harvard University}
\centerline{\it Cambridge, MA 02138 USA}
\smallskip
\centerline{and}
\smallskip
\centerline{\it School of Natural Sciences, Institute for Advanced Study}
\centerline{\it Olden Lane, Princeton, NJ 08540, USA}
\smallskip
\centerline{and}
\smallskip
\centerline{Edward Witten}
\smallskip
\centerline{\it School of Natural Sciences, Institute for Advanced Study}
\centerline{\it Olden Lane, Princeton, NJ 08540, USA}\bigskip
\baselineskip 18pt

\medskip

\noindent
By studying the partition function of $N=4$ topologically twisted
supersymmetric Yang-Mills on four-manifolds, we make an exact strong
coupling test of the Montonen-Olive strong-weak duality conjecture.
Unexpected and exciting links are found with two-dimensional rational
conformal field theory.
\Date{August, 1994}

\newsec{Introduction}

One of the most remarkable known quantum field theories in four  dimensions
is the $N=4$ supersymmetric Yang-Mills theory.  This theory
has the largest possible number of supersymmetries
for a four-dimensional theory without gravity.
It is believed
to be exactly finite and conformally invariant.

\nref\om{C. Montonen and D. Olive, Phys. Lett. B72 (1977) 117;
P. Goddard, J. Nyuts and D. Olive
, Nucl. Phys. B125 (1977) 1.}
A long-standing conjecture asserts that this theory has a symmetry exchanging
strong and weak coupling and exchanging
electric and magnetic fields.
This conjecture originated with work of
Montonen and Olive, who  \om\
proposed a symmetry with the above properties
and also exchanging the gauge group $G$ with the dual group $\widehat G$
(whose weight lattice is the dual of that of $G$).
It was soon realized that this duality was more likely to hold
supersymmetrically
\ref\wo{E. Witten and D. Olive, Phys. Lett. {\bf 78B} (1978) 97.}
and in fact the $N=4 $
theory was seen to be the most likely candidate
\ref\osb{H. Osborn, Phys. Lett. B83 (1979) 321.} since only in that
case the elementary electrons and monopoles have the same quantum numbers.
(It has recently been argued that an analog of Montonen-Olive duality
does hold for a certain $N=2$ theory with matter hypermultiplets
\ref\ws{N. Seiberg and E. Witten, ``Monopoles, Duality, And Chiral Symmetry
Breaking In $N=2$ Supersymmetric QCD,'' to appear.}.)

\nref\cardy{J. Cardy and E. Rabinovici, Nucl. Phys. {\bf B205} (1982) 1;
J. Cardy, Nucl. Phys. {\bf B205} (1982) 17.}
\nref\shapere{A. Shapere and F. Wilczek, Nucl. Phys.  {\bf B320}
(1989) 669.}
\nref\font{A. Font, L. Ibanez, D. Lust, and F. Quevedo,
Phys. Lett. {\bf B249} (1990) 35.}
\nref\rey{S.J. Rey, Phys. Rev. D43 (1991) 526.}
While Montonen-Olive duality was originally proposed as a ${\bf Z}_2$
symmetry involving the coupling constant only, the $N=4$ theory
has one more parameter that should be included, namely the $\theta$
angle.  As was originally recognized in lattice models \refs{\cardy,\shapere}
and string theory \refs{\font,\rey},
when the $\theta$ angle is included, it is natural
to combine it with the gauge coupling constant $g$ in a complex parameter
\eqn\nimbo{\tau ={\theta \over 2\pi}+{4\pi i\over g^2}.}
Then the ${\bf Z}_2$ originally proposed by Olive and Montonen
can be extended to a full $SL(2,{\bf Z})$ symmetry
 acting on $\tau$ in the familiar fashion
\eqn\kimbo{\tau\to {a\tau+b\over c\tau+d};}
here $a,b,c$, and $d$ are integers with $ad-bc=1$, so that
the matrix
\eqn\imbo{\pmatrix{a & b \cr c& d\cr}}
has determinant 1.
Indeed, $SL(2,{\bf Z})$ is generated by the transformations
\eqn\mbo{S=\pmatrix{ 0 & 1 \cr-1 & 0}}
and
\eqn\pimbo{T= \pmatrix{1 & 1\cr 0 & 1\cr}.}
Invariance under $T$ is the assertion that physics is periodic in
$\theta$ with period $2\pi$, and $S$ is equivalent at $\theta=0$ to the
transformation $g^2/4\pi \to (g^2/4\pi)^{-1}$
originally proposed by Montonen and Olive.

The difficulty in testing a conjecture that relates weak coupling
to strong coupling is, of course, that it is difficult to know what
happens for strong coupling.
Until now, tests of this conjecture have involved quantities that have no
quantum corrections and can
be determined exactly at the semiclassical level; one then checks
that the semiclassical results or formulas are invariant under
$SL(2,{\bf Z})$. For instance, the masses of stable
particles that saturate the Bogomol'nyi-Prasad-Sommerfeld (BPS) bound
are given exactly by the semiclassical result, as one can deduce
\wo\ from the structure of the supersymmetry algebra.
As explained by Sen in a recent survey \ref\seni{
A. Sen, preprint TIFR-TH-94-03 (hep-th/9402002).},
the $SL(2,{\bf Z})$ symmetry predicts the existence of BPS-saturated
multimonopole bound states.  Sen verified this
\ref\senii{A. Sen, preprint TIFR-TH-94-08 (hep-th/
9402032).}\ for the case of magnetic charge two
in an elegant calculation that gave some of the
most striking new evidence in many years
for the strong-weak duality conjecture.
The topological aspects of the generalization to arbitrary magnetic
charge have been demonstrated by Segal
\ref\segal{G. Segal, to appear.}.

\nref\sench{J. H. Schwarz and A. Sen, Phys. Lett. {\bf 312} (1993) 105.}
\nref\dix{L. Dixon, Lectures given at the 1987 ICTP Summer
Workshop in High Energy Physics
and Cosmology, Trieste, Italy,  1987. }
\nref\lvw{W. Lerche, C. Vafa and
N. Warner, Nucl. Phys. B324 (1989) 427.}
\nref\Greene{B.R. Greene
and M.R. Plesser, Nucl. Phys. B338 (1990) 15.}
\nref\Cand{P. Candelas, X.C. de la Ossa, P.S. Green and L. Parkes,
Nucl. Phys. B 359 (1991) 21.}
\nref\yau{{\it Essays
on Mirror Manifolds}, edited by S.-T. Yau, International
Press, Hong Kong 1992 .}
\nref\duff{M. Duff, Class.
Quantum Grav. {\bf 5} (1988) 189;
M. Duff and J. Lu,
Nucl. Phys. {\bf B354} (1991) 141; Phys. Rev. Lett. {\bf 66} (1991)
1402; Class. Quantum Grav. {\bf 9} (1991) 1; M. Duff, R. Khuri and
J. Lu, Nucl. Phys. {\bf B377} (1992) 281; J. Dixon, M. Duff and
J. Plefka, Phys. Rev. Lett. {\bf 69} (1992) 3009.}
\nref\strom{A. Strominger, Nucl. Phys. {\bf B343} (1990) 167}\
\bigskip
\noindent{\it Relation To String Theory}

The conjectured $SL(2,{\bf Z})$ symmetry of the $N=4$ theory gets
further appeal from the proposal that this is actually
a low energy manifestation of a similar symmetry in string theory.
To be precise, the conjecture \sench\
involves the compactification of the heterotic string theory
on a six-torus,
which gives a four-dimensional theory with $N=4$ supersymmetry;
the expectation value of the dilaton multiplet determines the low energy
parameters.
This theory is conjectured to have $SL(2,{\bf Z})$ symmetry acting on
the dilaton multiplet.
In this context, $SL(2,{\bf Z})$ symmetry has been called $S$-duality;
it is strikingly similar to the usual $R\leftrightarrow 1/R$ symmetry
of toroidal compactification of string theory known as $T$-duality.
The analogy helped motivate the original speculations about $S$-duality in
string theory.
$T$-duality has a conjectured generalization, mirror symmetry
\refs{\dix ,\lvw}, for
which there is ample evidence \refs{\Greene-\yau}.  In this
generalized sense $T$-duality relates
compactification on one manifold to compactification on another manifold.
It has been suggested that
string-fivebrane duality \refs{\duff ,\strom}
could exchange $S$ and $T$-duality, perhaps shedding light on the former;
see again \seni.

The existing evidence for $S$-duality in string theory
has been surveyed in \seni.  It largely
concerns the existence of quantities (like the low energy effective
action for the axion-dilaton system) that are unaffected by quantum
corrections and whose $SL(2,{\bf Z})$ symmetry can be verified at tree
level.  Also, there is the fascinating occurrence of duality of root
systems both in the heterotic string and in the Montonen-Olive conjecture.

The extension of $SL(2,{\bf Z})$ from field theory to string theory
greatly increases its potential significance.  In field theory,
this strong-weak duality would appear to be, at most, a curious phenomenon
applying to very special field theories.  In string theory, however,
if this symmetry appears under toroidal compactification, then -- as the
radius of the torus is arbitrary -- on thinking about the large volume limit
it would appear that the $SL(2,{\bf Z})$ must be a manifestation of some
property of the uncompactified theory (just as the same is believed to be
true for $T$-duality, which appears after compactification).
And if $S$-duality comes from a property of the uncompactified theory,
it must have some significance after any compactification.  Thus,
if valid in string theory, $S$-duality should have some implications
not just for special models but for the real world.

In field theory,
one has to be careful in calling $SL(2,{\bf Z})$ a
``symmetry'';
it not only changes the coupling parameters  but exchanges the gauge
group $G$ with the dual group $\widehat G$.  Only a subgroup of
$SL(2,{\bf Z})$ maps $G$ to itself.
In toroidal compactifications of
string theory, $SL(2,{\bf Z})$ is (if valid at all) a normal,
albeit for the most part spontaneously broken, symmetry (in fact a gauge
symmetry \seni) with
the unusual property of being valid only quantum mechanically.
That last fact suggests that $S$-duality may have a message that
is now hard to perceive,
especially since upon compactification to three dimensions, $S$ and
$T$ duality are apparently combined
into a bigger symmetry group
\ref\seniii{A. Sen, to appear.}, perhaps giving a unification  of $\hbar$
and $\alpha'$ - the parameters that control quantum mechanical and stringy
corrections.

Perhaps, as in the case of $T$-duality and mirror symmetry,
examples will be found of string theories in
which the strong coupling limit of one compactification is the weak
coupling limit of another.  A proper general statement may well
be simply that dilaton moduli space is compact and smooth
except for singularities
that correspond to weak coupling limits\foot{And perhaps also
orbifold singularities
where some discrete symmetries are unbroken.}; in particular, the same
string theory might have several very different-looking weak coupling
limits.

\bigskip
\noindent{\it Testing $S$-duality For Strong Coupling}

Despite the substantial evidence for $S$-duality,
it is frustrating that none of the existing tests of this strong coupling
symmetry really involve the strong coupling behavior.
The purpose of the present paper is to fill this gap by developing
a true strong coupling test of $S$-duality.

At the same time, the test  we carry out will involve
computations in the vacuum of the $N=4$ theory in which
the non-abelian gauge symmetry
is unbroken.  Existing tests have generally involved computations in
vacua in which the gauge symmetry is spontaneously broken to an abelian
subgroup.

To test $S$-duality for strong coupling involves finding quantities
that we can calculate for strong coupling.  Most strong coupling
calculations are out of reach, but exceptions are sometimes provided
by quantities that can be interpreted as correlation functions in a topological
field theory.
$N=2$ supersymmetric Yang-Mills theory has
a twisted version  that is a topological field theory
\ref\tqft{E. Witten, ``{Topological Quantum Field Theories},''
 Comm. Math. Phys. {\bf 117} (1988) 353.} and the same is true for $N=4$
\ref\yam{J. Yamron, ``{Topological Actions From Twisted Supersymmetric
Theories},'' Phys. Lett. {\bf B 213} (1988) 325.}.
The twisted theories coincide with the physical theory on a flat
manifold or more generally on a hyper-K\"ahler manifold, but not in general.

\nref\mukai{S. Mukai, Inv. Math. {\bf 77} (1984) 101.}
\nref\others{T. Nakashima, ``Moduli Of Stable Rank 2 Bundles With Ample
$c_1$ on $K3$ Surfaces,'' Archiv. der Mathematik {\bf 61} (1993) 100.}
\nref\bothers{Z. B. Qin, ``Moduli Of Stable Rank-2 Sheaves On $K3$-Surfaces,''
Manuscripta Mathematica {\bf 79} (1993) 253.}
\nref\yosh{K. Yoshioka, ``The Betti Numbers Of The Moduli Space Of
Stable Sheaves Of Rank 2 On ${\bf P}^2$,'' preprint, Kyoto University.}
\nref\yoshi{K. Yoshioka, ``The Betti Numbers Of The Moduli Space
Of Stable Sheaves Of Rank 2 On  A Ruled Surface,'' preprint, Kyoto Univer}
\nref\klyachko{A. A. Klyachko,  ``Moduli Of Vector Bundles And Numbers
Of Classes,'' Funct. Anal. and Appl. {\bf 25} (1991) 67.}
In this paper, we will consider just one of the twisted theories,
and show that its partition function for gauge fields in a given
topological class is the Euler characteristic of instanton moduli space.
Thus, to simplify a bit,
if $a_k$ is the Euler characteristic
of the moduli space of $k$-instantons,
and $q=\exp(2\pi i \tau)$, then the partition function is $Z(q)= \sum_ka_kq^k$.
We will test $S$-duality by examining the modular properties
of the function $Z(q)$, on various manifolds, mainly
for gauge groups $SU(2)$ and $SO(3)$.  For computing $Z(q)$ on various
four-manifolds,
we rely almost entirely on known mathematical results.
For $K3$, we rely on constructions by Mukai and others \refs{\mukai-\bothers};
for ${\bf CP}^2$
we use formulas of Yoshioka and Klyachko \refs{\yosh-\klyachko};
for the case of blowing up a point in a K\"ahler manifold we use
a formula of Yoshioka \yosh; and for ALE spaces we use formulas
of Nakajima \ref\nakajima{H. Nakajima,
``Instantons on ALE spaces, quiver varieties, and Kac-Moody
Algebras'', Tohoku Univ. preprint 1993;
``Gauge Theory on Resolutions of Simple Singularities
and Simple Lie Algebras'', Int. Math. Res. Notices,  (1994) 61.}.  All of these
formulas give functions with modular properties.

The organization of this paper is as follows.   In \S2 we discuss
the topologically twisted $N=4$ Yang-Mills theory and explain why, in
favorable conditions,
its partition function is the Euler characteristic of instanton moduli
space.  (More generally, one needs to consider a more complicated
equation that we describe in \S2.) This section
includes a self-contained
review of the relevant aspects of topological theories.

In \S3 we formulate more precisely the predictions of $S$-duality
that we are going to test.
We discuss some subtleties that may arise when discussing $S$-duality
on a compact four-manifold;
they have the effect that the partition function can transform as a modular
form rather than a modular function and that it can differ
by an overall factor of $q^{-s}$ from the generating function
of instanton Euler characteristics.
We then go on to sharpen the $S$-duality conjecture to
allow for non-abelian electric and magnetic flux \ref\thooft{G.
't Hooft, Nucl. Phys. B138 (1978) 1; Nucl. Phys. B153
(1979) 141.}.   (While this paper was in gestation, we received
a paper by Girardello, Giveon, Porrati, and Zaffaroni, who similarly
incorporated the discrete flux on the four-torus in the context
of $S$-duality \ref\giv{L. Girardello,
A. Giveon, M. Porrati, and A. Zaffaroni,
Phys. Lett. {\bf B334} (1994) 331.}.)
We propose a transformation law for the partition function computed
with fixed electric and magnetic fluxes.
We also determine some general constraints on exponents and singularities.

\nref\zag{D. Zagier,
C.R. Acad. Sci. Paris,
281A (1975) 879 . }
\nref\hirzag{F. Hirzebruch and D. Zagier,
Invent. Math. vol 36 (1976), 92.}
In \S4 we begin testing the predictions.
We first consider the case in which space-time is a $K3$  surface;
this is a particularly nice case, as it is a hyper-K\"ahler manifold so the
physical and twisted models coincide. Enough is known
about instanton moduli spaces on $K3$
for sufficiently many instanton numbers
 to allow a strong test of the $S$-duality.
As most of the instanton moduli spaces on $K3$ can be explicitly
exhibited as orbifolds, one can
use techniques from orbifold
constructions to compute its Euler characteristic.
We not only find that the partition function of $N=4$ super Yang-Mills
theory on $K3$ is a modular object, as expected from $S$-duality, but
in the midst of this calculation we unexpectedly
encounter the partition function
of bosonic strings!

We next consider ${\bf CP}^2$ using the formulas of Yoshioka and Klyachko.
We find that the topological partition function
has interesting modular properties but
does not quite converge well enough to be modular.
A natural
non-holomorphic modification makes it modular \refs{\zag,\hirzag}.
Presumably, what is going on is that there is a holomorphic anomaly
somewhat analogous to the one that arises
\ref\bcov{M. Bershadsky, S. Cecotti, H. Ooguri and C. Vafa,
Nucl.Phys.B405 (1993) 279; ``{Kodaira-Spencer
Theory of Gravity and Exact Results for Quantum String Amplitudes},''
preprint HUTP-93-A025.} in certain two dimensional models.
The two examples of $K3$ and ${\bf CP}^2$ enable
us to fix some important unknown constants that appeared in \S3.

Next we consider the case of the blow-up of a point on a
four-manifold (i.e. gluing in a copy of ${{\bf CP}^2}$ with opposite
orientation)
using the formula of Yoshioka
\yosh;  it turns out that under the blow-up the partition
function essentially is multiplied
by a character of the two-dimensional WZW model
of $SU(2)$ at level 1!  We have no idea why two-dimensional field theory
makes this appearance, but at any rate the
function involved is certainly modular.
(We propose a natural generalization  of the blowing-up formula
for groups of rank bigger than one, again involving WZW characters and
satisfying some non-trivial checks.)

Finally, following results of Nakajima,
we consider the $U(k)$ and $SU(k)$
theories on ALE spaces, which are non-compact hyper-K\"ahler manifolds.
Nakajima's formulas once again involve two-dimensional current algebra
in a beautiful and unexpected way.  His results appear not just
to incorporate $S$-duality but to go beyond what one would
expect from {\it field theoretic} $S$-duality; unfortunately
we do not understand the predictions of $S$-duality
on noncompact manifolds precisely enough to fully exploit them.

In \S5 we attempt to generalize these results using physical
arguments.
Imitating a strategy used recently in Donaldson theory
\ref\switten{E. Witten, ``{
Supersymmetric Yang-Mills Theory On A
Four-Manifold,}'' IASSNS-HEP-94/5.}, we restrict ourselves to K\"ahler
manifolds
with $h^{2,0}\not=0$, and make a massive perturbation of the
twisted $N=4$ theory that preserves part of the topological symmetry.
We propose a formula for the partition function of the $N=4$ theory on
these manifolds which beautifully obeys the various constraints.

We conclude
in \S6 with some comments on the relation to string theory.
On the one hand, we summarize the facts concerning the odd appearances  in
\S4 of
formulas from two-dimensional rational conformal field theory.
And we
pursue further the peculiar appearance of the (left-moving)
bosonic string partition
function in \S4; we explain that
a similar computation using the four-torus
instead of $K3$ gives the (left-moving) oscillator states of a fermionic
string.   In a way these mysterious observations generalize an observation by
Gauntlett and Harvey for the string ground state
\ref\harre{
J.P. Gauntlett and J.A. Harvey, unpublished; see also
hep-th/9407111.}.

\def\tilde{\widetilde}
\def\bar{\overline}
\nref\mq{V. Mathai and D. Quillen, ``Superconnections, Thom Classes,
and Equivariant Differential Forms,'' Topology {\bf 25} (1986) 85.}
\nref\atj{M. F. Atiyah and L. Jeffrey, ``Topological Lagrangians And
Cohomology,'' J. Geom. Phys. {\bf 7} (1990) 119.}
\nref\bs{L. Baulieu and I. M. Singer, ``Topological Yang-Mills Symmetry,''
Comm. Math. Phys. {\bf 117} (1988) 253.}
\nref\ugmit{E. Witten,``Introduction To Cohomological Field Theories,''
Int. J. Mod. Phys. {\bf A6} (1991) 2775.}
\nref\oldwit{E. Witten, ``The $N$ Matrix Model And Gauged WZW Models,''
Nucl. Phys. {\bf B371} (1992), 191, \S 3.3.}
\nref\distler{J. Distler, ``Notes On $N=2$ Sigma Models,'' hep-th-9212062.}
\nref\cgm{S. Cordes, G. Moore, and S. Ramgoolam, ``Large $N$  2D
Yang-Mills Theory \& Topological String Theory,'' hepth-9402107, \S6,7.}
\nref\thbl{M. Blau and G. Thompson,
Comm. Math. Phys. {\bf 152} (1993) 41;
Int. J. Mod. Phys. {\bf A 8} (1993) 573.}
\newsec{Twistings of Supersymmetric Yang-Mills Theory}

Before describing how topological field theories can be constructed
by twisting of $N=4$ super Yang-Mills theory, let us recall the situation
for $N=2$ \tqft.  $N=2$ super Yang-Mills
has a global symmetry group $SU(2)_I$.  The supercharges $Q_{\alpha i}$
and $\bar Q_{\dot\alpha j}$ transform in the two-dimensional representation
of this group.  Working on a flat ${\bf R}^4$, the rotation group
$K=SO(4)$ is locally $SU(2)_L\times SU(2)_R$.  Under $SU(2)_L\times SU(2)_R
\times SU(2)_I$, the supercharges transform as $({\bf 2},{\bf 1},{\bf 2})
\oplus ({\bf 1},{\bf 2},{\bf 2})$.

Now, as long as we are on flat ${\bf R}^4$, we could find an alternative
embedding of $K$ in $SU(2)_L\times SU(2)_R\times SU(2)_I$ and declare
this to be the rotation group.  This can be done by leaving $SU(2)_L$
undisturbed and replacing $SU(2)_R$ by a diagonal combination of $SU(2)_R
\times SU(2)_I$; we will call this diagonal combination $SU(2)'{}_R$.
The modified rotation group is hence $K'=SU(2)_L\times SU(2)'{}_R$.
When one departs from flat ${\bf R}^4$, either by considering a curved
metric or by working on a different four-manifold altogether,
one uses not the usual stress tensor $T_{ij}$ but a modified
stress tensor $T'_{ij}$ chosen so that the corresponding rotation
operators are in fact $K'$.  The ``twisted'' theory so defined
therefore coincides with the physical theory only when the metric is flat.

Under $K'$, the supercharges transform as $({\bf 2},{\bf 2})
\oplus ({\bf 1},{\bf 3})\oplus ({\bf 1},{\bf 1})$.  Let us call
the $({\bf 1},{\bf 1})$ element $Q$.  Its claim to fame is that
it obeys $Q^2=0$, and (roughly because it has spin zero in the sense of
$K'$), under the twisted coupling to gravity, it is
conserved on an arbitrary four-manifold $M$.
Moreover, one finds that
$T'{}_{ij}=\{Q,\Lambda_{ij}\}$ for some $\Lambda$; this means that if
$Q$ is interpreted as a BRST-like operator, only $Q$-invariant
observables being considered, then the coupling to the gravitational
field is a BRST commutator and the theory is a topological field theory.

The topological field theory constructed this way is quite interesting,
being equivalent to Donaldson theory.

\bigskip
\noindent{\it Generalization To $N=4$}

Now we come to the generalization to $N=4$.  From the above,
it is clear that the main point is to pick an homomorphism of $K$
into the global symmetry group of the theory to get a twisted Lorentz
group $K'$.
$N=4$ supersymmetric Yang-Mills theory in four dimensions has
global symmetry group $SU(4)$.
The possible homomorphisms of $K$ in $SU(4)$ can be described
by telling how the ${\bf 4}$ of $SU(4)$ transforms under $K$.
Thus they correspond simply to four-dimensional representations
of $K$.  To get a topological field theory, we need a representation
such that at least one component of the supercharge is a $K'$ singlet.
Up to an exchange of left and right, there are three four-dimensional
representations with this property:
(i) $({\bf 2},{\bf 2})$; (ii)
$({\bf 1},{\bf 2})\oplus ({\bf 1},{\bf 2})$; (iii)  $({\bf 1},{\bf 2})
\oplus ({\bf 1},{\bf 1})\oplus ({\bf 1},{\bf 1})$.  Correspondingly,
there are three topological field theories that one can consider.  The
theories corresponding to the last two of those three representations
were discussed some years ago by Yamron \yam.  Our interest in the
present paper will be in the theory determined by the representation
$({\bf 1},{\bf 2})\oplus ({\bf 1},{\bf 2})$.  Note that this embedding
of $K$ in $SU(4)$ commutes with a subgroup $F\cong SU(2)$ of $SU(4)$
that transforms the two copies of $({\bf 1},{\bf 2})$.  This becomes
a global  symmetry of the twisted theory.

The supercharges, which under $K\times SU(4)$
transform as $({\bf 2},{\bf 1},{\bf 4})\oplus ({\bf 1}, {\bf 2},
\bar{\bf 4})$, transform under $K'\otimes F$ as
$({\bf 2},{\bf 2},{\bf 2})
\oplus ({\bf 1},{\bf 3},{\bf 2})\oplus ({\bf 1},{\bf 1},{\bf 2})$.
Thus there
are two $K'$ singlets, say $Q$ and $Q'$.  They obey $Q^2=(Q')^2
=\{Q,Q'\}=0$.  They transform as a doublet of $F$.

The gauge bosons of the $N=4 $ theory are, of course, scalars under $SU(4)$.
The left and right handed fermions transform under $K\times SU(4)$
like the $Q$'s -- so under $K'\otimes F$, they transform as
$({\bf 2},{\bf 2},{\bf 2})\oplus ({\bf 1},{\bf 3},{\bf 2})
\oplus ({\bf 1},{\bf 1},{\bf 2})$.
If we ignore $F$, this is just two copies of the $K'$ representation
that appears in Donaldson theory.
The scalars of the $N=4$ theory transform in the six dimensional representation
of $K$; under $K'\times F$ they transform as $({\bf 1},{\bf 1},{\bf 3})
\oplus ({\bf 1},{\bf 3},{\bf 1})$.  In other words, in the twisted
theory these fields turn into three scalars (a triplet of $F$)
and a self-dual antisymmetric tensor (a singlet of $F$).  Of course, the
fermions and scalars all take values in the Lie algebra of the gauge
group.

With the scalars denoted as $v_y,\,y=1\dots 6$,
the bosonic part of the Lagrangian of the $N=4$
theory is
\eqn\plsk{L={1\over 2e^2}\int d^4x\Tr\left({1\over 2} F_{ij}F^{ij}
+\sum_{i=1}^6(D_iv_y)^2+\sum_{1\leq y<z\leq 6}[v_y,v_z]^2\right),}
plus a possible theta term
\eqn\lsk{{i\theta\over 8 \pi^2}\int\Tr F\wedge F.}
After twisting, there is an important curvature coupling term
that will emerge later.

If the twisted theory is formulated on a Kahler manifold $X$, some special
features arise.  The holonomy of the Riemannian connection on a Kahler
manifold is $SU(2)_L\times U(1)_R$ (for a generic oriented
Riemanian four-manifold it is  $SU(2)_L\times SU(2)_R$).  The twisting
involves the embedding in the global symmetry group $SU(4)$ of only $U(1)_R$,
not $SU(2)_R$.  The global symmetry group is the subgroup of $SU(4)$ that
commutes with the embedding of $U(1)_R$; it is isomorphic to
$SU(2)\times SU(2)\times U(1)$. Four supercharges instead of two are invariant
under
the twisted holonomy group $SU(2)_L\times U(1)'{}_R$, so  the twisted
$N=4$ theory on a Kahler manifold has four fermionic symmetries instead of two.
Indeed, one of these originates from each of the four underlying
supersymmetries
of the $N=4$ model.

\subsec{The Euler Class}
The goal in the rest of this
section is to explain why the partition function
of the twisted $N=4$ theory that was just described is,
under suitable conditions, the Euler characteristic of instanton moduli
space.  This will also help us understand the deviations from this formula
that occur under certain conditions.  Actually, the general background
is not new \refs{\mq -\ugmit} and the specific issues that lead
to the Euler characteristic have also been discussed previously
\refs{\oldwit - \thbl}, but we will
attempt to develop the subject here in such a way as to make this paper
as self-contained and readable as possible.  To do so, we will first
explain some very simple models, beginning with finite dimensional systems.

To begin with, we consider a compact\foot{Compactness
is not necessary if the section $s$, introduced presently, has a suitable
behavior at infinity.  A non-compact manifold with a suitable $s$ will,
eventually, be the main situation we study.}
oriented manifold $M$ of dimensions
$d=2n$, endowed with a real oriented
vector bundle $V$ of rank $d$.
We choose on $V$ a metric $g_{ab}$ (we write $(v,w)=g_{ab}v^aw^b$)
and an  $SO(d)$ connection $A$.
We will consider a system
with a topological symmetry $Q$ ($Q^2=0$) that carries charge
one with respect to a ``ghost number'' operator $U$.  There will be two
multiplets.  The first consists of local coordinates $u^i$ on $M$
(of $U=0$) together with fermions $\psi^i$ tangent to $M$, of $U=1$.
The transformation laws are
\eqn\translaw{\eqalign{ \delta u^i & = i\epsilon\psi^i \cr
                   \delta\psi^i & = 0.\cr}}
Here $\epsilon$ is an anticommuting parameter.  We also define
$\delta_0$ to be the variation with $\epsilon $ removed, so for instance
$\delta_0\phi^i=i\psi^i$.
The second multiplet consists of an anticommuting section $\chi^a$ of $V$
of $U=-1$, and a commuting section $H^a$ of $V$; $H$ has $U=0$.
The transformation laws are
\eqn\otherlaw{
\eqalign{\delta \chi^a & =  \epsilon H^a-\epsilon\delta_0
 u^iA_i{}^a{}_b\chi^b\cr
           \delta H^a    & =\epsilon\delta_0 u^i A_i{}^a{}_bH^b
-{\epsilon\over 2}\delta_0 u^i\delta_0 u^j F_{ij}{}^a{}_b \chi^b.
\cr}}
Of course, one could here substitute for $\delta_0 u^i$ from \translaw;
we have written the formula in this way to indicate that it is
a covariantized version of the more naive $\delta\chi^a=\epsilon H^a,\,
\delta H^a=0$.

The Lagrangian is to be $L=\delta_0 W$ for a suitable $W$.  Such a
$W$ is
\eqn\suitv{W ={1\over 2 \lambda }(\chi,H+2is)         }
with $s$ an arbitrary $c$-number section of $V$ and $\lambda$ a small
positive real number.  We get
\eqn\hurgo{
L= {1\over 2\lambda }(H,H-2is)+{1\over \lambda}g_{ab}\chi^a {\partial s^b
\over \partial u^i}\psi^i -{1\over 2\lambda}F_{ijab}\psi^i\psi^j\chi^a\chi^b
 .}
Now we want to do the integral
\eqn\uitv{
Z=\left({1\over 2\pi}\right)^d
\int d u\,d\psi\,d\chi\,dH\,e^{-L}.  }
(The factors of $2\pi$ correspond to the standard factor of $1/\sqrt{2\pi}$
for every bosonic variable in the Feynman path integral.)
This integral is guaranteed to be a topological invariant -- that is, to
depend only on $M$ and $V$
-- since the derivative of $L$ with respect to any of the other data
($\lambda$, $g$, $A$, and $s$) is of the form $\{Q,\dots\}$.
We will call it the partition invariant of the system.

As a first step to evaluate the integral, we integrate over $H$, getting
\eqn\putv{Z=\left({\lambda\over 2\pi}\right)^{d\over 2}
\int du\,d\psi\, d\chi\,
\exp\left(-{(s,s)\over 2\lambda}-{1\over\lambda}g_{ab}\chi^a{\partial s^b
\over \partial u^i}\psi^i+{1\over 2\lambda}F_{ijab}\psi^i\psi^j\chi^a\chi^b
\right).  }
To proceed, we first consider the case $s=0$.
The integral is then done by expanding the four fermi interaction, giving
\eqn\stanform{
Z=
\int_M{{\rm Pf}(F\wedge F\wedge \dots\wedge F)\over (2\pi)^{d/2}\cdot d!},}
with ${\rm Pf}(F\wedge \dots \wedge F)$ the Pfaffian on the $a,b$
indices.  The curvature integral in \stanform\ is a standard integral
representation for a topological invariant that is known as the Euler
class of $V$ (integrated over $M$); we will denote it as $\chi(V)$.
In case $V=TM$ is the tangent bundle of $M$,
$\chi(V)$ coincides with the Euler characteristic $\chi(M)$ of $M$.

Now we consider the case of $s\not= 0$, which is much closer to our
general interests in this paper.  The main idea is to consider the behavior for
$\lambda\to 0$.  In this limit, the integral is dominated by contributions from
infinitesimal neighborhoods of zeroes of $s$.
For the first basic case, we suppose that $s$ has only isolated
and non-degenerate zeroes $P_\alpha$.
In that case, near each zero one can choose
local coordinates on $M$ and a trivialization of $s$ so that
$s^a=f_au^a$ (no sum over $a$ here and in similar formulas below),
with some real numbers $f_a$.  Higher order terms are irrelevant for
small $\lambda$.  Then the contribution
of a particular zero is
\eqn\hudd{\left({\lambda\over 2\pi}\right)^{d/2}
\prod_{a=1}^d\int du^a\,d\psi^a\,d\chi^a\,\,
\exp\left(-{(f_au^a)^2\over2\lambda}+ {1\over\lambda}f_a\psi^a\chi^a\right)
=\prod_{a=1}^d{f_a\over |f_a|}=\pm 1.     }
The answer is thus
\eqn\udd{Z=\sum_{P_\alpha} \epsilon_\alpha, }
where
\foot{Another way to describe this is that $\epsilon_\alpha$
measures the relative orientation of $TM|_{P_\alpha}$
and $V|_{P_\alpha}$; regarding $ds|_{P_\alpha}$ as an isomorphism
between those spaces, $\epsilon_\alpha=\pm 1$ depending on
whether the orientations of the two spaces agree or disagree under
this isomorphism.}
\eqn\invfor{\epsilon_\alpha={\rm sign}\left(
\det\left({\partial s^a\over\partial u^i}\right)\right).}
Since the integral is independent of $s$, we learn by comparing
to the result for $s=0$ that the Euler
class of a bundle can be computed by counting the zeroes of a section,
weighted with signs; this is a standard theorem.

In our applications, we will actually need a hybrid of the two cases
considered above.  Suppose that $s=0$ on a union of submanifolds $M_\alpha$
of $M$, of dimensions $d_\alpha$.  We assume that the behavior of $s$ in
the normal directions to $M_\alpha$ is non-degenerate; this means that locally
one can pick coordinates $u^i,\,\,i=1\dots d-d_\alpha$
in the directions normal to $M_\alpha$ and a trivialization of $V$ such that
\eqn\locoform{
\eqalign{s^a = f^a{}_iu^i, \,\,\,&{\rm for}\,\,\,i,a,=1\dots d-d_\alpha\cr
           s^a=0, \,\,&{\rm  for}\,\,\, a>d-d_\alpha \cr}}
where $f^a{}_i$ (with $a,i=1\dots d-d_\alpha$) is  an invertible
$(d-d_\alpha)\times (d-d_\alpha)$
matrix.   It is convenient to regard $f$ as a $d\times d$ matrix whose
other components are zero.
Looking at the Lagrangian \hurgo, we see that this $f$ is the
``mass matrix'' for the fermions near $M_\alpha$. The ``massless
components'' of $\psi$ are those that are tangent to $M_\alpha$.  The
massless components of $\chi$ are in the above trivialization the
$\chi^a$ for $a>d-d_\alpha$.  That trivialization is valid only locally;
globally the massless components of $\chi$ are sections of a vector
bundle $V_\alpha$ over $M_\alpha$.

We will adopt the following terminology: we call the $\chi^a$
(which have ghost number $-1$) ``antighosts,'' and we  refer to
$V_\alpha$ as the vector bundle of antighost zero modes.

Now we will evaluate the integral for $Z$, in the limit of small $\lambda$.
The integral will be a sum of contributions from the various $M_\alpha$.
These contributions can be evaluated as follows.  Fixing a particular
$M_\alpha$, the integral over the ``massive modes,'' which roughly are
those ``normal'' to $M_\alpha$, proceeds precisely as in the
derivation of \hudd.  One gets a Gaussian integral with bosons
and fermions canceling up to sign, giving a factor of $\epsilon_\alpha=\pm 1$.
\foot{In the spirit of the last footnote,
one can describe
$\epsilon$ as a factor which, given orientations of $TM$ and $V$,
produces a relative orientation of $TM_\alpha$ and $V_\alpha$.
This is the data needed to fix the sign of $\chi(V_\alpha)$ below.}
Then one has the integral ``tangent'' to $M_\alpha$.  The Lagrangian
\hurgo\ has the property that if one sets all of the ``massive'' fields
to 0, one gets a Lagrangian of the same type, but with $M$ replaced
by $M_\alpha$, $V$ by $V_\alpha$, and $s$ by 0.
The integral over the ``massless''
fields is thus an integral of the type that we have already seen
in getting \stanform, and so equals $\chi(V_\alpha)$, the Euler class
of $V_\alpha$ (integrated over $M_\alpha$).  The final result is then
\eqn\hurf{Z= \sum_\alpha \epsilon_\alpha \,\,\chi(V_\alpha).     }

Another way to obtain the same result is to perturb $s$ to a nearby
section $\tilde s$ that has isolated zeroes.  For instance,
on each $M_\alpha$, pick a section $s_\alpha$ of $V_\alpha$ with only
isolated zeroes.  Regard $s_\alpha$ as a section of $V|_{M_\alpha}$; extend
it in an arbitrary fashion to a section of $V$ that vanishes outside
a small tubular neighborhood of $M_\alpha$ (a neighborhood disjoint
from $M_\beta$ for $\beta\not=\alpha$).  Then $\tilde s=s+\epsilon
\sum_\alpha s_\alpha$ for sufficiently small $\epsilon$
vanishes precisely on points on
$M_\alpha$ on which $s_\alpha=0$.
This gives a check on \hurf\ in the following sense: evaluating
\hurf\ by using the zeroes of $s_\alpha$ to compute $\chi(V_\alpha)$ one
gets the same result that one gets by
evaluating \udd\ for the section $\tilde s$.

\subsec{Counting Solutions Of An Equation}

It is illuminating to consider a very special case of this:
counting the solutions of an equation.  We will
start with an elementary example.  Consider a single
real variable $u$ and an equation
\eqn\sinfr{ u^2-a= 0 , }
with real $a$.  Obviously, the number of solutions is not a topological
invariant; there are two for positive $a$, and none for negative $a$.

To put this in the above format, take $M$ to be the $u$-axis, $V$ to be
a one dimensional trivial bundle, and $s$ to be the section of $V$
\eqn\vinfr{s(u) =u^2-a. }
We want to compute the partition invariant of this system.  According
to \hudd, the result is
\eqn\urfo{Z=\sum_\alpha \epsilon_\alpha, }
with $\alpha$ running over the zeroes of $s$, and $\epsilon$ the sign
of $df/du$ at a given zero.  For $a<0$, $Z=0$ since there are no zeroes.
For $a>0$, there are two zeroes, at $u=\pm\sqrt a$.  $Z$ still vanishes
since $\epsilon=\pm 1$ for $u=\pm \sqrt a$.

The moral is, of course, that while the total number of solutions
of an equation is not a topological invariant, the number
of solutions weighted with signs (or in general, with multiplicities,
if one encounters degenerate solutions) is such an invariant.

Suppose, however, that we want to find an integral formula
that counts {\it without signs}
the total number of solutions of an equation.
This cannot really be done, as is clear
from the above example.  But there is a partial substitute which we will
explain first in the above special case.
Double the degrees of freedom, adding a new variable $y$
and replacing the $u$ axis by the $u-y$ plane.
Take $V$ to be a two dimensional trivial bundle, and let $s$ be the section
of $V$ given by the two functions
\eqn\mippol{
\eqalign{ s_1 & = u^2-a-y^2  \cr
            s_2 & = 2uy. \cr}}
Now we consider the system of equations $s=0$, that is $s_1=s_2=0$.
The partition invariant $Z$ is a topological invariant
of this system which is the number
of solutions of the equations
{\it weighted by sign}.  Let us compute this number -- which
of course can be defined as above by an integral formula --
for various $a$.

Suppose first that $a>0$.  The equations $s_1=s_2=0$ have the
two solutions $u=\pm \sqrt a$, $y=0$.
The result of including $y$ is that $\epsilon=1$ for both solutions.
Indeed, the determinant in \invfor\ is always positive
(as $\partial s_1/\partial u$ and $\partial s_2/\partial y$ have
the same signs at each root of the equations).  So $Z=2$.

What about $a<0$?  There are again two solutions, now at $u=0$,
$y=\pm \sqrt{- a}$.  One can verify that again the contributions are $+1$
and so  $Z=2$.

The reason that, for $y=0$, $\epsilon=+1$ at each zero
is that
\eqn\ensur{s_2=y{\partial s_1\over \partial u}+O(y^2).  }
This ensures that for zeroes with
$y=0$ the determinant in \invfor\ is positive.
The conclusion is that {\it not for all sections $s$, but for those
sections that obey \ensur\
and vanish only at $y=0$}, $Z$ is equal to the total number
of solutions of the ``original equations''
$s_1(u,y=0)=0$.\foot{In  this
statement  and similar statements below,
we  assume the zeroes are nondegenerate; otherwise
one must count the multiplicities.}

\bigskip
\noindent{\it The General Story}

The generalization of this is as follows.  Suppose that one is interested
in counting {\it without signs} the solutions of some equations
\eqn\gunk{s^a(u^i)=0,\, a=1\dots n}
in $n$ variables $u^1\dots u^n$.
The $u^i$ are, in general, coordinates on some manifold $M$ and, globally,
the $s^a$ define a section of a vector bundle $V$.
Introduce another set of variables
$y_a, \, a=1\dots n$, and extend $s^a$ to arbitrary functions
$s^a(u^i,y_b) $ such that\foot{Geometrically, let $\widehat M$ be the total
space of the bundle $V\to M$, and $\widehat V$ the pullback of $V$ to
$\widehat M$.  The extended functions $s^a(y^i,y_b)$ define a section
of $\widehat V\to\widehat M$.  Other formulas below have analogous
interpretations.}
\eqn\ung{s^a(u^i,0)=s^a(u^i)}
Let $h_i(u^j,y^a)$
be any $n$ additional functions such that
\eqn\unk{h_i(u^j,y_b)=\sum_ay_a{\partial
s^a\over \partial u^i}+O(y^2).}
Consider the system
of equations
\eqn\yurmo{ s^a=h_i=0.}
Every solution of the original system \gunk\ gives by taking $y^j=0$
a solution of the extended system \yurmo.  Solutions of this kind
have $\epsilon=+1$ (since $\partial h_i/\partial y_a=\partial s^a/\partial
u^i$ and the determinant in \invfor\ is positive).  If these are the only
solutions,
then \udd\ reduces to
\eqn\reddu{Z=\sum_{P_\alpha}1= N,}
with $N$ the total number of solutions of the original equation.
Thus {\it if all solutions of the extended system are at $y^i=0$,
then the number of solutions of the extended system, weighted by sign,
is the same as the total number of solutions of the original system}.
Under this restriction, therefore, we do get an integral representation
for the {\it unweighted} number of solutions of the original equation.

Of course, the utility of all this depends on finding
an interesting situation in which there is a suitable
vanishing theorem.  This paper will be based on such a situation.

\bigskip
\noindent{\it{Generalization}}

More generally, we will need the following variant of the above
construction.  Consider a system of $d'$ equations
\eqn\kk{s^a=0,\,\,\,a=1\dots d'}
for $d$ variables $u^1\dots u^d$, with $d'<d$.  If everything
is sufficiently generic, the solutions will consist of disjoint, smooth,
compact manifolds
$M_\alpha$ of dimension $d-d'$ and nondegenerate in the sense of \locoform.
Introduce $d'$ new variables $y_b$, and extend the $s^a$ to
functions $s^a(u^i,y_b)$.
Introduce
$d$ additional functions
$h_i$ such that $h_i=\sum_by_b(\partial s^b/\partial u^i)+O(y^2)$.
Consider the system
\eqn\longsyste{s^a=h_i=0 }
of $d+d'$ equations for $d+d'$ unknowns $u^i,\,y_a$.

The partition invariant $Z$ is the
number of solutions of this system weighted by signs.
The solutions that have $y=0$
are simply the solutions of the original system $s^a=0$, and thus
consist of the union of manifolds $M_\alpha$. Suppose that there is
a vanishing theorem that ensures that all solutions of \longsyste\
are at $y=0$. Then
$Z$ can be evaluated using \hurf, and is
\eqn\ungo{Z=\sum_\alpha \chi(V_\alpha), }
where $V_\alpha$ is the bundle of antighost zero modes along $M_\alpha$.
The signs $\epsilon_\alpha$ are all $+1$ because of cancellation
between $u$ and $y$.

Moreover, in this situation, $V_\alpha$ has a special interpretation.
Since there are two sets of equations $s^a$ and $h_i$, there are
two sets of antighosts, say $\chi^a$ and $\tilde \chi_i$.
The assumption that the $M_\alpha$ have the expected dimension
$d-d'$ and are nondegenerate
means that the original antighosts $\chi^a$ have no zero modes.
As for the $\tilde \chi_i$, they can be analyzed as follows:
they are cotangent to the original target space $M$, and
their
zero modes are cotangent to the space $M_\alpha$ of classical solutions.
So $V_\alpha$ is the cotangent bundle of $M_\alpha$,
and  the Euler class $\chi(V_\alpha)$ is the same
as the Euler characteristic $\chi(M_\alpha)$.
Hence we can rewrite \ungo\ as
\eqn\tungo{Z=\sum_\alpha\chi(M_\alpha) =\chi({\cal W}),}
with ${\cal W}=\cup_\alpha M_\alpha$
the space of solutions of the original equations $F(u)=0$.
So under the above-stated restrictions, the Euler characteristic
of the space of solutions of a system of equations can be given an
integral representation.

\subsec{Gauge Invariance}

The situation that we really want is a gauge invariant
version of the above.  So let us explain how to incorporate
a symmetry group in the finite dimensional model.  First we consider
the general construction that counts solutions weighted by signs
and then the more special construction that (given a vanishing theorem)
eliminates the signs.

To incorporate a group action in these models, we assume
that a compact Lie group $G$ acts on $M$ and $V$ (preserving all the data
such as the metric $g$ on $V$ and the section $s$).
If $G$ has dimension $t$, we take the dimension of
$M$ to be $d=2n+t$ and the rank of $V$ to be $\tilde d=2n$.
Thus
\eqn\dimcond{d-\tilde d - t= 0}

We introduce a field $\phi$, in the adjoint representation
of $G$, with ghost number $U=2$.  We write $\phi=\sum_{x=1}^t \phi^xT_x$
with $T_x$ a basis of the Lie algebra of $G$.
The action of $T_x$ on the manifold $M$ is described by a vector field
$U_x{}^i$, and the lifting of $T_x$ to act on the bundle $V$
is described by the action on a section $\chi: \delta \chi^a=U_x{}^iD_i
\chi^a+Y_x{}^a{}_b\chi_b$, with some $Y$. (Of course $D_i$ is the
covariant derivative with respect to the connection $A_i$ on $V$.)

For the fermionic symmetry, we take
\eqn\deltaphi{\delta\phi = 0.}
The transformation laws of other fields are as follows.
\translaw\ is modified to
\eqn\anslaw{\eqalign{ \delta u^i & = i\epsilon\psi^i \cr
                   \delta\psi^i & = \epsilon \phi^x U_x{}^i \cr}}
while \otherlaw\ is replaced by
\eqn\notherlaw{
\eqalign{\delta \chi^a & =  \epsilon H^a-\epsilon
\delta_0 u^iA_i{}^a{}_b\chi^b\cr
           \delta H^a    & =\epsilon\delta_0 u^i A_i{}^a{}_bH^b
-{\epsilon\over 2}\delta_0 u^i\delta_0 u^j F_{ij}{}^a{}_b \chi^b+i\epsilon
\phi^x Y_x{}^a{}_b\chi^b.
\cr}}  (This is just the gauge-covariant version of $\delta\chi=\epsilon
H,\,\,\delta H=-\epsilon[\phi,\chi]$, which is analogous to \anslaw.)
It is no longer the case that $Q^2=0$; rather, $Q^2$ is equal to a gauge
transformation with generator $\phi^xT_x$.  (This structure gives
a model of what mathematically is called equivariant cohomology; see
\refs{\mq,\atj}.)

To make it possible to write a Lagrangian, we introduce another multiplet
$(\bar \phi,\eta)$, in the adjoint representation of $G$, with
ghost number $U=(-2,-1)$, and a transformation law analogous to the above:
\eqn\nurmo{\eqalign{\delta \bar\phi & = i\epsilon \eta\cr
                \delta\eta & = i\epsilon[\phi,\bar\phi].\cr}}
We also pick a $G$-invariant metric $g_{ij}$ on $M$.

Now, set
\eqn\neww{W={1\over 2\lambda}(\chi,H+2is) +{1\over \lambda'}
\bar\phi^xg_{ij}U_x{}^i\psi^j
+W',}
with $\lambda'$ a new small parameter and
$W'$ consisting of possible non-minimal terms.
Then define the Lagrangian
\eqn\newl{\eqalign{
L=\delta_0 W = &{1\over 2\lambda }(H,H-2is)+{1\over \lambda}g_{ab}
\chi^a {\partial s^b
\over \partial u^i}\psi^i -{1\over 2\lambda}F_{ijab}\psi^i\psi^j\chi^a\chi^b\cr
&-{i\over 2\lambda}\chi^a\chi^b\phi^xY_{x\,ab}
+{i\over\lambda'}\bar\eta^xg_{ij} U_x{}^i\psi^j+{1\over\lambda'}\bar\phi^x
g_{ij}U_x{}^iU_y{}^j\phi^y+\cr &{i\over 2\lambda'}\bar\phi^x(\partial_kU_{x\,i}
-\partial_iU_{x\,k})\psi^k\psi^i+\delta_0 W'.\cr}}
Notice that $\delta L = \delta^2 W= \phi^xT_x(W)=0$, as $W$ is gauge-invariant.

Now we wish to study an integral
\eqn\wishto{Z={1\over {\rm Vol}(G)\cdot (2\pi)^d(-i)^t}
\int d\phi\, d\bar\phi\,d\eta \,du\,d\psi\,d\chi\,dH
\,\,e^{-L}.}
Notice that, while all other degrees of freedom are in bose-fermi
pairs and so have a natural measure, $\phi$ is unpaired.  To make
sense of the integration measure in \wishto, we pick a translation-invariant
measure $d\phi$ on the Lie algebra of $G$.  This determines a measure
on the group manifold, and by ${\rm Vol}(G)$ we mean the volume of $G$
with that measure.  The choice of measure therefore cancels out of
the ratio $d\phi/{\rm Vol}(G)$, so \wishto\ has no unspecified or arbitrary
normalization.  It is convenient to use the chosen measure on the Lie
algebra to define separately $d\bar\phi$ and $d\eta$ (whose product, in any
case, is naturally defined without any choices).  Similarly, we take the
Riemannian metric $g_{ij}$ on $M$ to define the separate
$u$ and $\psi$ measures
-- and measures on any subspaces of $u$'s or $\psi$'s.

The standard BRST argument shows that
the integral in \wishto, if sufficiently
well convergent, is a deformation invariant and depends only on the manifold
$M$ and bundle $V$, and the $G$ action on them.
There is one basic case in which this invariant is easy to determine.

That is the case
in which $G$ acts freely on $M$.
When that happens, it is possible to reduce the gauge invariant problem
on $M$ to an ordinary problem, without gauge invariance, on the quotient
$M'=M/G$.  The bundle $V$ and section $s$ will be replaced by the objects
$V'$ and $s'$ over $M'$ which pull back to $V,s$ over $M$.  The
integral \wishto\ will reduce to the one on $M'$ that counts
-- with signs -- the number of solutions
of $s'=0$ on $M'$.  Or equivalently, it counts with signs
the number of solutions
of $s=0$ on $M$, up to gauge equivalence.

To justify these claims, note first that the statement that $G$ acts
freely on $M$  implies
that for any complex-valued $\phi^x\not= 0$,
the vector field $\theta=\phi^xU_x{}^i$ has no zeroes anywhere on $M$
and therefore the $\phi$ kinetic energy in \newl, which is
$|\theta|^2$, is strictly positive.  Thus there are no $\phi$ or
$\bar\phi$
zero modes.
(If $G$ does not act freely at least locally,
there are instead $\phi$ and $\bar \phi$ zero modes, at some points on
$M$.)

Turning this around, the expression $g_{xy}=U_x{}^iU_y{}^j
g_{ij}$ is for each $G$ orbit in $M$ a positive definite metric on the Lie
algebra of $G$; it also determines a metric on the $G$ manifold.
This is simply the metric that comes from the fact
that (i) $G$ acts freely on $M$, so the orbits are copies of $G$;
(ii) the orbits are embedded in $M$ and get an induced metric from
the Riemannian metric on $M$.
The ratio
of the induced measure to the one that was chosen in defining \wishto\ will
be called $\sqrt{\det(g_{xy})}$.  (That is really the definition of
$\det(g_{xy})$; as $g_{xy}$ is a quadratic form rather than a matrix,
its determinant only makes sense as a number if there is a pre-existing
measure to compare to.)
If we call the volume of $G$ with
the metric induced from the embedding ${\rm Vol}'(G)$, then of course
\eqn\nishto{{\rm Vol}'(G)=\sqrt{\det(g_{xy})}\cdot {\rm Vol}(G).}

Since the $\phi,\bar\phi$ kinetic energy is nondegenerate,
one can perform the Gaussian integral over $\phi$ and $\bar\phi$.
It gives a factor of
\eqn\funnyfac{(2\pi\lambda')^{t}\det(g_{xy})^{-1}.}
Similarly, one can integrate over $\eta$, which appears linearly in the
Lagrangian.  This gives a factor
of
\eqn\fivefac{\left({-i\over\lambda'}\right)^t\delta(g_{ij}U_x{}^i\psi^j).}

The delta function in \fivefac\ has the following interpretation.
The vectors $U_x{}^i$ generate the $G$ action and so are tangent to the
$G$ orbits on $M$.  The delta function in \fivefac\ thus projects
onto the components of $\psi$ that are normal to the group orbits.
The surviving components can be interpreted as giving a section of the
pullback to $M$ of the tangent bundle of $M'=M/G$.

In fact, we can divide by the free action of $G$ and reduce what is left
of \wishto\ to an integral on $M'$.
{}From integrating over the $G$ orbits, we get a factor of ${\rm Vol}'(G)$,
which is given in \nishto.
The delta
function in \fivefac\ is $\sqrt{\det {g_{xy}}}\cdot \delta(\tilde\psi)$
where $\tilde\psi$ are orthonormal components of $\psi$ tangent to the
$G$ orbits.   With the Riemannian measure for $\tilde\psi$,
\eqn\longo{\int d\tilde\psi\,\,\delta(\tilde\psi)=1.}

In this process of eliminating components of $u$ tangent to the $G$ orbits
by dividing by $G$, and eliminating components of $\psi$ tangent
to the orbits by using the delta function, the factors of $\lambda'$
and $\det (g_{xy})$ and extra factors of $2\pi$ cancel out.
What remains is a $Q$-invariant integral on $M'$ of the standard
type, with the standard measure.  $V$ and $s$ ``go along
for the ride'' in the above manipulations, and so are simply
replaced on $M'$ by the objects
$V'$ and $s'$ that pull back to $V$ and $s$ on $M$.

So we can carry over all of our analysis of \uitv.  When $G$ acts
freely on $M$, the invariant $Z$ defined by the integral in
\wishto\ simply counts, with signs, the solutions of $s'=0$ on $M'$,
or, equivalently, the gauge orbits of solutions of $s=0 $ on $M$.

\bigskip
\noindent{\it Counting Solutions Without Signs}

If we want to find a way to count gauge orbits of solutions
{\it without signs}, we must imitate the special construction that led
to \reddu\ and \tungo.

We recall that in that discussion, we started with fields $u^i,\,i=1\dots d$
and
equations $s^a(u)=0,\,\,a=1\dots d'$, with $d'<d$.
Then we added dual variables $y_a$, extended the $s^a$ to possibly
depend on $y$,  and added dual equations
$h_i=0$.
Among other things, these steps gave a system with equally many fields
and equations.  When the appropriate vanishing theorem holds,
the partition function was the Euler characteristic
of the space $\tilde M$ of solutions of the original equations $s^a(u)=0$.

In the $G$-invariant case, that is not quite right, because (assuming
$G$ has dimension $t$ and acts freely) by dividing by $G$ one could
remove $t$ degrees of freedom from $u$, so that there are effectively
only $d-t$ fields to begin with.
So to balance the fields and equations,
one would need not $d$ but $d-t$ dual equations $h$.
Moreover, instead of the $h$'s being dual to the tangent space of
$M$, they should be dual to the pullback to $M$ of the tangent space
of $M'=M/G$.  The latter condition will ensure that -- after we descend
to $M'$ by dividing by $G$ -- we will arrive at the construction that we
have already analyzed.

A suitable set of $d-t$ dual equations can be constructed as follows.
Start with any $G$-invariant set of $d$ functions $h_i$ such that
\eqn\hobo{h_i=y_a{\partial s^a\over\partial u^i}+O(y^2).}
Let
\eqn\letthem{L_x= U_x{}^ih_i.}
Let $\Pi$ be the projection operator (using the metric on $M$) onto
the subset of $h$'s for which $L=0$, and let
\eqn\newones{\tilde h=\Pi( h).}
The desired set of $d-t$ equations is $\tilde h=0$.

Suppose that a vanishing theorem ensures that the solutions of $s=\tilde h=0$
are all nondegenerate and have $y=0$.
The partition invariant for the system consisting of fields $u^i,y_a$,
equations $s=\tilde h=0$, and symmetry group $G$
can in that case be evaluated by dividing by $G$ and using \tungo.
It equals the Euler
characteristic of ${\cal W}/G$, where ${\cal W}$ is the space of
solutions of the original equations $F(u)=0$ for $u\in M$.

\bigskip
\noindent{\it Locality}

In field theory, however, the projection operator $\Pi$ may be nonlocal
and for that reason its use is best avoided.
Instead of using this projection operator to reduce the number of
equations, one can increase the number of fields, as follows.
The assertion that $\Pi (h)=0$ is equivalent to the assertion that
there exists an adjoint-valued function $C^x$ on $M$ such that
\eqn\norf{h_j +C^x U_x{}^ig_{ij}= 0 . }
Let us include the $C^x$ as additional fields of $U=0$.  Their ghost
number one partners will be called $\zeta^x$.
The multiplet is the standard one $\delta C=i\epsilon\zeta,
\,\,\delta \zeta=i\epsilon[\phi,C]$.
Let $k_i$ be any functions such that
\eqn\oggo{k_j=h_j +C^x U_x{}^ig_{ij}+O(y^2,Cy,C^2).}

Consider the system of equations $s^a=k_i=0$ for fields $u,y,C$, with
group action $G$.
Suppose it is the case that there is a vanishing theorem ensuring
that the solutions are all at $y=C=0$.
Then the partition invariant $Z$ for this system,
upon integrating out the $(C,\zeta)$ multiplet, reduces to the
partition function for the system with fields $u,y$ and equations
$s^a=\tilde h_i=0$.  This reflects the fact that the equations
$k=0$ with $C$ present are equivalent to the equations $\tilde h=0$
with $C$ absent.
Under these conditions, we can again invoke \tungo\ and conclude
\eqn\longo{Z=\chi({\cal W}/G),}
with, again, ${\cal W}$ the space of solutions of the original equations.

We should stress that the derivation of this formula has assumed
that $G$ acts freely on $M$ and that ${\cal W}$ is a smooth nondegenerate
compact manifold of the expected dimension.  When these assumptions
fail, the integral must be examined more closely.

\subsec{Gauge Theories In Four Dimensions}

At last we have assembled the needed tools, and we turn to our
real interest -- four dimensional gauge theories.

We take $X$ to be an oriented four-manifold with local coordinates $x^i$,
$i=1\dots 4$.  We
pick a finite dimensional gauge group $G_0$, and a $G_0$ bundle $E$ with
a connection $A$ and curvature $F_{ij}=\partial_iA_j-\partial_jA_i+[A_i,A_j]$.
The curvature can be decomposed in self-dual and anti-self-dual
pieces,
\eqn\fplus{F_{ij}{}^{\pm}={1\over 2}\left(F_{ij}\pm {1\over 2}
\epsilon_{ijkl}F^{kl}\right)
.}
We take $G$ to be the group of all gauge transformations of this bundle,
acting on $A$ in the standard fashion, $D_i\to h^{-1}D_ih$, where
$D_i=\partial_i+A_i$.

If $G_0$ is connected and simply-connected, the bundle $E$ is determined
topologically by a single integer called the instanton number.
For $G_0=SU(N)$ this is
\eqn\defk{k={1\over 8\pi^2}\int_M\Tr\, F\wedge F,}
with $\Tr$ the trace in the $N$ dimensional representation.
$k$ can be an arbitrary integer.\foot{For any compact simple $G_0$,
one defines $k$ by a curvature integral analogous to \defk, normalized
so that if $G_0$ is replaced by the universal cover $\widehat G_0$ of its
identity component, then $k$ is an arbitrary integer.}  If $G_0$ is not
simply connected, $k$ is still defined, but may not be integral;
for instance, for $G_0=SO(3)$, $k\in {1\over 4}{\bf Z}$.

\nref\ahs{M. F. Atiyah, N. Hitchin, and I. M. Singer,
``Self-Duality In Riemannian Geometry,'' Proc. R. Soc. Lond.
P{\bf A362} (1978) 425.}
To apply the above constructions, we take $M$ to be the space of connections
on $E$.  We take $V$ to be the bundle of self-dual two forms with
values in the adjoint-representation of $E$.  A natural section
of $V$ is given by the $+$ part of the curvature:
\eqn\natsection{s(A)= F^+(A).}
A zero of this section is called an instanton.
The space ${\cal W}$ of instantons -- solutions of $s=0$ -- is
infinite dimensional because of gauge equivalence.  However \ahs,
the moduli space of instantons ${\cal M}={\cal W}/G$ is finite
dimensional.  Under assumptions analogous to the ones that we have
made in discussing the finite dimensional examples, the dimension
of the moduli space is \ahs\
\eqn\dimm{{\rm dim}({\cal M})=4k h(G_0)-{\rm dim}(G_0)(1+b_2{}^+).}
Here $b_2{}^+$ is the dimension of the space of self-dual harmonic
two forms, and $h(G_0)$ is the dual Coxeter number of $G_0$ (equal to $N$ for
$G_0=SU(N)$).

In finite dimensions, under our usual assumptions, ${\dim({\cal M})}$
is equal to $d-\tilde d -t$,  with $d$ and $t$ the dimensions
of $M$ and $G$ and $\tilde d$ the rank of $V$.  In the gauge theory
problem, $d$, $\tilde d$, and $t$ are all infinite, but
the difference $d-\tilde d-t$ makes sense as the index of a certain
elliptic operator that appears in the moduli problem \ahs,
and still equals $\dim({\cal M})$.

If $E$ is chosen so that ${\rm dim}({\cal M})=0$, we are in the
situation in which one can try to count, with signs, the number of
points in ${\cal M}$.  From our finite dimensional
discussion, we know exactly how to formulate an integral that
will compute this quantity.
We need a system with the following supermultiplets.

At ghost numbers $0,1$ we have the gauge fields $A_i$ and
ghosts $\psi_i$.  The ghosts have the gauge and Lorentz
quantum numbers of the fields
and so are a one-form with values in the adjoint representation.
At ghost numbers $-1,0$, we have antighosts
$\chi$ and auxiliary fields $H$; they have the quantum numbers of the
equations, and so are self-dual two-forms with values in the adjoint
representation.
At ghost number $2$, we have a field $\phi$ with the quantum numbers
of a generator of gauge transformations -- that is, $\phi$ is a scalar
field with values in the adjoint representation.
At ghost numbers $-2,-1$ are the conjugate $\bar\phi$ of $\phi$
and its fermionic partner $\eta$.

The propagating bosonic fields are thus $A,\phi,\bar\phi$ -- a gauge
field and a complex scalar in the adjoint representation.  This is
precisely the bosonic part of the field content of $N=2$ super
Yang-Mills theory.  The fermionic fields $\psi$, $\chi$, and $\eta$,
transforming as $({\bf 2},{\bf 2})$, $({\bf 1},{\bf 3})$,
and $({\bf 1},{\bf 1})$ under $SU(2)_L
\times SU(2)_R$, similarly coincide with the fermions of the topologically
twisted $N=2$ theory, as described at the outset of this section.

The bosonic part of the action can be read off
from \newl.  In doing so, we set $\lambda=\lambda'=e^2$ with $e$ the
gauge coupling.
We also note that $U_y{}^j\phi^y$ is the change in the field under
a gauge transformation generated by $\phi$ so in the gauge theory
case is determined by the formula $\delta A_i=-D_i\phi$.  Finally,
after eliminating the auxiliary field $H$, the bosonic part of the action is
\eqn\bosac{L ={1\over 2e^2}\int_X\Tr\left( |F^+|^2 +|D_i\phi|^2\right).}
If we bear in mind that
\eqn\osac{\int_X|F^+|^2={1\over 2}\int_X\Tr F_{ij}F^{ij}+{1\over 4}
\int_X \epsilon^{ijkl}\Tr F_{ij}F_{kl},}
we see that -- except for the last term, which is a topological invariant,
a multiple of the instanton number $k$ -- the
bosonic part of the action \bosac\ is just the standard kinetic energy.
{}From $N=2$ super Yang-Mills theory, we are still missing a bosonic
interaction
\eqn\bosint{{1\over 2e^2}\Tr [\phi,\bar\phi]^2.}
This will appear if for $W'$ in \neww\ we take
\eqn\wprime{W'={1\over 2e^2}\Tr \,\,\eta [\phi,\bar\phi].}

The whole topological Lagrangian \newl\ is indeed, in this situation,
the twisted version of $N=2$ super Yang-Mills theory
that we recalled at the beginning of this section.
The topological meaning of this theory
is now clear, at least
when the bundle $E$ is such that the expected dimension of instanton
moduli space is zero: this theory computes with signs the number
of instantons, up to gauge transformation.

For other $E$, the partition
function of the theory vanishes because of ghost counting.  However,
for ${\rm dim}({\cal M})>0$, there are interesting BRST-invariant
operators that can have non-trivial correlation functions.  They are
described in \tqft.  Their
correlation functions are in fact the celebrated Donaldson invariants
of four-manifolds.

\bigskip
\noindent{{\it Euler Characteristic Of Moduli Space}}

Closer to our interests in this paper is to find a way to eliminate
the minus signs and compute, when ${\rm dim}({\cal M})=0$, the total
number of instanton solutions, up to gauge transformations.  More
generally, when ${\rm dim}({\cal M})>0$, we want to compute the
Euler characteristic of instanton moduli space.

{}From our finite dimensional discussion, we know how to do this as well:

(1) We introduce conjugate fields with the quantum numbers of the equations.
In the present problem, the conjugate
fields are a self-dual two-form $B^+$ with values in the adjoint
representation.  They come with
ghosts $\tilde \psi^+$ with the same gauge and Lorentz quantum numbers.

(2) We introduce additional fields $C$ with the quantum numbers of the
gauge generators.  In the present problem, $C$ is a scalar field with values
in the adjoint representation of $G$.  $C$ comes with a ghost $\zeta$
with similar quantum numbers.

(3)  We extend the original equations $F^+=0$ to have possible dependence
on the new fields.  Success depends on whether the extension can be chosen
so that, eventually, a suitable vanishing theorem will hold.  In the
present case, a suitable choice is to modify the section $s$ to
\eqn\news{s_{ij}=F^+{}_{ij}+{1\over 2}[C,B^+{}_{ij}]+{1\over 4}
[B^+_{ik},B^+_{jl}]g^{kl}.}
The additions are needed to ultimately get a vanishing theorem -- and
compare to twisted $N=4$ super Yang-Mills theory --  as we
will see.

(4)  Finally, one needs conjugate equations.  Their general structure
is given in \oggo; up to first order in $B$ and $C$, they are uniquely
determined, but the higher order terms are arbitrary.  In the present
problem, the higher order terms are best set to zero.  So the conjugate
equations, found by interpreting \oggo\ in the present situation,
are $k=0$ with
\eqn\defk{k_j=D^iB^+{}_{ij}+D_jC.}
Associated with the conjugate equations, one adds new antighosts
$\tilde\chi_j$, and new auxiliary fields.

Let us examine the field content of the theory.
The bosonic fields are the gauge field $A$, a self-dual two-form $B$,
and three scalars $C,\phi,\bar\phi$.  The fermionic fields are
two self-dual two-forms, $\chi$ and $\tilde\psi$, two vectors $\psi$
and $\tilde \chi$, and two scalars, $\eta$ and $\zeta$.  This is precisely
the field content of a certain topologically twisted version of
$N=4$ super Yang-Mills theory, as described at the outset of this
section.

Now let us work out the bosonic part of the action \newl.
Apart from comparing to the $N=4$ theory, this will enable us to see
the conditions for a vanishing theorem.  The crucial terms
are the square of the section $(|s|^2+|k|^2)/2e^2$.
Integration by parts and use of the Jacobi identity with some slightly
delicate cancellations
leads to the following identity:
\eqn\usefulid{\eqalign{
{|s|^2+|k|^2\over 2e^2}&={1\over 2e^2}\int_Xd^4x\sqrt g
\Tr\left(\left(F^+{}_{ij}
+{1\over 4}[B_{ik},B_{jl}]g^{kl}+{1\over 2}[C,B_{ij}]\right)^2+\right.\cr
&~~~~~~~~~\left.
+\left(D^jB_{ij}+D_iC\right)^2\right) \cr
& =
{1\over 2e^2}\int_Xd^4x\sqrt g
\Tr\left(F^+{}_{ij}{}^2+{1\over 4}(D_lB_{ij})^2+(D_iC)^2
+{1\over 16}[B_{ik},B_{jk}][B_{ir},B_{jr}]\right.
\cr
&~~~~~~~~~\left.  +
{1\over 4}[C,B_{ij}]^2
+{1\over 4}B_{ij}\left({1\over 6}(g_{ik}g_{jl}-g_{il}g_{jk})R+W^+{}_{ijkl}
\right)B_{kl}\right), \cr}}
with $R$ the scalar curvature of $X$ and $W^+$ the self-dual part of the
Weyl tensor.

\bigskip
\noindent{\it The Vanishing Theorem}

Now let us look for a suitable vanishing theorem.
Of course, if there is no vanishing theorem, the theory still
exists.  It is just harder to study, though conceivably richer.
But to study the theory, it is certainly important to understand
whatever vanishing theorems do exist.

According to our general discussion,
the appropriate vanishing theorem would assert
that the solutions
of $s=k=0$ all have $B=C=0$; then it follows that the topological
partition function associated with the equations $s=k=0$ has for
its partition function the Euler characteristic of instanton moduli space.

The most obvious inference from
\usefulid\ is that if
the metric is such that
\eqn\poppo{\sum_{ijkl}
B_{ij}\left({1\over 6}(g_{ik}g_{jl}-g_{il}g_{jk})R+W^+{}_{ijkl}
\right)B_{kl}>0}
for any non-zero $B$,
then any solution of $s=k=0$ has
\eqn\huzzo{B=D_iC=F^+=0.}
This is slightly less than we hoped for because we learn only that
$C$ is covariantly constant, not zero.  However, the condition $D_iC=0$
has the following significance: it means that $C$ is covariantly constant
and generates a gauge transformation that leaves the gauge field invariant,
so that the gauge group $G$ does not act freely on the space of solutions
of $s=k=0$.\foot{A
gauge field that admits a nonzero solution $C$ of $D_iC=0$ is
called reducible.  Such a gauge field can be interpreted
as a connection with values in the subgroup $G_0'$ of the gauge group
$G_0$ that leaves $C$ invariant.  For instance, for $G_0=SU(2)$, $G_0'$ will
be the abelian group $U(1)$.}
Instanton moduli space ${\cal M}$ is singular at such points
and our general assumptions fail there.
Moreover, when ${\cal M}$ is singular, one would want to specify
exactly what one means by the Euler characteristic.
Our general formal arguments really need an extension
(which we do not know how to give) when such singularities occur.

At least informally, though, \huzzo\ can be described by saying that
when \poppo\ is positive definite, the argument identifying the partition
function with $\chi({\cal M})$ is valid if one treats singularities of
${\cal M}$
properly. This is to be contrasted with the generic situation
that would prevail in the absence of a vanishing theorem: then ${\cal M}$ might
be perfectly smooth, yet there might be solutions of $s=k=0$
with $B,C\not= 0$ having nothing to do with instantons.

The vanishing theorem that we have just stated, which is a nonlinear
version of the one in \ahs, applies, for instance,
to the four-sphere with its standard metric (for which $W=0$ and $R>0$).
However, there is a severe topological
limitation on its applicability,
namely $b_2{}^+$ (the dimension of the space of self-dual harmonic
two forms) must vanish.  To see this, note the following identity
for a (neutral) self-dual two-form $w$:
\eqn\nuseful{\int_X(D^iw_{ij})^2={1\over 4}\int_X\left((D_iw_{jk})^2+
w_{ij}\left({1\over 6}(g_{ik}g_{jl}-g_{il}g_{jk})R+W^+{}_{ijkl}
\right)w_{kl}\right).}
(This identity was part of the derivation of
\usefulid.)
If $w$ is harmonic, $D^iw_{ij}=0$, so if the quadratic form in \poppo\
is strictly positive, then $w=0$.  So if that quadratic form is
positive, then $b_2{}^+=0$.

Examples with $b_2{}^+=0$ are very restricted and will not be useful
in this paper because our computations will all involve gauge groups that
are locally isomorphic to a product of $SU(2)$'s. For $SU(2)$ the dimension
of instanton moduli space is $8k-3(1+b_2{}^+)$, and so is odd if
$b_2{}^+=0$.\foot{For $SO(3)$, $4k$ is integral and the same statement
holds.}  The partition function is therefore zero for $SU(2)$
if $b_2{}^+=0$.\foot{If instanton moduli space is an odd dimensional
smooth compact manifold (without boundary),
its Euler characteristic vanishes.
Normally, those assumptions are too optimistic.  However, the
Euler characteristic appears in our formulas as a curvature integral
-- this is explicit in \stanform\ -- and so the Euler characteristic
in the sense we want
vanishes when ${\cal M}$ is of odd dimension.}
Manifolds with $b_2{}^+=0$ might be of interest with
other gauge groups such as $SU(3)$.

We will therefore need some further vanishing theorems, and we will
discuss several variants.  As a preliminary, let us consider the important
case in which the quadratic form in \poppo\ is positive {\it semi}-definite.
Then  from \nuseful\ we learn that a harmonic self-dual two-form
$w$ is covariantly constant.  The existence of such a nonzero $w$ reduces
the holonomy group of $X$, and there are two possibilities:
(1) If $b_2{}^+=1$, so there is essentially a single $w$, the holonomy
group is reduced to $U(2)$; then $X$ is K\"ahler and $w$ is the K\"ahler form.
(2) If $b_2{}^+>1$, the holonomy is reduced still further.  The only
possibility is that $X$ is hyper-K\"ahler and $b_2{}^+=3$.
We will later discuss more precise vanishing theorems for such manifolds.

The following simple fact will be useful.
Though the right hand side of \usefulid\ is not manifestly positive
(unless one is given some information about the curvature of $X$),
the equation itself shows that the right hand side is non-negative and
vanishes when and only when $s=k=0$.  Now the right hand side of \usefulid\
is invariant under
\eqn\funsym{\tau:C\to -C.}
(This is part of the $SU(2)$ global symmetry of the $N=4$ super
Yang-Mills theory mentioned in the introduction.)
Hence, given any solution of $s=k=0$, we get
a new solution by replacing $C$ by $-C$.  It follows that any solution
of $s=k=0$ has
\eqn\unsym{D_iC=[C,B]=0,}
{\it without any assumption on the curvature of $X$}.
Thus, either $C=0$ or $C$ generates a gauge transformation that
acts trivially on both the gauge connection and $B$.

If the gauge group is (locally) $SU(2)$ (or a product of $SU(2)$'s)
 we can make a more precise
statement since if $C\not= 0$, it breaks $SU(2)$ to an abelian subgroup
$U(1)$.  If $C\not= 0$, then $[C,B]=0$ implies $B$ lies in the same $U(1)$
so $[B_{ij},B_{kl}]=0$, and hence
$s=0$ implies that $F^+=0$.  But if $b_2{}^+>0$ (which we may as well
assume, if the gauge group is $SU(2)$), then for a generic metric on
$X$, there are no abelian instantons.\foot{The first Chern class of an
abelian instanton is a two-dimensional cohomology class of
$X$ that (i) is integral, and (ii) is an eigenstate of the Hodge $*$ operator
with eigenvalue $-1$.  (The minus sign is because we take the instanton
equations to be that $F^+=0$, so that the non-zero part of $F$ is $F^-$.)
If $b_2{}^+>0$ (which means that the subspace
of $H^2(X)$ with eigenvalue $+1$ of $*$ is non-empty) then for a
generic metric on $X$, there are no non-zero cohomology classes obeying
conditions (i) and (ii).}
  So for such gauge groups,
we can assume that $C=0$.

\bigskip
\noindent{\it Comparison To $N=4$}

Before resuming the discussion of vanishing theorems,
it will be helpful to make a more precise comparison of the
topological theory that we have been considering so far to
$N=4$ super Yang-Mills theory.

Apart from terms involving $\phi$ and $\bar\phi$, the bosonic
part of the action in the topological theory
is simply the right hand side of
\usefulid.  If we work on flat ${\bf R}^4$ and set $B_{0i}=B_i$,
the right hand side of \usefulid\ is
\eqn\mcgur{{1\over 2e^2}\int_X\Tr\left(F^+{}_{ij}{}^2+
(D_iB_{j})^2+(D_iC)^2
+\sum_{i<j}[B_{i},B_{j}]^2+\sum_j[C,B_{j}]^2\right).}
This has an $O(4)$ symmetry rotating $C,B_i$; it is the subgroup of
the underlying $O(6)$ symmetry of the $N=4$ theory that does not
act on $\phi,\bar\phi$.  In fact, \mcgur\ is precisely the bosonic
part \plsk\
of the action of the $N=4$ theory with two of the scalars
-- $\phi$ and $\bar \phi$ -- set to zero, and a $\theta$ term added.

The bosonic terms involving $\phi$ can be found in a fashion similar
to our discussion of the $N=2$ theory. The $\phi$ kinetic energy
arises as in the discussion of \bosac.  The other bosonic interactions,
namely
\eqn\huj{{1\over 2e^2}\int_Xd^4x\sqrt g\Tr\left([C,\bar\phi][C,\phi]
+[B_i,\bar\phi][B_i,\phi]-{1\over 4}[\phi,\bar\phi]^2\right)}
originate as in \bosint\ by adding to $W'$ some terms with
the structure $\Tr\,\,\eta_T[T,\bar\phi]$ where $T$ ($=\,\bar\phi,\,
C$, or $B$)
is a bosonic field that transforms into $\eta_T$ under the fermionic symmetry.
In this way one gets a topological theory whose bosonic part
on flat ${\bf R}^4$ agrees precisely with $N=4$
super Yang-Mills theory.  The fermionic part of the action similarly
coincides on flat ${\bf R}^4$ with the $N=4$ theory; it could hardly
be otherwise given the supersymmetry.

\bigskip
\noindent
{\it Vanishing Theorems On K\"ahler Manifolds}

On a general four-manifold $X$, the topological theory differs from the
$N=4$ theory by the twisting that shifts the spins and by the curvature
coupling on the right hand side of \usefulid.  Generally, these couplings
break the $O(4)$ symmetry that we noted in \mcgur, leaving only the
${\bf Z}_2$ in \funsym\ and a similar ${\bf Z}_2$ acting on $B$.
There is an important case in which a larger subgroup
survives.  This is the case that $X$ is a K\"ahler manifold.
In that case, one can naturally decompose the self-dual two-form $B$
into components of type $(2,0)$, $(1,1)$, and $(0,2)$.  We will
write the $(1,1)$ piece as $b\,\omega$, with $\omega$ the K\"ahler form and
$b$ a scalar field,
and call the $(2,0)$ and $(0,2)$ pieces $\beta$ and $\bar \beta$.

Of the $SO(4)$ symmetry of \mcgur\ on a flat manifold, an $O(2)$
that rotates $b$ and $C$ survives on a K\"ahler manifold.  In fact,
$b$ and $C$ are both scalars and have kinetic energy of the same form.
Also,
the $(1,1)$ piece of $B$ is in the kernel of \poppo\
(this is more or less obvious
from \nuseful) so the curvature term does not spoil the symmetry
between $b$ and $C$.

Hence the arguments that we gave above for $C$ carry over to $b$,
and for instance, on a K\"ahler manifold, with gauge group locally
a product of $SU(2)$'s, we can assume that $b=0$ in a solution
of $s=k=0$, since we have proved that assertion for $C$.

Now let us analyze the situation for the $(2,0)$ and $(0,2)$ part of
$B$.  For these components, \poppo\ collapses to a positive
multiple of $R\,\,\Tr\,\, \beta\bar\beta$, and so is positive if the scalar
curvature $R$ is
positive.  Thus, in that case $\beta=0$.
Even if $R$ is zero rather than positive, \usefulid\
implies that $D_iB_{jk}=0$, so that if not
zero $B$ is covariantly constant.
For gauge group locally a product of $SU(2)$'s, it follows (unless $A$
is gauge equivalent to the trivial connection) that
$[B,B]=0$ and hence (if $s$ vanishes) $F^+=0$.\foot{If $A$
is trivial, the stated conclusions still hold since
then $F^+=0$ and $s=0$ implies that $[B,B]=0$.}
But for a generic
K\"ahler metric, there are no abelian instantons (Kahler manifolds
always have $b_2{}^+>0$ since the Kahler form is self-dual), contradicting the
fact that $B\not= 0$ forces the connection to be abelian.

So we conclude that, for a K\"ahler metric with $R\geq 0$ and
gauge group locally a product of $SU(2)$'s, the desired vanishing theorem
holds and the partition function of the topological theory is the
Euler characteristic of instanton moduli space.
This important vanishing theorem applies to examples such as $K3$,
${\bf CP}^2$, and blowups of ${\bf CP}^2$ at a small number of points.

\bigskip
\noindent{\it More General K\"ahler Manifolds}

What about more general K\"ahler manifolds?  We still have $b=C=0$.
The curvature $F^+$ can be usefully decomposed on a K\"ahler manifold
in pieces $F^{p,q}$ of types $(p,q)$, with $(p,q)=(2,0),\,\,(1,1),$ or $(0,2)$.
The equations $s=k=0$ give first
\eqn\firste{F^{2,0}=F^{0,2}=0,}
that is, the connection $A$ endows the bundle $E$ with a holomorphic
structure; second
\eqn\seconde{\bar D\beta=0,}
that is, $\beta$ is a holomorphic section of ${\rm End}(E)\otimes K$
(with $K$ the canonical bundle of $X$),
and finally
\eqn\thirde{\omega\wedge F+\left[\beta,\bar\beta\right]=0}
($\omega$ is the Kahler form, of type $(1,1)$; only the $(1,1)$ part of
$F$ contributes in the equation.  Interpreting $\beta$ as $ (2,0)$ form,
both terms in the equation are $(2,2)$ forms.)
Analogy with other somewhat similar problems (such as the ``Higgs bundle''
equations \ref\hitchin{N. Hitchin, ``The Self-Duality Equations On
A Riemann Surface,'' Proc. London Math. Soc. {\bf 3} 55, 59 (1987).})
suggests that the last
equation can be interpreted holomorphically
as a kind of stability condition for the pair $(E,\beta)$.
If so, a determination of contributions -- if any -- to the partition
function from solutions with $\beta\not= 0$ should be quite accessible.

The following is a severe constraint.
The above equations have the obvious $U(1)$ symmetry
\eqn\ffip{\beta\to e^{i\theta}\beta}
(which is, again, a survivor of the $SO(4)$ symmetry of \mcgur).
The contributions of solutions with $\beta\not= 0$ to the topological partition
function would equal the number of gauge orbits of
such solutions, weighted by signs,
if the number is finite.  If there is instead a manifold ${\cal W}$
of such solutions, the contribution (according to equation \hurf)
is $\pm \chi(V)$ with $V$ the bundle of antighost zero modes.
The Euler class $\chi(V)$ of a $U(1)$-equivariant bundle $V$ can be computed
by summing over fixed points of the $U(1)$ action.
Thus, the only solutions with $\beta\not= 0$ that really have to be
considered are those that are invariant under \ffip, up to a gauge
transformation.

That is only possible if the gauge connection is reducible.
For gauge group $SU(2)$, for instance, the only fixed points with
$\beta\not=0$ are abelian
configurations, with a connection of the form
\eqn\jin{A=\pmatrix{* & 0 \cr 0 & * \cr}}
and $\beta $ of the form
\eqn\bbeta{\beta = \pmatrix{ 0 & 0 \cr * & 0 \cr}.}
Thus, the bundle $E$ is $E\cong L \oplus L^{-1}$ with $L$
a holomorphic line bundle; $\beta$ is a holomorphic section
of $K\otimes L^{-2}$.
Equation \thirde\ reduces in this situation
to
\eqn\girde{\omega\wedge F=\beta\wedge \bar\beta }
with now $F$ the curvature of the connection on $L$.
Since the right hand side is positive, this equation requires
that $[\omega]\cdot c_1(L)>0$, where $[\omega]$ is the Kahler class
and $c_1(L)$ the first Chern class of $L$.  Conversely, if $L$
is a line bundle with $[\omega]\cdot c_1(L)>0$ and $\beta$ is
a holomorphic section of $K\otimes L^{-2}$, then a standard convexity
argument shows formally
 that there is a unique metric on $L$ giving a solution of
\girde.

We can now uncover a qualitative consequence of the failure of
the vanishing theorem on some Kahler manifolds.  Of course, when
the vanishing theorem holds, solutions of the equations are instantons
and necessarily have $c_2(E)>0$.  But connections of the form \jin\
have
\eqn\goldfish{c_2(E)=-c_1(L)^2,}
and this can be negative.  For instance, on a minimal surface
of general type,
\foot{The condition roughly means that the line bundle $K$ is very positive.
``Most'' two dimensional compact complex manifolds are of this type
or obtained from such manifolds by blowing up points.}
we can obey the conditions if we take $L=K^{1/2}$
\foot{That is, $L$ is a line bundle such that $L\otimes L\cong K$.
Such a line bundle only exists globally if $X$ is a spin manifold,
so only if $X$ is spin does the bundle $E=L\oplus L^{-1}$ exist and
contribute
for the $SU(2)$ theory.  The corresponding $SO(3)$ bundle ${\rm ad}(E)
\cong L^{2}\oplus {\cal O}\oplus L^{-2}\cong
K\oplus {\cal O}\oplus K^{-1}$ (here ${\cal O}$ is a trivial line
bundle) always exists and contributes to the $SO(3)$ theory.}
and $\beta=1$.  (In fact, in that case one can pick a Kahler metric
such that $[\omega]=c_1(K)$ and solve \girde\ very explicitly
with $F$ a positive multiple of $\omega$.) This solution
has instanton number
\eqn\neginno{c_2(E)=-{1\over 4}c_1(K)^2.}

Conversely, on a minimal surface of general type, if we pick
a Kahler metric with $[\omega]=c_1(L)$, the instanton
number of a solution of \girde\ is bounded below by \neginno.
\foot{The argument was explained to us by D. Morrison.}
The Hodge index theorem states that the intersection pairing on
$H^{1,1}(X)$ is of ``Lorentz signature'' $(+--\dots -)$.
On a minimal surface of general type, $c_1(K)^2>0$
and hence
$c_1(L)=\lambda c_1(K)+\alpha$,
where $\lambda $ is a real number,
$\alpha\cdot c_1(K)=0$, $\alpha\cdot \alpha<0$.  The fact
that $K\otimes L^{-2}$ has a nonzero holomorphic section $\beta$
implies that $\lambda\leq 1/2$; the fact that $[\omega ]\cdot c_1(L)>0$
implies $\lambda>0$.  These conditions together imply $c_1(L)^2\leq
c_1(K)^2/4$.  This will be useful in \S5.

One more small extension of the vanishing theorems will be helpful
in \S4.
 $K3$ with a generic complex structure
has no non-trivial holomorphic line bundles.
Let $X$ be such a generic $K3$ surface with one point blown up.
On $X$ there is an exceptional divisor $D$ produced by the blow up;
the canonical bundle is $K={\cal O}(D)$, with $c_1(K)^2=-1$.
Any line bundle on $X$ is of the form $K^{\otimes n}$ for some integer
$n$.  Any Kahler form $\omega $ on $X$
has
\eqn\ithas{[\omega]\cdot c_1(K)=\omega\cdot [D]>0;}
the right hand side is just the area of $D$ in the Kahler metric.
Let us now show that \girde\ has no non-trivial solutions on $X$.
The line bundle $L$ would have to be of the form $K^{\otimes n}$
as those are the only line bundles.  For $K\otimes L^{-2}$ to have
a non-zero holomorphic section, one needs $n\leq 0$.
But in view of \ithas, $[\omega]\cdot c_1(L)>0$ requires
$n>0$.  So we get a vanishing theorem on $X$: one can compute
via instantons.  Essentially the same argument holds for $K3$ with any
number of points blown up.

\subsec{Singularities Of Instanton Moduli Space}

In at least one respect, the above discussion is misleading.
We have constantly assumed that the moduli space ${\cal M}$
of solutions of the original equations is compact and non-singular.
For moduli spaces of instantons, those assumptions are unrealistic.

Compactness fails because instantons can shrink to zero size.
The Euler characteristic entered our problem as a kind of curvature
integral, beginning with \stanform.  Only for a compact manifold
(or for connections of very special type)
does such an integral reproduce the Euler characteristic as defined
topologically.  To facilitate computations, one would hope that
the curvature integral can be interpreted as the Euler characteristic
of some compactification of ${\cal M}$.  At least for $X$ a Kahler
manifold, there is a natural compactification of ${\cal M}$ by
stable sheaves in algebraic geometry; we will very optimistically
use this compactification, since our computations will be based on
results borrowed from mathematicians who used it.  We do not
know how to justify this assumption.

Physically, the only obvious place that a compactification of ${\cal M}$
would come from is string theory.  To the extent that ${\cal M}$ can
be interpreted as a space of classical solutions of string theory,
the good ultraviolet behavior of string theory should lead to a natural
compactification.

Instanton moduli space may also have singularities; these arise
at points in ${\cal M}$ where $C$ or $B$ has a zero mode.
For a generic metric on $X$, ${\cal M}$ is smooth.  Nevertheless,
the zero modes and singularities that occur in a one-parameter family
of metrics are important in verifying the formal arguments
for topological invariance of the theory.

\subsec{Self-Conjugacy}

One point that may puzzle the reader is that the twisted $N=4$ system
has $N=2$ topological symmetry and an $SU(2)$ global symmetry;
but we have so far discussed it as an $N=1$ topological system,
and (in discussing the vanishing theorems)
we exhibited only a small piece of the global symmetry.
Here we will fill this gap in generality.  It is convenient
to do so in the general context of the whole class of models that we have
discussed in this section.

First we consider
the general construction that counts solutions of a system of
equations weighted by signs.
The fields at various ghost numbers are as follows:
\eqn\allthefields{\eqalign{
                  U=2: &  ~~~~~ \phi^x\cr
                  U=1: & ~~~~~\psi^i   \cr
                  U=0: & ~~~u^i,~ H^a \cr
                  U=-1:& ~~~\chi^a, ~\eta^x \cr
                  U=-2: &  ~~~~~\bar\phi^x .\cr} }
Here $(u^i,\psi^i)$ are a multiplet of fields and ghosts;
$(\chi^a,H^a)$ are a multiplet associated with the equations;
and $\phi$,
$\bar\phi$, and $\eta$ are fields associated with the symmetry group $G$.

Now we consider the more detailed construction that eliminates
signs when a vanishing theorem holds.  In this case, there
are three sets of multiplets containing  fields: the ``original'' multiplets
$(u^i,\psi^i)$,
the ``dual'' multiplets $(y_a,\tilde\psi_a)$,
and the multiplet $(C^x,\zeta^x)$ associated with the
symmetries.  In addition,
there are auxiliary multiplets $(\chi^a,H^a)$ associated with the original
equations and $(\tilde\chi_i,\tilde H_i)$
associated with the dual equations.  The
other fields $\phi^x,\bar\phi^x,\eta^x$
are unchanged from  the general picture in \allthefields.  So we get
this setup:
\eqn\callthefields{\eqalign{
                  U=2: &  ~~~~~~~~~ \phi^x\cr
                  U=1: & ~~~~~~\psi^i,\tilde\psi_a,\zeta^x   \cr
                  U=0: & ~u^i,y_a,C^x, H^a,\tilde H_i \cr
                  U=-1:& ~~~~~~\chi^a,\tilde\chi_i, \eta^x \cr
                  U=-2: &  ~~~~~~~~~\bar\phi^x .\cr} }

Now, \callthefields\ differs from the more general
structure \allthefields\ in being self-conjugate
in the following sense.  The fields at ghost number $U=-1$ have the
same quantum numbers as the fields at ghost number $1$ (if we bear in mind
that there are metrics $g_{ij}$ and $g_{ab}$ that can be used to raise
and lower indices), and likewise the quantum numbers are the same
for ghost number 2 and $-2$.
In this self-conjugate case, instead of our usual BRST-like operator
$Q$ of $U=1$, we obviously could define a similar operator
$Q'$ of $U=-1$.  It will soon be clear that we can take
$Q^2=(Q')^2=\{Q,Q'\}=0$, up to gauge transformation.

One can actually ask for more.  Define an $SU(2)$ action
on the fields in \callthefields\ such that $\phi,C,\bar\phi$ make
a three-dimensional representation,  $\psi,\tilde\chi$  and $\tilde\psi,\chi$
make up two different two-dimensional representations, and the other
fields are invariant.  One can arrange so that the pair
$Q,Q'$ transform in a two-dimensional
representation of this $SU(2)$.

To implement this, even in a superfield language, is really quite
easy.  Introduce a doublet of anticommuting variables $\theta^A$, $A=1,2$,
transforming in a two-dimensional representation of $SU(2)$.  Arrange
$\phi,C,\bar\phi$ into an $SU(2)$ triplet $\phi_{AB}=\phi_{BA}$.
The supersymmetry transformations are to be
\eqn\suptobe{Q_A=i{\partial\over\partial\theta^A}-\theta^B[\phi_{AB},~\cdot~]
,}
with $[\phi_{AB},~\cdot~]$ denoting the infinitesimal $G$ transformation
generated by the Lie algebra element $\phi_{AB}$.
Obviously, $\{Q_A,Q_B\}=0$ up to a gauge transformation.

Form superfields
\eqn\tisup{T^i=u^i+i\theta^A\psi_A{}^i+{i\epsilon_{AB}\theta^A\theta^B\over 2}
\tilde H^i,}
with $\psi^i$ and $\tilde\chi^i$ being the components of $\psi_A{}^i$,
and
\eqn\yisup{Y^a=y^a+i\theta^A \tilde\psi_A{}^a+{i\epsilon_{AB}\theta^A\theta^B
\over 2}H^a,}
with $\tilde\psi^a$ and $\chi^a$ being the components of $\tilde\psi_A{}^a$.
The transformation laws given earlier for the fields in those multiplets
can be summarized by
\eqn\sumtrans{\eqalign{\delta T^i & =-i\epsilon^A\{Q_A,T^i\} \cr
                      \delta Y^a-\delta  T^iA_i{}^a{}_bY^b & = -i\epsilon^A
\{Q_A,Y^b\}. \cr}}
If we combine $\eta,\zeta$ as an $SU(2)$ doublet $\eta_A$,
then the transformation laws for the fields $\phi_{AB}$ and $\eta_A$
associated with the gauge symmetry are
\eqn\nugtrans{\eqalign{\delta\phi_{AB}& = i
\epsilon_A\eta_B+\epsilon_B\eta_A\cr
      \delta\eta_B & = -{1\over 2}\epsilon_A\left[\phi_{BC},\phi^{CA}\right].}}

Unfortunately, we do not know of a general description of the possible
$Q_A$-invariant Lagrangians.  In the case of gauge theory, of course,
the standard $N=4$ Lagrangian is one.

\newsec{Predictions of Strong-Weak duality}
\def\tilde{\widetilde}
\def\hat{\widehat}
\def\bar{\overline}
In this section we formulate precisely the predictions of strong weak duality
that we are going to test.  Consider $N=4$ super Yang-Mills theory
with gauge group $G$\foot{We are making a small change of notation
relative to \S2, where the finite dimensional gauge group was called
$G_0$, and the name $G$ was reserved for
the group of local gauge transformations, that is
(roughly) the group of maps of space-time to $G_0$.}
 on flat Euclidean four-dimensional space.
Then the fact that
the energy-momentum tensor $T_{\mu \nu}$ is invariant under $S$-duality
 implies that if we consider the same theory on a curved background it
should still respect $S$-duality.  The simplest object to compute
is the partition function of the theory. This will in general depend
on the manifold, its metric $g_{\mu \nu}$, and the gauge coupling constant
and $\theta $ angle which we combine as $\tau={\theta
\over 2\pi}+
{4\pi i\over g^2}$.  One would expect from $S$-duality that (up to some
universal factors) the partition function
$Z_M(\tau, \bar \tau ,g_{\mu \nu},G)$ should
transform under $\tau \rightarrow -{1/\tau}$
 to the same object for the dual group $\widehat G$
\foot{Recall that $\widehat G$ is the group whose weight lattice
is dual to that of $G$.};
the transformation
from $G$ to $\widehat G$ is part of the original Montonen-Olive
conjecture.

To test any prediction of $S$-duality, we need to be able
to compute exact (or at least strong coupling)
quantities in the theory, as $S$-duality
relates weak to strong coupling.
Unfortunately the partition function $Z_M$
in general is too difficult to compute.\foot{The case that $M$
is a hyper-K\"ahler manifold is an exception as we see later.}
Here is where topological twisting
becomes helpful.  First of all, the twisted theory should
still be $S$-dual
since twisting, as discussed in the last section,
basically means introducing background
fields which couple to  the $SU(4)$ global symmetry
current
of the theory, and those currents are $S$-dual.
The theory
being topological means in particular that,
barring anomalies,\foot{As we will find later, such
anomalies do occur in certain cases, as in Donaldson
theory. However the discussion of modular properties
of the partition function in the rest of this section still holds.}
the partition function
is a holomorphic function of $\tau$ and is independent of the metric
on $M$.  Thus we would formally expect the
partition function on $M$ to depend only on
$\tau$ and the group chosen; we write it therefore as $Z_M(\tau , G)$.
As discussed in the previous section, $Z_M$ is determined
(when the appropriate vanishing theorem holds) by the Euler
characteristics of instanton moduli spaces.
Suppose for simplicity that $G$ is connected and simply-connected.
Then $G$-bundles on $M$ are classified by a single integer, the instanton
number.
If ${\cal M}_n$ is the
moduli space of instantons of instanton number $n$, $\chi$
denotes the Euler characteristic, and $q=\exp(2\pi i\tau)$,
then the partition function would be\foot{If $G$ is not connected
and simply-connected, the classification of bundles is finer, as we discuss
later, and the instanton number $n$ may not be an integer.  However,
we still write the following formula, with the sum now running over
all topological types of $G$-bundles.  The discussion below is
valid with obvious modifications.
In any case, we systematically examine these issues later.}

\eqn\minssum{Z_M={1\over \#Z(G)}\sum_n\chi({\cal M}_n) q^n}
($\#Z(G)$ is the number of elements of the center $Z(G)$ of $G$;
this factor is present because $Z(G)$ acts trivially on the space
of connections and one divides by it in performing the Feynman path
integral.)
Under the $S$ transformation,
$\tau \rightarrow -{1\over \tau}$, how should $Z_M$ transform?
The most naive
guess is that simply
$$Z_M(-1/\tau , G)=Z_M (\tau ,{\widehat G}),$$
in other words that $Z_M$ is strictly invariant under $S$-duality.

However, two natural generalizations of this come to mind.
One is the possibility that instead of being modular invariant,
$Z_M$ might transform like a ``modular form,''
\eqn\dsdu{Z_M({-1/\tau},G) =\pm \left({\tau \over i}\right)^{w/2}
        Z_M(\tau ,{\widehat G})} %
for some $w$ (the factor of $i$ in the denominator
is there to guarantee $S^2=1$; this leaves room for an extra overall
$\pm$ sign).
One other familiar fact suggests a modification
of $S$-duality in curved space:
the leading power of $q$ in a modular object
is not always 0.  For example the Dedekind
$\eta$-function has an integral expansion multiplied
by $q^{1\over 24}$.  In string theory this comes from a shift in
the zero point of the energy.
Similarly here we might expect a shift in the zero point of the instanton
number, as a result of which
the formula for the partition function should
be modified by an overall multiplicative factor to be
\eqn\inssum{Z_M={q^{-s}\over \#Z(G)}\sum_n\chi({\cal M}_n) q^n}
for some $s$.

To give at least some explanation of how such subtleties
could arise, note that
even if a Lagrangian $L$ is $S$-dual in flat space and has an $S$-dual
extension to curved space, we have to ask
precisely what extension of $L$ to curved space is $S$-dual.
For instance, as we saw in \S2, the BRST symmetry of the twisted
theory requires the presence of some curvature couplings that one might
not have guessed.  Even if all $q$-number terms are known in the extension
of $L$ to curved space (in the twisted theory they are all determined
by BRST invariance, modulo BRST commutators), one can still add $c$-number
terms.  Thus, if $L_1$ is one extension of the theory to curved space,
another is
\eqn\lonep{L'{}_1=L_1+\int_Md^4x \sqrt g\left(e(\tau) +f(\tau)R
+f(\tau)R^2+\dots\right),}
where the terms are local operators constructed from the metric,
and we have to ask whether it is $L_1$ or $L'{}_1$ that is $S$-dual.
We have here made an expansion in local
operators because we assume that the statement ``the theory is $S$-dual
in curved space'' means that there is a {\it local} Lagrangian which
is $S$-dual in curved space.

Taking $L_1$ to be the twisted $N=4$ theory as formulated in \S2,
topological invariance means that the
$c$-number terms must themselves be topological invariants.  The only
topological invariants of a four-manifold that can be written
as the integral of a local operator are the Euler characteristic $\chi$
and the signature $\sigma$.  Thus, the unknown $c$-number terms must
be of the form $e(\tau)\chi+f(\tau)\sigma$, with $e$ and $f$ being
unknown functions of $\tau$.  If this is so, then $Z_M$ defined
as in \inssum\ would fail to be $SL(2,{\bf Z})$-invariant by
a universal $\chi$ and $\sigma$-dependent factor, and with suitable
$e$ and $f$, this could lead to the subtleties suggested above.
We now however have the additional information that we should
expect the modular weight $w/2$ and the instanton shift $s$ to be linear
functions of $\chi$ and $\sigma$:
\eqn\rolf{\eqalign{ w & = a\chi +b\sigma \cr
                    s & = \alpha\chi+\beta\sigma,\cr}}
with universal constants $a,b,\alpha$, and $\beta$.
For the physical $N=4$ theory, the coefficients $b$ and $\beta$ would
have to be zero as $\sigma$ is odd under parity (which is a symmetry
of the physical model when $\theta=0$).  For the twisted
theory, parity is violated explicitly and it is not clear
that $b$ and $\beta$ should vanish.  However, it will turn out that they do.

A further subtlety arises if one does not require that $s$ be
integral.  Then, although \minssum\ is strictly invariant
under
\eqn\ulg{T=\pmatrix{1 & 1 \cr 0 & 1\cr},}
which corresponds to $\theta\to\theta+2\pi$ or
$\tau\to \tau+1$, \inssum\ would
change under $\tau$ by an overall phase.
One could interpret this as a kind of global gravitational anomaly
in the $2\pi$ periodicity in $\theta$ -- or perhaps better,
as a clash between $S$-duality and the $2\pi$ periodicity.
If this possibility is realized, then $Z_M$ is an object
somewhat like the Dedekind $\eta$ function -- transforming under
$SL(2,{\bf Z})$ almost like a modular form of some particular weight,
but with some additional phase factors.
We will see that all of these possibilities are realized;
the zero of instanton number is shifted by a multiple of $\chi$, which
is not always integral so one gets the phases just mentioned,
and there is also a modular weight that is a multiple of $\chi$.


It is convenient sometimes to get
rid of the modular weight.  This can be done by multiplying
$Z_M$ by $\eta^{-w}$ to get
$$\hat Z_M =\eta^{-w} Z_M$$
Under $S$-duality we have
$$\hat Z_M \rightarrow \pm \hat Z_M$$
We denote the shift from integer power
$q$ expansion in $\hat Z_M$ by
$q^{-{c/24}}$.
This definition is motivated by a similar appearance of the central charge
$c$ in two-dimensional conformal field theory.
In principle  $c$ depends on both $M$ and $G$.
Note that $c$ is related to $s$ by
\eqn\vardef{c=24 s+w=a' \chi+b' \sigma}
where $a',b'$ can be written in terms of $a,b,\alpha ,\beta $.

\subsec{Fractional Instanton Numbers}

The original Montonen-Olive conjecture states that under
$S:\tau\to -1/\tau$, the gauge group $G$ is replaced by the
dual group $\widehat G$. With the exception of $G=E_8$, it is
impossible for both $G$ and $\widehat G$ to be simply-connected,
and therefore we are all but forced to discuss the phenomena that arise
for non-simply-connected groups.

Before being general, let us discuss the important special case
that $G=SU(2)$; then the dual group is $\hat G=SU(2)/{\bf Z}_2=SO(3)$.
Of course, $\pi_1(SO(3))={\bf Z}_2$.

$SU(2)$ or $SO(3)$ bundles on the four-sphere are both classified by
a single integer, the instanton number.  That is not so on a more
general four-manifold.  The basic difference between $SU(2)$ and $SO(3)$
bundles arises first
in two dimensions.  On the two-sphere ${\bf S}^2$, bundles with
any gauge group $G$ are classified by $\pi_1(G)$, so $SU(2)$ bundles
are trivial, but there are two types of $SO(3)$ bundle, labeled by
${\bf Z}_2$.
The non-trivial $SO(3)$ bundle can be described explicitly as follows.
Let $L$ be the basic $U(1)$ magnetic monopole bundle of magnetic charge
one.  Then the $SU(2)$ bundle $F=L^{1/2}\oplus L^{-1/2}$ does not exist,
since $L^{1/2}$ -- which would be a line bundle of magnetic charge $1/2$ --
does not exist. However, the $SO(3)$ bundle that would be derived
from $F$ (for instance, by taking the symmetric part of $F\otimes F$
to construct spin one from spin $1/2$) is
\eqn\ponl{
E=L\oplus {\cal O}\oplus L^{-1}}
(${\cal O}$ is a trivial
bundle), and does exist.  The fact that $E$ exists but there is no associated
$SU(2)$ bundle makes it clear that $E$ is non-trivial.

Now, if $E$ is an $SO(3)$ bundle on a four-manifold
$M$, then for every two sphere $S$ in $M$, we can assign an element
$\alpha(S)\in {\bf Z}_2$ that measures whether the restriction of $E$ to $S$
is trivial or not.
$\alpha(S)$ is what 't Hooft
\thooft\ called the non-abelian magnetic flux through
$S$.  One justification for the terminology is that
if $S$ is the boundary of a three-manifold, then
automatically $\alpha(S)=0$, as one would expect for magnetic flux.
Mathematically, the association $S\to \alpha(S)$ (or a slight elaboration
of it if $M$ is not simply-connected), is an element of
$H^2(M,{\bf Z}_2)$ which is called the second Stieffel-Whitney class
of $E$, $w_2(E)$.
The $SO(3)$ bundles that are associated with $SU(2)$
bundles are precisely those for which $w_2=0$.

For instance, if $E$ is of the form \ponl\ for some line bundle $L$,
then $\alpha(S)$ is the mod two reduction of $\langle S,c_1(L)\rangle$
(the latter is the pairing of $S$ with the first Chern class $c_1(L)$;
it vanishes if $S$ is a boundary).  Therefore, $w_2(E)$ in this
situation is the mod two reduction of $c_1(L)$, a fact that will
be used in \S5.

Since $\pi_1(SO(3))={\bf Z}_2$ and $\pi_3(SO(3))={\bf Z}$ are the
only non-trivial homotopy groups of $SO(3)$ in dimensions low enough
to matter, an $SO(3)$ bundle $E$ on a four-manifold $M$ is classified by
two invariants, which are $v=w_2(E)$ and the instanton number $k$.
However, these are correlated in a perhaps surprising fashion; $k$
is not an integer in general, but rather
\eqn\insno{k=-{v\cdot v\over 4}~~{\rm mod}~~1.}
Here $v\cdot v$ is defined as follows: one lifts $v$ to an integral
cohomology class $v'$ and then interprets $v\cdot v$ as the usual cup product
$v'\cdot v'$;
it is evident that $v\cdot v/4$ is independent of the lifting modulo 1.

At least for simply connected $M$, \insno\ can be proved as follows.
After lifting $v$ as above, pick a line bundle $L$ with $c_1(L)=v'$
and let $E'=L\oplus {\cal O}\oplus L^{-1}$.  Thus $w_2(E')=w_2(E)$.
The $SU(2)$ bundle associated with $E'$ would be $L^{1/2}\oplus L^{-1/2}$
and its instanton number would be $-c_1(L)^2/4=-v\cdot v/4$.
\foot{The fact that this $SU(2)$ bundle may not really exist is immaterial;
the instanton number is defined by a curvature integral which can be taken
either in the spin one-half representation or -- with an extra
factor of $1/4$ -- in the spin one representation.  Writing down
the formal expression for $E'$ simply lets us use a short-cut:
for an $SU(2)$ bundle $N\oplus N^{-1}$ with $N$ a line bundle, the
instanton number is $-c_1(N)^2$.}
Thus we have proved \insno\ for a bundle with the same $w_2$ as $E$.
However, any two $SO(3)$ bundles $E$ and $E'$
on $M$ with the same $w_2$ become isomorphic if a point $p$
is deleted from $M$ (since then the classification of bundles involves
the homotopy groups $\pi_k(SO(3))$ for $k\leq 2$, and $\pi_1$ is the only
non-trivial one).  They differ therefore by a topological twist
that is localized near $p$, but the localized topological twists
are the ones that can be defined on the four-sphere, and shift the
instanton number by an integer.

\subsec{The Group And The Dual Group}

\insno\ implies
that the instanton numbers in the $SO(3)$ theory,
for which we sum over all bundles with arbitrary $w_2$,
take values in ${\bf Z}/4 $ but not in general in ${\bf Z}$.
Therefore, while the $SU(2)$ theory on a four-manifold is
invariant under $\tau\to \tau+ 1$, the $SO(3)$ theory is in general only
invariant under $\tau\to\tau+4$.  There is one situation in which
this is improved slightly: if $M$ is a spin manifold, then $v\cdot v$
is even for all $v$; and hence
the $SO(3)$ instanton numbers take values
in ${\bf Z}/2$, so the $SO(3)$ theory is invariant under $\tau\to\tau+2$.

Now we can understand better the precise implications of $S$-duality.
The modular transformation $T:\tau\to\tau+1$ maps the $SU(2)$
theory to itself (perhaps with an anomalous phase, as discussed above).
The transformation $S:\tau\to -1/\tau$, according to Montonen and Olive,
maps $SU(2)$ to $SO(3)$.  A tranformation by $T^4$ then maps the $SO(3)$
theory to itself, and subsequent transformation by $S$ will map us back
to $SU(2)$.  So the $SU(2)$ theory will be mapped to itself by the
operation $ST^4S$.
The $SU(2)$ theory should therefore be transformed to itself by the subgroup
of $SL(2,{\bf Z})$ generated by the matrices
\eqn\nogoood{\eqalign{T & =\pmatrix{1 & 1 \cr 0 & 1}\cr
                     ST^4S& =\pmatrix{1 & 0\cr 4 & 1\cr}.\cr}}
These matrices generate the subgroup of $SL(2,{\bf Z})$ consisting
of matrices whose lower left entry is congruent to $0$ modulo 4.
This subgroup is known as $\Gamma_0(4)$.
In the case of a spin manifold, one would get the subgroup of $SL(2,{\bf Z})$
generated by $S$ and $ST^2S$; this is the group $\Gamma_0(2)$ of matrices
whose lower left entry is congruent to 0 modulo 2.

Therefore, the prediction of $S$-duality is that the $SU(2)$
(or $SO(3)$) partition function $Z_M$ is modular for $\Gamma_0(4)$,
or $\Gamma_0(2)$ for spin manifolds.
$Z_M$ is not quite a modular form for these groups, but transforms
with extra phases,
because we have allowed
a shift in the instanton number in multiplying by
$q^{-s}$ and moreover because
\dsdu\ is not quite the conventional
definition of the transformation of a modular form.
We can get rid of most of the phases by studying the function
\eqn\dff{f_M=\eta^{-w+c} Z={\eta^c {\hat Z} }}
We see that $f_M$ has
weight $c/2$ and  is invariant under $\tau \rightarrow \tau+1$.
It is still not quite a standard modular form unless
$c$ is an integer, but this will turn out to be the case in our examples.
If $c$ is an odd integer, we get what is called a modular form of
half-integral weight.
\foot{These are objects that
transform as $\theta^c$ under $\Gamma_0(4)$,
where $\theta$ is the usual theta function.
To check that $f_M$ has the appropriate $q$ expansion near the third cusp
of $\Gamma_0(4)$ one needs the sharpened version discussed later
of the $S$-duality
conjecture.  $f_M$ may still fail to be
a modular form in the usual sense if it has poles at some cusps,
which may occur -- as we discussed in connection with
\girde -- if bundles of negative instanton number contribute
due to a failure of the vanishing theorem.}

\nref\koblitz{N. Koblitz, {\it Introduction To Elliptic Curves And
Modular Forms}, (Springer-Verlag, 1984).}
Let us note some aspects of
the fundamental domains of each of these two groups.
These fundamental domains parametrize
  the inequivalent values of
$\tau$
if $S$-duality is valid.
The fundamental domain of $\Gamma_0(2)$ has two cusps and a ${\bf Z}_2$
orbifold point.\foot{See
\koblitz, p. 231, solutions to exercises 12 and 14, for
this and subsequent assertions.}  One cusp (say at $\tau=\infty$) corresponds
to the instanton expansion for $SU(2)$ and one (say at $\tau=0$)
corresponds to the instanton expansion for $SO(3)$.  The ${\bf Z}_2$
orbifold point has $\theta=\pi$ from the $SU(2)$ viewpoint.
The fundamental domain of $\Gamma_0(4)$
is a two-fold cover of that of $\Gamma_0(2)$; this eliminates
the orbifold point and creates a third cusp.
While again two of the cusps correspond to the instanton expansions
of the $SU(2)$ and $SO(3)$ theories, the meaning of the third cusp
is more mysterious.  We will make a proposal later: the expansion
near the third cusp is the instanton expansion of the $SO(3)$ theory
restricted to bundles $E$ for which $w_2(E)=w_2(M)$.  Here $w_2(M)$
is the second Stieffel-Whitney class of the tangent bundle of $M$
and vanishes precisely if $M$ is a spin manifold; when that occurs
two of the three cusps of $\Gamma_0(4)$ are equivalent and we get
back to the $\Gamma_0(2)$ picture.

\bigskip
\noindent{\it Generalization To $SU(N)$}

For $SU(N)$, the story is very similar.  The dual group is now
$SU(N)/{\bf Z}_{ N}$.  As $\pi_1(SU(N)/{\bf Z}_N) ={\bf Z}_N$,
an $SU(N)/{\bf Z}_N$ bundle has in addition to the instanton number
an additional invariant $v$ taking values in $H^2(M,{\bf Z}_N)$.
One can think of this as a ${\bf Z}_N$-valued magnetic flux.
The instanton number $h_v$
for a bundle characterized by a magnetic flux $v$ is not necessarily
an integer but obeys
\eqn\insf{h_v={v\cdot v \over 2N}-{v\cdot v\over 2}
\qquad {\rm mod }\ 1 }
Note that this is invariant under $v\to v+Nv'$ and so is
independent of the choice of lifting
of $v$ to an integral class.

To justify \insf, we follow the same steps as in the $SO(3)$ case;
it is enough to prove the result for some bundle with given $v$.
Picking a line bundle $L$ with $c_1(L)=v$ mod $N$, we let $F$
be the ``$SU(N)$ bundle'' $F=\oplus_{i=1}^NL^{a_i}$,
where (for instance) $a_i=1/N$ for $i<N$ and $a_N=-(N-1)/N$.
Though $F$ may not
exist as an $SU(N)$ bundle, the corresponding $SU(N)/{\bf Z}_N$ bundle
$E'$ does.  Its ``magnetic flux'' is $v$.
The instanton number of $E'$ (which can be computed
as if $F$ exists) is
\eqn\nurgo{v^2\cdot\left({1\over 2N}-{1\over 2}\right), }
and this gives \insf.

The vector $\vec a=(a_1,\dots a_N)$ used above, with
$a_i\in {\bf Z}_N$, $a_i-a_j\in {\bf Z}$, and $\sum_i a_i=0$,
can be interpreted as a vector in the weight lattice $\Gamma_w$
of $SU(N)$.  $\Gamma_w$ contains the root lattice $\Gamma$ as a sublattice
of index $N$ with $\Gamma_w/\Gamma\cong {\bf Z}_N$.
If we replace $\vec a$ by an arbitrary vector in $\Gamma_w$,
the flux $v$ of the bundle $E$ would be $[\vec a]\cdot c_1(L)$,
and the instanton number would be $-c_1(L)^2\vec a^2/2$ modulo 1
(with $[\vec a]$ the coset of $\vec a$ in $\Gamma_w/\Gamma\cong {\bf Z}_N$,
and $\vec a^2$ the length squared of $\vec a$).
This is similar to conformal field theory, where $SU(N)$ characters
at level one are theta functions of the $\Gamma$ cosets in $\Gamma_w$,
and for the character derived from a given coset, $L_0=\vec a^2/2$ modulo
1.  The analogy would be perfect for a one dimensional $H^2(M,{\bf Z})$
generated by $e$ with $e^2=-1$ (see the discussion below).

Just as in conformal field theory, this
way of writing things leads at once to the generalization for arbitrary
simply laced groups.  The magnetic flux $v$ takes values in
$H^2(M,\Gamma_w/\Gamma)$, and the instanton numbers are equal modulo 1 to
$-v\cdot v/2$ where the $v\cdot v$ is computed
using the tensor product of the natural inner
product on $H^2(M,{\bf Z})$
and the inner product on the weight lattice $\Gamma_w$.

\subsec{Sharpening the $S$-duality Conjecture}

Let us consider the theory with
$SU(N)/{\bf Z}_N$ gauge group.  As we have discussed,
an $SU(N)/{\bf Z}_N$ bundle on $M$ has a ``magnetic'' invariant
$v \in H^2(M,{\bf Z}_N)$.
Let $b_2$ be the second Betti number of $M$.  If for simplicity we
suppose that $M$ is simply-connected (or at least that there is no
$N$-torsion in $H^2(M,{\bf Z})$), then the magnetic flux takes
$b_2 ^N$ distinct possible values.
In the $SU(N)/{\bf Z}_N$ theory we sum over all these
possibilities. We will consider the partition
function of the theory with a fixed magnetic flux
vector $v$, that is summing over all values of the instanton
number with a fixed $v$. Let us denote this partition function by $Z_v$ where
we hide the $M$ dependence to avoid too many
labels. It is natural to ask how the individual $Z_v$'s transform under
$SL(2,{\bf Z})$. It will also be natural to study how ${\hat Z_v}
=\eta^{-w}{ Z_v}$ transforms.
  The $T$-transformation is particularly easy.
{}From \insf\ and from the fact that we  have shifted by an
overall $q^{-s}=q^{-{c\over 24}+{w\over 24}}$ we have
$$\qquad Z_v(\tau+1)={\rm exp}(2\pi i(-s +h_v)) \ Z_v(\tau)$$
\eqn\ttran{{\hat Z_v}(\tau+1)={\rm exp}(2\pi i(-{c\over 24} +h_v))
\ {\hat Z_v}(\tau)}
where $h_v$ is given by $\insf$.  More subtle
is of course the $S$ transformation.

It has been shown by
't Hooft \thooft\ that
in the Hamiltonian formulation,
analogous to the ${\bf Z}_N$-valued magnetic flux,
it is natural to introduce also
${\bf Z}_N$-valued electrical
fluxes.  Moreover he showed that they are Fourier transforms
of one another.  In our setting this means that
the path integral with electrical flux $w$ is
$$Z^{elec.}_w={\rm const.}
\sum_{v} {\rm exp}\left({2 \pi i v\cdot w\over N}\right)Z_v$$
where again the $v\cdot w$ is the inner product on $H^2(M)$.
It is natural to extend the conjecture of strong/weak duality
to include not only
the exchange of ordinary
electric and magnetic flux but
also the statement that 't Hooft's electric and
magnetic fluxes get exchanged under $S$ transformation.  This means in
particular that the transformation laws are
$$\qquad Z_u(-1/\tau)= \pm N^{-b_2/2}
\left({\tau \over i }\right)^{w\over 2}
\sum_{v} {\rm exp}({2 \pi i v\cdot u\over N})Z_v(\tau)$$
\eqn\stran{{\hat Z_u}(-1/\tau)=\pm N^{-b_2/2}
\sum_{v} {\rm exp}({2 \pi i v\cdot u\over N})\hat Z_v(\tau).}
The proportionality constant was fixed by requiring $S^2=1$,
which leaves a $\pm$ sign ambiguity in fixing $S$.  The proof that $S^2=1$
requires using the fact that by Poincar\'e duality the cup
product on the integral lattice $H^2(M,{\bf Z})$ is
self-dual.  In particular one uses
\eqn\useq{\sum_{v} {\rm exp}({2\pi i v\cdot u\over N})=N^{b_2}\delta_{u,0}}

The partition function for the
$SU(N)$ theory is the same as the contribution of zero magnetic
flux to the $SU(N)/{\bf Z}_N$ partition function, times an elementary factor:
\eqn\tuff{Z_{SU(N)}= N^{-1+b_1}Z_0}
Here $b_1$ is the first Betti number of $M$. The prefactor  in \tuff\
arises
as follows.  The volume of the group of $SU(N)$ gauge transformations
is (i) bigger than that of $SU(N)/{\bf Z}_N$ by a factor of $N$ because
constant gauge transformations by an element of the center of $SU(N)$
are non-trivial in $SU(N)$ but trivial in $SU(N)/{\bf Z}_N$, but (ii)
smaller by a factor of $N^{b_1}$ because an $SU(N)/{\bf Z}_N$
gauge transformation, in traversing a loop in $M$, may make a non-trivial
loop in $\pi_1(SU(N)/{\bf Z}_N)$, a  possibility that is absent for $SU(N)$.
(For simplicity we assume that $H_1(M,{\bf Z})$ is torsion-free or
at least that the torsion is prime to $N$, so that the ordinary first
Betti number $b_1$ enters this assertion.)  So the volume of the gauge
group for $SU(N)$ is $N^{1-b_1}$ times that of $SU(N)/{\bf Z}_N$,
and the $SU(N)$ partition function is obtained by dividing by this
factor the contribution $Z_0$ of bundles with zero flux to the $SU(N)/{\bf
Z}_N$ partition function.

In the $SU(N)/{\bf Z}_N$ theory we have to sum over all
allowed bundles with equal weight, which means summing
over all allowed magnetic flux vectors. So
$$Z_{SU(N)/Z_{\bf N}}=\sum_v Z_v$$
Thus, the original Montonen-Olive relation between
$SU(N)$ and $SU(N)/{\bf Z}_N$, with some slight
correction factors, is a consequence of \stran:
$$Z_{SU(N)}({-1/\tau})=\pm N^{-1+b_1-{b_2\over 2}}
\left({\tau\over i}\right)^{w\over 2}Z_{SU(N)/{\bf Z}_N}(\tau )$$
\eqn\pred{=\pm N^{-\chi/2}\left({\tau\over i}\right)^{w\over 2}Z_{SU(N)/{\bf
Z}_N}(\tau )}

\bigskip
\noindent{\it The Third Cusp}

Using \stran, we can determine the meaning of the third
cusp for the $SU(2)$ or $SO(3)$  theory.
The $SL(2,{\bf Z})$ transformation $ST^2S$
takes the $SU(2)$ cusp to the other cusp.
 Since the partition function at the $SU(2)$ cusp
is proportional to $Z_0$, applying $ST^2S$ to it
with the help of \stran\ and \ttran\ (and setting $N=2$) we get
\eqn\nuddy{Z_0\rightarrow {1\over 2^{b_2} }\sum_{v,u}(-1)^{v\cdot (v+u)}Z_u
={1\over 2^{b_2}}\sum_{u,v}(-1)^{v\cdot (w_2(M)+u)}Z_u=
Z_{w_2(M)}.}
Here we have used the Wu formula (see \ref\krd{S. Donaldson
and P.Kronheimer,{\it The Geometry of Four Manifolds}, Oxford
University Press (1990), p. 6.} for a quick proof in the simply-connected
case) which asserts that
$v^2=v\cdot w_2(M)$ mod 2 for every vector $v\in H^2(M,{\bf Z}_2)$,
where $w_2(M)$ is the second Stiefel-Whitney class of $M$.
Thus the expansion at the third cusp gives the partition function
with $v=w_2(M)$.

\bigskip
\noindent{\it An Interesting Self-Dual Group}

The transformation property
\stran\ can be in part tested (and was originally guessed) by
the following considerations.
Consider the group
$${SU(N)\times SU(N)\over {\bf Z}_N}$$
where the ${\bf Z}_N$ is embedded diagonally in the product of the centers
of the two $SU(N)$'s.  This is a self-dual group. Its weight lattice is
the sublattice of the product
of the two $SU(N)$ weight lattices given by the condition
that the difference of weights is in the root lattice;
this is a self-dual lattice.  Note that since
the ${\bf Z}_N$ is embedded diagonally in $SU(N)\times SU(N)$,
an $({SU(N)\times SU(N)})/{\bf Z}_N$ bundle is an $(SU(N)/{\bf Z}_N)^2$ bundle
such that the two magnetic flux vectors are equal.
Allowing for the possibility of different
coupling constants for the two $SU(N)$'s, the partition function of
the $SU(N)\times SU(N)/{\bf Z}_N$ theory is
$$Z_{SU(N)\times SU(N)/{\bf Z}_N}=N^{-1+b_1}\sum_v Z_v(q_1)Z_v(q_2)$$
where we have denoted the instanton counting parameters as $q_1$ and $q_2$.
(The prefactor has the same origin as in \tuff.)
Since the group is self-dual,
$S$-duality says that $Z$ should be invariant (up to
a factor associated with the modular weight)
under simultaneous transformations $\tau_i\to -1/\tau_i$ for $i=1,2$.
In order to
have a nice action also for $T$,
it is convenient to choose $\tau_2=\bar \tau_1$.  This in
particular implies that the $\theta$ angles have opposite
signs and therefore the fractionality of the instanton number
\insno\ cancels between the two groups; so the partition
function is invariant under $\tau \rightarrow
\tau+1$.  With this choice
the partition function is therefore $SL(2,{\bf Z})$-invariant up to a factor
involving the modular weight; in what follows we divide by
a power of $|\eta|^{2w}$ to remove this factor.
The partition function so corrected is
$$\hat
Z_{SU(N)\times SU(N)/{\bf Z}_N}
=N^{-1+b_1}\sum_{v}|\hat Z_{v}(q)|^2$$
So we learn that the ``partition functions'' $\hat Z_v$
transform as a unitary representation of $SL(2,{\bf Z})$.
This is a consequence of equations \ttran\ and \stran, as we will
verify below; that is an important check on our ansatz.

\bigskip
\noindent{\it Analogy With Rational Conformal Field Theory}

This structure is reminiscent of rational conformal field theory,
with the $\hat Z_v$ playing the role of the conformal blocks.
To pursue the analogy, consider for simplicity the
case that $H^2(M,{\bf Z})$ is one dimensional with the lattice
generated by a vector $e$ with $e^2=-1$. (This situation arises for
${\bf CP}^2$ with the opposite of the usual complex orientation or
in connection with the blow-up of a point on a complex surface.)
Then the
partition functions are
$Z_{re}$ for $r=0,...,N-1$.

This is reminiscent of the partition function
of a two-dimensionsal
rational conformal theory with $N$ blocks.  In fact the $T$
transformation \ttran\ in this case
is exactly the same as for the $SU(N)$ level
1 WZW model; this model has
one block for each conjugacy class of $SU(N)$ and the conformal weight
for each one is given (mod 1) by $w^2/2$ where $w$ is a weight vector
in that conjugacy class -- precisely the quantity
appearing in \insf.  Therefore they
have the same $T$ matrix (up to at most an overall phase
coming from $c$).  Meanwhile, $S$
is
strongly constrained from the condition that $(ST)^3=1$
 and in fact the transformation law under $S$
in \stran\ coincides
(up to an overall sign) with that of the $SU(N)$ WZW theory at level 1.

This does not necessarily imply that the $\hat Z_{re}$'s
equal the characters of $SU(N)$ at level 1,
but only that they transform the same way.
In \S4, we will see that in a particular case -- involving
blow-ups and $G=SU(2)$ -- these functions do coincide.

\bigskip
\noindent{\it Verification Of Modular Behavior}

We now return to the general case of an arbitrary four-manifold $M$
(with no $N$-torsion in $H^2(M,{\bf Z})$, for simplicity).
We wish to verify that the formulas \ttran\ and \stran\ give
a unitary representation of $SL(2,{\bf Z})$.
The non-trivial point is to verify that $(ST)^3=1$.
(Given the known structure of $T$, this equation highly constrains
$S$ and under some assumptions on $M$ uniquely determines it; in
that sense our ansatz in \stran\ can nearly be derived from conventional
$S$-duality.)

In verifying that $(ST)^3=1$, we will also find some interesting
restrictions on $c$.
In particular we will find that $c=(N-1)\chi \  {\rm mod}\ 4$
for $ SU(N)$.
Later by studying certain examples we will prove
that for $SU(2)$, $c=\chi$ thus suggesting that the $SU(N)$
answer is also $c=(N-1)\chi$.  The same mod 4 condition can
be shown for simply laced groups with $(N-1)$ replaced
by the rank of the group.  For non-simply laced group $G$,
 it is tempting given the analogy with  conformal
field theory to conjecture that $c=c_1(G)\chi $ where
$c_1(G)={\rm dim}G/(1+h(G))$
 is the central charge of the WZW
theory with target $G$  at level 1 ($h(G)$ is the dual Coxeter number of  $G$).

The $\{ \hat Z_v\} $, form an $N^{b_2}$
dimensional representation of $SL(2,{\bf Z})$ that is described explicitly
in \ttran\ and \stran. One can use those formulas to check the statement
$(ST)^3=1$.  When $H^2(M,{\bf Z})$ is one-dimensional the fact that
the partition functions transform the same way as the
characters of $SU(N)$ at level 1 implies that
$(ST)^3=1$ (with a suitable choice of $c$).   It is easy
to generalize this to the arbitrary case.
 Using the self-duality of $H^2(M,{\bf Z})$ (and in particular
using \useq )
we find that $(ST)^3=1$ holds
up to an overall factor, which disappears if
\eqn\nurto{\left({\rm exp}(2\pi i c/24)\right)^3=\pm
N^{-b_2/2}\sum_v (-1)^{v^2}\omega^{v^2\over 2}}
where $\omega ={\rm exp}(2\pi i/N)$.  We will refer to the right
hand side of \nurto\ (with the $+$ sign) as $A_N(L)$.  $A_N(L)$
 depends on $N$ and on the lattice $L=H^2(M,{\bf Z})$ (with its canonical
quadratic form).  We would like to compute $A_N(L)$ for an arbitrary
self-dual lattice $L$  and verify \nurto. \foot{How to compute such
sums
is explained in \ref\serre{J.-P. Serre, {\it A Course in Arithmetic},
New York, Springer-Verlag (1973).}, as was pointed out to us by Dick Gross.}
 It is easy to prove the following properties
for $A_N(L)$:
$$\big| A_N(L) \big|=1$$
$$A_N(-L)=\bar{A_N(L)}$$
\eqn\multi{A_N(L_1\oplus L_2)=A_N(L_1)A_N(L_2).}
Here $-L$ denotes the same lattice lattice as $L$ except
with the opposite
sign for the quadratic form.
Let $I^{\pm}$ denote the one dimensional lattices
generated by a vector $v$ with $v^2={\pm 1}$.
Then \serre\ every indefinite odd
self-dual lattice\foot{An odd lattice is simply a lattice such
that $v\cdot v$ is odd for some $v$.}
is a direct sum of the form
$$r_+ I^+ \oplus r_- I^-$$
So we learn from \multi\ that for such lattices
\eqn\signa{A_N(L)=A_N(I^+)^{r_+}A_N(I^-)^{r_-}
=A_N(I^+)^{r_+-r_-}=A_N(I^+)^{\sigma (L)}}
where $\sigma =r_+-r_-$ is the signature of the lattice.
Moreover, it is known \serre\ that every self-dual lattice
can be transformed to an odd, indefinite self-dual lattice
by taking a direct sum,
if necessary, with  $I^+$ or $I^-$.  Therefore using \multi\ we learn
that \signa\ is true for all lattices, and we are left just with computing
$A_N(I^+)$:
$$A_N(I^+)={1\over \sqrt{N}}\sum_{s=0}^{N-1}(-1)^s {\rm exp}(2\pi i s^2/2N)$$
This sum can be computed using the Poisson resummation technique (the sum is
a simple generalization of so-called ``Gauss sums''
(see appendix 4
of \ref\husmil{D. Husemoller and J. Milnor, {\it Symmetric Bilinear
Forms}, New York, Springer-Verlag (1973).})).  We learn
\eqn\gnm{A_N(I^+)={\rm exp}{-2i\pi (N-1)\over 8}}
which implies
\eqn\prc{\exp{2\pi i c\over 8}=\pm {\rm exp}{-2i \pi (N-1) \sigma \over 8}}
and so
$$c=-(N-1) \sigma+4 \epsilon \qquad {\rm mod}\ 8$$
where $\epsilon =0,1$ depending on the $\pm$ in \prc .
We can rewrite $c$ as
$$c=(N-1)\chi -(N-1)(\chi +\sigma) +4\epsilon \qquad {\rm mod} \ 8$$
Note that if $N$ is odd, then $N-1$ is even; and
since $\chi+\sigma=0\ {\rm mod}\ 2$, $(N-1)(\chi +\sigma)=0\ {\rm mod}
\ 4$.
For $N$
even, the partition function
vanishes  unless $\chi+\sigma =0$ mod $4$ (otherwise
instanton moduli space is odd dimensional
and its Euler characteristic in the relevant sense is zero).
Thus in either
case for a non-trivial partition function we have $(N-1)(\chi+\sigma)
=0 \,\,\,{\rm mod} \,\,\, 4$, and so
\eqn\cchi{c=(N-1) \chi \qquad {\rm mod}\ 4}

We will see in \S4 that at least for $N=2$,
this equation for $c$ is true identically
and not just mod 4.  This will in particular imply
\eqn\epi{\epsilon ={(N-1)(\chi+ \sigma)\over 4}\qquad {\rm mod}\  2}
Note that we have thus determined the $\pm$ sign in \stran\
to be $(-1)^\epsilon$.  In the case of $SU(2)$ note
that $(-1)^\epsilon =(-1)^{\nu}$ where $\nu ={\chi+\sigma \over 4}$.
We will use this fact later on.

\newsec{Testing the Predictions of $S$-duality}

In this section we will test the predictions made in the
previous section in a small but satisfying set of examples.
The examples we consider, mostly with gauge group $SU(2)$,
are, in turn, $K3$, ${\bf CP}^2$, blow ups
of K\"ahler manifolds,
and ALE spaces.  These examples not only
provide strong coupling
tests of $S$-duality but also fix the quantities $c$ and $w$ which
were not completely fixed in \S3 to be for $SU(2)$
$c=\chi$ and $w=-\chi$.  We will also find a little bit of a  surprise
in the context of our discussion of ${\bf CP}^2$.  We find that the
$S$-duality predicts that there are holomorphic anomalies
(a dependence of the partition function on topologically trivial
observables)
in the topologically twisted $N=4$ super Yang-Mills theory.
These holomorphic anomalies are somewhat reminiscent
of holomorphic anomalies that arise for certain two-dimensional
topological theories \bcov .  There also may be a relation to a sort
of anomaly that arises in Donaldson theory on certain manifolds.
Both in the case of blowups and for ALE spaces, the analogy with
rational conformal field theory that we have seen already will reappear.

\subsec{$N=4$ Yang-Mills on $K3$}

For our first example, we study the partition function of $N=4$
Yang-Mills theory on $K3$.  This turns out to be computable because
of the existence of nice constructions of instantons on $K3$.
It is also a particularly nice example because -- as it is hyper-K\"ahler
-- the physical model coincides with the topological model if one
orients $K3$ correctly (so that holomorphic
vector bundles are instantons).
Furthermore, the vanishing theorem of \S2  applies to K3, so we
can expect to compute just in terms of instantons.

In general,
the dimension of the $SU(2)$ instanton moduli space ${\cal M}_k$
of instanton number $k$ is
\eqn\diminst{{\rm dim}{\cal M}_k =8k-{3\over 2}(\chi +\sigma)}
where $\chi$ and $\sigma $ are the Euler characteristic and the
signature of the manifold respectively.  For $K3$, one has $\chi=24$ and
$\sigma =-16$, so
$${\rm dim}{\cal M}_k=8k -12$$

If $E$ is an $SU(2)$ instanton bundle on $M=K3$ with instanton number $k$,
one can
seek \refs{\mukai-\others} a convenient description of
 $E$ by finding a line bundle
$L$ on $K3$ such that the index of the $\bar\partial $ operator
coupled to $L^{-1}\otimes E$ is   1.\foot{These matters were explained
to us by P. Kronheimer.}  It will then be generically
true that $H^0(M,L^{-1}\otimes E)$ has (up to scalar multiple) a single
holomorphic section $s$.  The number of zeroes of $s$, counted with
multiplicity, is equal to the second Chern class of $L^{-1}\otimes E$,
which (with $L$ and $E$ as stated) turns out to be $2k-3$.  Thus, if $L$
exists,
one has a natural way to extract from $E$  a configuration of $2k-3$
points on $K3$, namely the zeroes of $s$.
Conversely (when an appropriate $L$ exists), given
a configuration of $2k-3$ points on $K3$, one can reconstruct a unique
$E$, using a process of extension of sheaves due originally to Serre.

A configuration of $2k-3$ points on the four-manifold $M$ depends on
$8k-12$ real parameters.  This number equals the dimension of instanton
moduli space as given above.  In fact it is true in this situation
that the instanton moduli space can be identified with the space of
configurations of $2k-3$ distinct (but unordered) points on $K3$, and is,
in particular, hyper-K\"ahler.  If the points
are permitted to coincide, one gets a compact  hyper-K\"ahler
manifold with orbifold singularities.
A resolution of the singularities of that moduli space, preserving
the hyper-K\"ahler structure, gives the
algebrogeometric compactification of instanton moduli space.
Such resolutions  all have the same Betti numbers and Euler
characteristic and  can be studied by standard orbifold methods that we use
below.

A line bundle $L$ with the properties needed for the above construction
exists if and only if $k$ is odd and a suitable complex structure is
picked on $K3$.  (For different $k$'s one needs different complex structures.)
The restriction on the complex structure will not cause difficulties,
because the partition function we are trying to compute is independent
of the complex structure.  The restriction to odd $k$ will result
in the appearance of an extra function below.

The construction just sketched can also be carried out for $SO(3)$
bundles with non-zero $w_2$.  In this case, a suitable $L$ always
exists (if the complex structure on $K3$ is suitably chosen).  The
instanton number $k$ might be half-integral ($K3$ is a spin manifold
so a denominator of 4 cannot arise), but $2k-3$ is always integral
and the moduli spaces with $w_2\not=0$ have the Euler characteristic
of (a hyper-K\"ahler resolution of)
the symmetric product of $2k-3$ copies of $K3$.

Even though we do not have any convenient description of
the moduli spaces with $w_2=0$ and even $k$, we will be able
from the facts stated above both to test
$S$-duality and to use it to predict
the Euler characteristics in the missing cases.

\bigskip
\noindent{\it $SO(3)$ Bundles On $K3$}

We will need to know some facts about $SO(3)$ bundles on $K3$.
Of course, in addition to the instanton number, such a bundle $E$
is classified topologically by the ``magnetic flux'' $v=w_2(E)$.
As $H^2(K3,{\bf Z})$ is 22 dimensional (and torsion-free),
$v$ can take $2^{22}$ values.  In the $SO(3)$ theory, we must sum over them.

There is no need to study separately $2^{22}$ possibilities because
$K3$ has a very large diffeomorphism group which permutes the possible
values of $v$.  One obvious diffeomorphism invariant of $v$
is the value of $v^2$ modulo $4$; if it is 0 we call $v$ even and
if it is 2 we call $v$ odd.  If $v$ is odd, it is certainly non-zero,
but for $v$ even there is one more obvious invariant: whether $v$ is
zero or not.
It turns out that up to diffeomorphism, the invariants just stated
are the only invariants of $v$.
So on $K3$ there are really three partition functions to compute, namely
the partition functions for $v=0$, $v$ even but non-zero, and $v$ odd.
We call these
$Z_{0}, Z_{even}$, and $Z_{odd}$.  Similarly, we write $n_0$, $n_{even}$,
and $n_{odd}$ for the number of values of $v$ that are, respectively, trivial,
even but non-trivial, and odd.

Let us count how many $v$'s of each type there are.
It turns out that for this purpose,
the intersection form on $H^2(K3,{\bf Z})$
can be replaced by the sum of 11 copies of
\eqn\humpo{H=\pmatrix{0 & 1\cr 1 & 0 \cr}.}
This makes the combinatorics of counting the different kinds
of bundle straightforward -- in fact,
equivalent to the combinatorics of counting
the number of even and odd spin structure on a Riemann surface
of genus $11$, except that one must separate out the case of zero flux.
The result is
$$n_0=1 \qquad {\rm trivial \ type}$$
$$n_{even}={2^{22}+2^{11}\over 2}-1\qquad {\rm even \ type}$$
$$n_{odd}={2^{22}-2^{11}\over 2}\qquad {\rm odd\ type}$$
So in particular the $SU(2)$ and $SO(3)$ answers are
$$Z_{SU(2)}={1\over 2}Z_0$$
$$Z_{SO(3)}=Z_{0}+n_{even}Z_{even}+n_{odd}Z_{odd}$$
(The factor of $1/2$ in the first equation comes from the factor
of $1/\# Z(G)$ in \inssum.)

\bigskip
\noindent{\it Euler Characteristic Of A Symmetric Product }

According to the description sketched above of the instanton moduli
spaces on $K3$, we need to calculate the Euler characteristic
of a symmetric product of $K3$'s, that is, of the quotient
$(K3\times K3\times \dots \times K3)/{\bf S}_n$ of the product
of $n$ copies of $K3$ by the group ${\bf S}_n$ of permutations of
$n$ objects; ${\bf S}_n$ acts by permuting the factors.
The hyper-K\"ahler condition means that one can use
the orbifold description \ref\orb{L. Dixon, J. Harvey, C. Vafa and
E. Witten, Nucl. Phys. B261 (1985) 678; Nucl. Phys. B274 (1986)
285.}\ of the cohomology of
the quotient of a manifold by a finite group.
The result we need has been computed before
\ref\Gott{L. G\"ottsche, Math. Ann. 286 (1990) 193.
} and has even been obtained using the orbifold
formula \ref\hirz{F.
Hirzebruch and T. H\"ofer, Math. Ann. 286 (1990) 255.
}.  However, we will include a derivation
for completeness.  It is actually convenient, and no more difficult,
to apply the orbifold formula to the symmetric product of an arbitrary
manifold $M$ (the orbifold formula is only expected to agree with
the cohomology of a resolution of the symmetric product if $M$ is
hyper-K\"ahler or at least Calabi-Yau).
In the computation, we will aim to describe the cohomology of the
symmetric product, and not just compute the Euler characteristic.
This will make the meaning of the result clearer.

First we recall the general construction
of the cohomology, in the
orbifold sense,
of a quotient $X/\Gamma$, with $X$ a manifold acted on by a discrete
group $\Gamma$.
It is constructed as
\eqn\inon{H^*(X/\Gamma)=\oplus_\gamma H_\gamma}
where $\gamma$ runs over conjugacy classes in $\Gamma$ (for each
conjugacy class we pick a representative that we also call $\gamma$)
and $H_\gamma$
is the cohomology in the sector ``twisted'' by $\gamma$.  This is obtained
as follows.  Let $X^\gamma$ be the subset of $X$ left fixed by $\gamma$.
Let $N_\gamma$ be the subgroup of $\Gamma$ that commutes with $\gamma$.
Let $H^*(X^\gamma)$ be the cohomology of $X^\gamma$ and let
$H^*(X^\gamma)^{N_\gamma}$ be the part of $H^*(X^\gamma)$ that is
invariant under $N_\gamma$.  Then
\eqn\minon{H_\gamma =H^*(X^\gamma)^{N_\gamma}.}

In the case at hand, $X$ is the product $M^n=M\times M\times \dots M$
of $n$ copies of $M$, and $\Gamma$ is the group ${\bf S}_n$
of permutations of the factors.  We also want to take the sum over
$n$.  It is convenient, as we will see, to introduce an operator $L_0$
that acts on $H^*(M^n/{\bf S}_n)$ with eigenvalue $n$, and to write
the sum over $n$ as
\eqn\hipo{{\cal H}=\oplus_{n=0}^\infty q^nH^*(M^n/{\bf S}^n)}
where the powers of $q$ are a formal way to keep track of the $L_0$ action.
${\cal H}$ will turn out to be a kind of Fock space.

First let us work out using the definition in \minon\ the contribution
of the untwisted  sector.  For $\gamma=1,$ $M^\gamma=M^n$.  The
cohomology of $M^n$ is $H^*(M^n)=H^*(M)^{\otimes n}$,
the tensor product
of $n$ copies of the cohomology of $M$.  Also, $N_1={\bf S}_n$, the
full permutation group.  So  $H_1$ is the ${\bf S}_n$-invariant
part of $\left(H^*(M)\right)^{\otimes n}$.  This can be interpreted
as follows.  If we think of an element of $H^*(M)$ as a
``one particle state'' and an element of $H^*(M)^{\otimes n}$
as an ``$n$ particle state,'' then ${\bf S}_n$ invariance means
that we should think of the $n$ particles as
being identical bosons or fermions\foot{That is, the cohomology
classes of even and odd dimension correspond to bosons and fermions
respectively.}
and impose bose and fermi statistics.  If therefore we pick
a basis $w^a$ of the cohomology of $M$, and introduce a corresponding
set of ``creation operators'' $\alpha^a_{-1}$ acting on a ``Fock vacuum''
$|\Omega\rangle$, then the contribution of the untwisted sectors
to \hipo\ is a Fock space generated by the $\alpha^a_{-1}$.

We now want to show that the inclusion of twisted sectors results
instead in a
Fock space generated by a ``stringy'' set of
oscillators $\alpha^a_{-n}$ $n=1,2,3,\dots,$
which have $L_0=n$ just as one would expect in string theory.
We have to remember that every element of ${\bf S}_n$ can be decomposed
in disjoint cycles, for instance a permutation of six objects
might take the form of a one-cycle, a two-cycle, and a three-cycle,
often denoted $(1)(23)(456)$. The conjugacy classes in ${\bf S}_n$
are labeled by giving the number $n_l$ of $l$-cycles for $l=1,2,3\dots$.
The $n_l$ are arbitrary non-negative integers
except for the obvious restriction
\eqn\ponms{n=\sum_lln_l.}
$N_\gamma$ is of the form
\eqn\pms{N_\gamma=\prod_{l=1}^\infty N_\gamma(l)}
where $N_\gamma(l)$ is the subgroup of $N_\gamma$ consisting of permutations
that act non-trivially only on objects in the $l$-cycles of $\gamma$.
This factorization, which holds because the order of an object
under $\gamma$ is invariant under $N_\gamma$, is reflected in a similar
factorization of ${\cal H}$.

It is convenient to consider first the contributions of $l$-cycles
of a fixed $l$.  Suppose that $\gamma$ consists only of $l$-cycles,
say $k$ of them.  Then of course $n=kl$, and
$M^n$ is a product of $n$ copies of $M$, divided into $k$ sets
of $l$ such copies; each set is permuted cyclically by $\gamma$.
The fixed point set $(M^n)_\gamma$
is a product of $k$ copies of $M$, one for each $l$-cycle,  so its
cohomology is $H^*(M)^{\otimes k}$.  $N_\gamma$ again
acts on $H^*(M)^{\otimes k}$
by permuting the factors, and again this means that if we think of
the elements of $H^*(M)$ as ``one  particle states,'' then we should
impose bose and fermi statistics on the ``$k$ particle states''
in $H^*(M)^{\otimes k}$.  Also, we want to sum over $k=0,1,2\dots.$
So if we introduce a new set of creation
operators $\alpha^a_{-l}$, then the contribution to ${\cal H}$
from conjugacy classes consisting of $l$-cycles only is the Fock
space generated by the $\alpha^a_{-l}$ acting on the Fock vacuum.
There is one novelty: $\alpha^a_{-l}$ has $L_0=l$ because of the relation
$n=kl$, or if you will because taking the fixed points of $\gamma$
collapsed all the copies of $M$ in an $l$ cycle
into a single copy, the diagonal.

Now it is easy to put the results together. Once we sum over $n$
the restriction in \ponms\ loses its force and the $n_l$ are independent.
The fixed point set of a general permutation $\gamma$ contains a factor
of $M$ for each $l$-cycle in $\gamma$ of any $l$.
So the cohomology $H^*(M^n_\gamma)$ has a factor of $H^*(M)$ for each
cycle.
The symmetry
group $N_\gamma$  acts separately, according to \pms, on the cycles
of different order.  So bose and fermi statistics are only imposed
on ``particles'' of the same $l$.  The net effect of this is that
if we introduce the creation operators $\alpha^a_{-n}$ for all
positive integers $n$, then ${\cal H}$ is the Fock space generated
by the $\alpha^a_{-n}$, as claimed above.

For instance, if $b_+$ and $b_-$ are the dimensions of the bosonic
and fermionic subspaces of $H^*(M)$, then we can write a simple
result for the generating functional of the number of states,
\eqn\gs{\sum q^n\ {\rm dim}(H^*(M^{n}/{\bf S}_n))={\prod_{n=1}^\infty
(1+q^n)^{b_-}
\over \prod_{n=1}^\infty(1-q^n)^{b_+}}}
We can also write a simple result for the generating function
of the Euler characteristics,
\eqn\eul{\sum q^n \chi (M^{n}/{\bf S}_n)
={ \prod_{n=1}^\infty (1-q^n)^{b_-}
\over \prod_{n=1}^\infty (1-q^n)^{b_+}}={1\over
\prod_{n=1}^\infty (1-q^n)^{\chi(M)}}}
Happily, up to elementary factors, those formulas are modular.
In particular the generating function
for the Euler characteristic after multiplying by $q^{-\chi/24}$ becomes
\eqn\eull{G(q)=
q^{-\chi/24}\sum q^n \chi (M^{n}/{\bf S}_n)={1\over \eta^{\chi}} }
where $\eta $ is the Dedekind eta-function.

\bigskip
\noindent{\it Instantons On $K3$}

Now we return to the computation of Euler characteristic of
instantons on $K3$.  For $K3$, equation \eull\ gives
$$G(q)={1\over \eta^{24}},$$
which is none other than the partition function of left-moving modes
of the bosonic string!

We can now calculate the Euler characteristic of the moduli
spaces of instantons on $K3$.
It is easier to consider first the $SO(3)$ bundles with nonzero
$w_2$ because then the description by a symmetric product
of $K3$'s work for all values of the instanton number.
We recall that if the instanton number is $k$, then the number
of points is $n=2k-3$, so for even (or odd) bundles only
odd (or even) $n$ contributes.
Also the formula $n=2k-3$ shows that adding another copy of $K3$
adds only $1/2$ to the instanton number, so if we want to count
instantons with powers of $q$, we must count points on $K3$ with
powers of $q^{1/2}$.
So we can evaluate the partition functions
\eqn\dns{Z=q^{-s}\sum_{n}q^n\chi({\cal M}_n).}
If we pick $s=2$, we get
\eqn\mcn{\eqalign{Z_{even}(q)&={1\over 2}\left(G(q^{1/2})+G(-q^{1/2})
\right)\cr
Z_{odd}(q)&={1\over 2}\left(G(q^{1/2})-G(-q^{1/2})\right).\cr}}
These formulas are modular (of weight $-12$),
giving finally our first strong coupling test of
$S$-duality.  In writing \mcn, we added or subtracted
$G(q^{1/2})$ and $G(-q^{1/2})$ to project onto terms with an even or
odd number of copies of $K3$.

The formula $\dim{\cal M}_k=8k-12$ for the dimension of instanton
moduli space shows that the smallest value of $k$ for which ${\cal M}_k$
is non-empty\foot{Generically, and with an exception discussed below.}
is $k=3/2$.  The leading behavior
$Z_{odd}\sim q^{-1/2}$ of $Z_{odd}$ comes from an instanton of $k=3/2$;
its contribution is shifted from $q^{3/2}$ to $q^{-1/2}$ by the overall
factor of $q^{-2}$ that comes from picking $s=2$.
Similarly, the leading behavior $Z_{even}\sim 1$
of $Z_{even}$ is the shifted contribution from $k=2$.

The contribution of the trivial $SO(3)$ bundle on $K3$ can be treated
similarly, with one difference: since the description of the moduli
spaces by symmetric products of $K3$ is in this case only valid for
odd instanton number, we have to add an unknown function $F(q^2)$
contributing to the terms with even instanton number.
The partition function is thus of the form
\eqn\kthr{Z_{0}(q)=F(q^2)+{1\over 2}(G(q^{1/2})+G(-q^{1/2}))}
where we have multiplied by the same factor of $q^{-2}$.

Now we will make a precise test of $S$-duality in its sharpened
form of equation \stran.
This predicts the transformation law under $\tau\to -1/\tau$ to be
$$Z_{0}\rightarrow 2^{-11}\left({\tau \over i}\right)^{w/2} Z_{SO(3)}$$
$$Z_{even}\rightarrow 2^{-11}\left({\tau \over i}\right)^{w/2}\big[
Z_{0}+(2^{10}-1)Z_{even}-
2^{10}Z_{odd} \big]$$
\eqn\prdk{Z_{odd}\rightarrow 2^{-11}\left({\tau \over i}\right)^{w/2}\big[
Z_{0}+(-2^{10}-1)Z_{even}+2^{10}Z_{odd}\big]. }
To obtain these formulas from \stran, one needs to evaluate, for given
$v$, the sums
\eqn\runn{\sum_{u\,\,odd}(-1)^{v\cdot u},~~\sum_{u\,\,\,even}(-1)^{v\cdot u}.}
This is easily done using the fact that modulo two the
intersection form on $H^2(K3,{\bf Z})$ is equivalent to
11 copies of $H$, and picking a convenient choice of $v$.

Since we have incomplete information about $Z_{0}$,
we can eliminate it from the above to get the prediction that
under $\tau\to -1/\tau$,
$$Z_{even}-Z_{odd}\rightarrow ({\tau \over i})^{w/2}(Z_{even}-Z_{odd})$$
which is indeed the case
$$Z_{even}-Z_{odd}=G(-q^{1/2})\rightarrow \tau^{-12}G(-q^{1/2}).$$
In particular, the modular weight is $-12$, and $w=-24$.
But we can go further and determine $F(q^2)$. Using the
modular transformation properties
$$G(-q^{1/2})\rightarrow \tau^{-12}G(-q^{1/2})$$
$$G(q^{1/2})\rightarrow 2^{-12} \tau^{-12}G(q^2)$$
$$G(q^{2})\rightarrow 2^{12} \tau^{-12}G(q^{1/2})$$
we find that {\it all} of the equations \prdk\ are satisfied
if and only if
$$F(q^2)={1\over 4}G(q^2)$$
That there exists an $F$ which satisfies all three
equations at the same time is a very precise further test of
$S$-duality.

Note that the leading behavior of $F(q^2)$ for small $q$ is
$F(q^2)\sim q^{-2}+{\rm constant}$.
Thus the partition function $Z_0$ for zero flux has an expansion
\eqn\snons{Z_0\sim {1\over 4q^2}+O(1).}
The contribution of order $q^{-2}$ must be interpreted as the contribution
of the trivial connection, shifted from $q^0$ to $q^{-2}$ because
we have multiplied the whole series by $q^{-2}$.  There is indeed
one exception to the statement that instantons generically do not
exist for $8k-12<0$: the trivial connection always exists, even though
one might think that generically it shouldn't.  Evidently, it contributes
to the twisted $N=4$ theory, and (for $SO(3)$) its contribution has the
somewhat mysterious value $1/4$.  On the other hand, $8k-12$ is
negative for $k=1$, and for $k=1$ there are in fact generically no instantons.
That explains the absence of a term of order $q^{-1}$ in \snons.

Of course, we can reassemble our results into formulas for the $SU(2)$
and $SO(3)$ partition functions.
The $SU(2)$ partition function, allowing for the usual factor of $1/2$,
is
\eqn\plon{Z_{SU(2)}={1\over 8}G(q^2)+{1\over 4}G(q^{1/2})+
{1\over 4}G(-q^{1/2})}
This transforms as a modular form of $\Gamma_0(2)$, as it should,
since it is invariant under $T$ and transforms correctly under $ST^2S$.
This formula makes predictions for the Euler characteristics of the moduli
spaces with $w_2=0$ and even instanton number; the predictions
contain mysterious denominators of $1/4$ which must somehow
reflect the singularities of these moduli spaces.  The $SO(3)$ formula
is similarly
\eqn\glon{Z_{SO(3)}={1\over 4}G(q^2)+2^{21}G(q^{1/2})+2^{10}G(-q^{1/2}).}
and again has modular properties for  $\Gamma_0(2)$.

Since $w=-24$ and $s=-2$, the ``central charge'' defined in
\vardef\ is $c=24$.  This agrees with the
general arguments discussed in the last section \cchi\
which gave $c=\chi=24$ mod 4.  We see that the equality
is true even if we remove the mod $4$ condition.
We also see that at least in this case $w=-\chi=-24$.  Since in
general $c$ and $w$ are linear combinations of $\chi$ and $\sigma$
we need one other example with a different
ratio of $\chi$ and $\sigma$ to completely fix $c$ and $w$ in general;
we will find that the coefficients of $\sigma$ vanish (as one might
have expected
from parity conservation were it not that parity is explicitly
broken by the twisting used to construct a topological field theory).

Before leaving $K3$, we will point out a fairly natural
guess for the partition function for the
$SU(N)$ model on $K3$.
We will state the guess for the partition function $Z_0$ of bundles
with zero magnetic flux.  Contributions of the other bundles are determined by
modular transformations.  Our guess is
$$Z_0={1\over N^2}G(q^N)+{1\over N}\left[G(q^{1/N})+G(\omega q^{1/N})+...+
G(\omega^{N-1} q^{1/N})\right]$$
where $\omega ={\rm exp}(2\pi i/N)$.  The formula has some attractive
properties (but considerations
of its modular properties as well as some considerations explained
at the end of \S5.3
suggest that it may be valid, if at all, only for $N$ prime).
For $SU(N)$ the dimension of instanton moduli space on $K3$ is
$4kN-4(N^2-1)$, so after the trivial instanton at instanton
number 0, instantons appear first at $k=N$.  That agrees with
the fact that in the above formula
 $Z_0\sim q^{-N}/N^2+O(1)$ with a gap between $q^{-N}$
and $q^0$.  The modular transform of the formula for $Z_0$ can also
be seen to give the right gaps for bundles with non-zero flux.

The above  guess suggests that the  instanton moduli spaces for integral
instanton numbers that are not 0  mod
$N$ (or perhaps only those
 prime to $N$, in view of what we find in \S5.3),
 can be identified with  appropriate symmetric products of $K3$'s.
At least for instanton numbers 1 mod $N$ there are some indications of
this \ref\kronh{P. Kronheimer, private commmunication.}.
Our guess is also compatible with $c=(N-1)\chi$
in accord with \cchi, and with $w=-\chi$ (in other
words this guess suggests that the modular weight is independent
of the gauge group for a simple group; this is also supported
by the analysis discussed below on ALE spaces).

\subsec{${\bf CP}^2$}

We now consider the case of ${\bf CP}^2$, again with gauge
group $SU(2)$ or $SO(3)$.  ${\bf CP}^2$ is  another case in which
the vanishing theorems of \S2 apply, so it would seem
that we can compute from instantons only.  We will actually run into
a surprise, a kind of anomaly that affects holomorphy.

As $H^2({\bf CP}^2)$ is one-dimensional,
an $SO(3)$ bundle $E$ has two possible values of $v=w_2(E)$.
There is $v=0$, with $v^2=0$; and there is $v\not= 0$, with $v^2=-1$ modulo 4.
Accordingly, there are two partition functions,
\eqn\snln{\eqalign{Z_0& = q^{-s}\sum_{n}\chi({\cal M}_{0,n})q^n\cr
              Z_1 & = q^{-s}\sum_{n}\chi({\cal M}_{1,n})q^n,\cr}}
where ${\cal M}_{0,n}$ and ${\cal M}_{1,n}$ are, respectively,
moduli spaces of
bundles with $v=0$ and $v\not= 0$, and instanton number $n$.
To study these functions, we rely on formulas
of Yoshioka and Klyachko \refs{\yosh,\klyachko}.  Yoshioka determined
a formula for $Z_1$, while Klyachko determined formulas
for the related functions
\eqn\relfn{Y_i(q)=q^{-s}\sum_{n=0}^\infty \chi({\cal M}^{(0)}{}_{i,n})q^n,}
where ${\cal M}^{(0)}{}_{i,n}$ is the {\it uncompactified} moduli space
of instantons of instanton number $n$ and magnetic flux determined by $i$.

Klyachko's formula for $Y_1$ is
\eqn\nurg{Y_1(q)=3\sum_{n=1}^\infty H(4n-1)q^{n-{1\over 4}},}
where $H(m)$ is the Hurwitz function that equals the number
of equivalence classes of integral binary quadratic forms of discriminant
$m$, weighted by their number of automorphisms.\foot{$H(m)$ vanishes
unless $m$ is congruent to $0$ or $-1$ modulo 4.}
A formula of Yoshioka\foot{This follows from the Weil conjectures
and Theorem 0.4 of \yosh.}
shows that for any K\"ahler manifold
of Euler characteristic $\chi$, if we take $s=-\chi/12$ then
\eqn\rulln{Z_v(q) = {Y_v(q)\over\eta(q)^{2\chi}}}
for appropriate non-zero $v$; the formula applies for
$v\not= 0$ on ${\bf CP}^2$.  (The fact that the modular function $\eta$
appears here should hopefully be related to $S$-duality,
though we do not know how.)  Combining these formulas, one has
\eqn\ullno{Z_1(q)={3\over \eta(q)^6}\sum_{n=1}^\infty H(4n-1)q^{n-{1\over 4}}.}
(Yoskioka also obtained directly \yosh\ a formula for $Z_1$
which must coincide with this formula
-- we checked this for the first few terms -- but the modular properties
are easier to see in \ullno.)   Since the modular properties
of \ullno, which will be discussed presently, would be ruined
by multiplying by a power of $q$, the value of $s$ is the one
required to make the formula \rulln\ modular, or
\eqn\uncon{s=-{\chi\over 12},}
which is the same result that we found for $K3$.
Similarly, as the modular weight of $Z_1$ is $-3/2$ (in a suitable
sense that will be explained) we can
determine in general (using also the $K3$ result) that the modular
weight is $-\chi/2$ or that
\eqn\nuncon{w=-\chi.}
We will have a further check on \uncon\ and \nuncon\ when we consider
blowups.

Klyachko's formula for $Y_0$ has an extra term relative to
\nurg, but Yoskioka's analog of \rulln\ for $v=0$ is also more
complicated\foot{This formula is unpublished and we are grateful
to Yoshioka for providing it to us.}; combining them, it appears
that $Z_0$ is the obvious analog of \ullno:
\eqn\dullno{Z_0={3\over \eta^6}\sum_{n=0}^\infty H(4n)q^n.}
The series in \ullno\ and \dullno\ are known as Eisenstein series
of weight 3/2.  For a relatively elementary introduction to such series
see \koblitz, section IV.2, especially p. 194; see also
\ref\miyake{T. Miyake, {\it Modular Forms}, New York, Springer-Verlag (1989)
.}.  The simplest Eisenstein
series of half-integral weight for $\Gamma_0(4)$  are defined by
series such as
\eqn\simper{E_{k/2}(z)=\sum_\gamma j(\gamma,z)^{-k}.}
Here the sum runs over certain cosets of $\Gamma_0(4)$, and
$j(\gamma,z)^{-k}$ generalizes the factor $(mz+n)^{-k}$ that
appears in the conventional Eisenstein series
\eqn\defjo{G_k(z)=\sum_{m,n}(mz+n)^{-k}.}
For $k$ an odd integer $\geq 5$, \simper\ defines a modular form
of weight $k/2$ for $\Gamma_0(4)$.  However, for $k=3$, the sum
in \simper\ does not converge quite well enough to define a holomorphic
modular form.  To define a modular object, one regularizes the sum,
replacing \simper\ with
\eqn\himper{E_{3/2}(z,s)=\sum_\gamma j(\gamma,z)^{-3}|j|^{-2s}.}
With this regularization, $E_{3/2}(z,s)$ transforms as a modular
form of weight $3/2$ (for any fixed $s$) but is not holomorphic in $z$.
One can actually take the limit $s\to 0$; the function
$E_{3/2}(z)=E_{3/2}(z,0)$ that one obtains this way is modular
but not holomorphic.  The story is just analogous for the ordinary
Eisenstein series \himper\ for $k=2$, which similarly does not quite converge
well enough to define a holomorphic modular form.

As was discovered twenty years ago by Zagier
\zag\ (see also \hirzag ),
the process just described, starting with a slightly different
Eisenstein series of weight $3/2$, gives such functions as
\eqn\gufus{\eqalign{f_0=&\sum_{n\geq 0}3H(4n) q^{n}+
6{\tau_2}^{-1/2}\sum_{n\in Z} \beta (4\pi n^2\tau_2 ) q^{-n^2}\cr
f_1=&\sum_{n>0} 3H(4n-1)q^{n-{1\over 4}}+
6{\tau_2}^{-1/2}\sum_{n\in Z} \beta (4\pi (n+{1\over 2})^2
\tau_2 ) q^{-(n+{1\over 2})^2},\cr}}
where $q=e^{2\pi i \tau}$, $\tau_2={\rm Im}(\tau)$, and
$$\beta (t)={1\over 16 \pi}\int_1^\infty u^{-3/2} {\rm exp}(-u t)\ du$$
Thus, the modular status of \ullno\ and \dullno\ is clear:
these functions are the ``holomorphic part'' of the
non-holomorphic modular functions in \gufus\ (divided
by $\eta^6$).   The ``holomorphic part''
can be more formally defined by
considering the limit of $\bar \tau \rightarrow \infty$ while
keeping $\tau$ fixed.  It is easy to see that the
additional non-holomorphic terms vanish in this limit.

Moreover\foot{According to an unpublished argument kindly explained
to us by D. Zagier.}
under $\tau\to- 1/\tau$, these functions transform
as
\eqn\sonsp{\pmatrix{f_0(-1/\tau) \cr f_1(-1/\tau)\cr} =\left({\tau\over i}
    \right)^{3/2}\cdot\left(-{1\over \sqrt 2}\right)\pmatrix{ 1 & 1\cr
                                                              1 & -1\cr}
                                 \pmatrix{f_0(\tau) \cr f_1(\tau)\cr}.}
This is exactly the transformation law predicted in \stran.
Since $f_0$ is invariant under $T$
and $f_1$ under $T^4$, it follows, for
instance, that $f_0$ has the expected modular transformation law
under the group $\Gamma_0(4)$ generated by $T$ and $ST^4S$.

\bigskip
\noindent{\it Holomorphic Anomaly?}

Thus, all is well except that the $Z_v$ are not holomorphic.
This is reminiscent of the holomorphic anomaly in certain topological
two-dimensional models \bcov .
The situation there is as follows:  One has an untwisted physical
theory with coupling constants $\tau $ and $\bar \tau$, parametrizing
some moduli space,
appearing in the
action.  In the physical theory $\bar \tau $ is the complex conjugate
of $\tau$.  One then considers the twisted version and shows that
the $\bar \tau$ dependence
 in the action is of the form ${\bar \tau}\{ Q,...\}$, and
so formally the path integral is independent of $\bar \tau$.  One
then takes the unphysical limit $\bar \tau \rightarrow \infty$
keeping $\tau$ fixed to argue
that the twisted theory computes the Euler characteristic
of a  certain moduli space.  However, under certain circumstances \bcov,
the formal independence of the path integral from $\bar \tau$
is anomalous and  the twisted theory has nice modular
properties only when $\bar \tau$ is chosen
to be the complex conjugate of $\tau$.  Nevertheless the topological
answer is recovered, and modularity lost, by taking
${\bar \tau}\rightarrow \infty$ and keeping $\tau $ fixed.
This is very similar to the above situation for the $N=4$
twisted Yang-Mills on ${\bf CP}^2$, with the obvious substitutions.
In particular the S-duality conjecture seems to suggest
that the twisted physical theory on
${\bf CP}^2$ has $\bar \tau$ dependence.

Another precedent, perhaps equally relevant, is Donaldson theory
which  (as Donaldson discovered in one of his early papers
\ref\don{S. Donaldson, ``Irrationality And The $h$-Cobordism Conjecture,''
Jour. Diff. Geom. {\bf 26} (1987) 141.})
fails to produce topological invariants
on manifolds with $b_2{}^+=1$. ${\bf CP}^2$ has $b_2{}^+=1$
(though in Donaldson theory it is not a typical example with $b_2{}^+=1$
since also $b_2{}^-=0$).

The anomaly in Donaldson theory comes from abelian connections where
zero modes appear for the fields called $\phi,\,\bar\phi$ in \S2.
Though this anomaly has not been given its proper physical expression,
it very plausibly involves the essential failure of compactness
of field space due to the flat directions in the $\phi$ potential.
(Those flat directions are suppressed when the gauge field has $SU(2)$
holonomy.)
The condition $b_2{}^+=1$ for the anomaly looks like it might be
natural from this point of
view, since for $b_2{}^+>1$ there are extra
fermion zero modes along the flat directions.

The behavior of the twisted $N=4$ theory on ${\bf CP}^2$ suggests that
it too may have an anomaly associated with abelian configurations
when $b_2{}^+=1$.  In fact, the instanton numbers of abelian
configurations on ${\bf CP}^2$ are $-n^2$ for $v=0$ and
$-(n+1/2)^2$ for $v=1$.
These are precisely the exponents that appear in the anomalous,
non-holomorphic sum in \gufus.
The failure of holomorphy of the $f_i$ can be summarized in the
elegant equations
\eqn\anom{\eqalign{
{\partial\over \partial\bar \tau} f_0&={3\over 16 \pi i}\tau_2^{-3/2}
{\sum_{n\in
Z}{\bar q}^{ n ^2}}\cr
{\partial\over \partial\bar \tau} f_1&={3\over 16 \pi i}\tau_2^{-3/2}
{\sum_{n\in
Z}{\bar q}^{ (n+{1\over 2})^2}},\cr}}
which give further encouragement for seeking a field theoretic
explanation of the anomaly.

The formulas of Yoskioka \yoshi\
for instantons on rational ruled surfaces (which all have $b_2{}^+=1$) strongly
suggest that on this whole class of manifolds there is some sort of
anomaly generalizing the one that we have found experimentally
on ${\bf CP}^2$.

\subsec{Blow Ups}

Our next example will involve the behavior under blowing
up a point on a K\"ahler manifold.  Thus, we let $M$ be
a K\"ahler manifold, and $\hat M$ the new manifold obtained
by blowing up a point $p\in M$.  Topologically, the effect of the
blow-up is to glue in into $M$ a copy of $\overline{{\bf CP}}^2$ (that is,
${\bf CP}^2$ with the opposite of the usual complex orientation).

The effect on the second homology is that the lattice $L$ of $M$
is replaced by the lattice
\eqn\itscomp{\hat L= L \oplus I^-}
for $\hat M$
(where $I^-$, as in \S3,
is the one dimensional lattice with quadratic form $-x^2$).
If $E$ is an $SO(3)$ bundle on $\hat M$, then $\hat v=w_2(E)$
can be decomposed as
\eqn\jimcon{\hat v= v\oplus r}
under the decomposition
\itscomp.

In equation \stran, we proposed a precise formula for how the
partition functions transform under $S$-duality.  Since $\hat L$
is a direct sum, the formula implies that the action of $SL(2,{\bf Z})$ on
$Z_{\hat v}=Z_{v,r}$ is the product of a representation acting
on the $v$ label and a representation acting on the $r$ label.
Thus, the action of $SL(2,{\bf Z})$ is compatible with the possibility
that the $Z$'s have a factorization as
\eqn\unson{Z_{\hat M,\hat v}=Z_{M,v}Z_r }
where for clarity we write the manifold $\hat M$ or $M$ explicitly, but
the function $Z_r$ could be ``universal,'' independent of $M$.

The possible $Z_r$'s can be analyzed just as in \S3.3,
where we considered a manifold with the intersection form $I^-$;
if a factorization such as \unson\ holds, that discussion can be applied
to the blowup of an arbitrary $M$.
In that discussion, we pointed
out that (for $SU(N)$, not just $SU(2)$) the modular transformations
of the $Z_r$ are those of the level one characters of the $SU(N)$
WZW model.  However, the modular transformations alone do not
imply that a factorization such as \unson\ holds, or if it does
that the $Z_r$ are the WZW characters.

These assertions are, however, true, at least in one important case,
according to another formula of Yoshioka (\yosh,
proposition (0.3)).  He demonstrates
that under blowup of a K\"ahler manifold,
the factorization of \unson\ holds for $SU(2)$ and $r=0$, with
\eqn\zoooo{Z_0={\theta_0(q)\over\eta(q)^2},}
where
\eqn\xonx{\theta_0(q)=\sum_{n\in {\bf Z}}q^{n^2}.}
This is indeed $1/\eta$ times the appropriate $SU(2)$ character
at level 1.

Under blowup, the Euler characteristic of a manifold is increased
by 1.  On the other hand, the function in \zoooo\ behaves for small
$q$ as $q^{-1/12}$ and transforms as a modular form of weight
$-1/2$.  This gives a further check on our earlier finding that
the ``zero point'' of the instanton number is $-\chi/12$ and
the modular weight of the partition function is $-\chi/2$.

$S$-duality implies that the factorization
\unson\ must also hold for $r=1$ and that
\eqn\zooom{Z_1={\theta_1(q)\over\eta(q)^2},}
where
\eqn\xonxo{\theta_1(q)=\sum_{n\in {\bf Z}+1/2}q^{n^2}.}

\subsec{ALE Spaces}

For our last example, following Nakajima \nakajima,
we consider instantons on ALE (asymptotically locally Euclidean)
spaces.  In many ways this is the richest example, as precise
information is available for $U(k)$ or
$SU(k)$ gauge group, not just $SU(2)$,\foot{The partition function
of the $U(k)$ theory would vanish on a compact four-manifold because of
fermion zero modes in the $U(1)$ factor,
but that is not so on the ALE spaces, where there are no normalizable
zero modes. Nakajima's results are most elegantly stated for $U(k)$.}
and one
encounters WZW models at arbitrary levels, not just level 1.  Unfortunately,
it is hard to exploit the examples fully as one does not know enough
about the implications of $S$-duality on non-compact manifolds.

The ALE spaces $X_\Gamma$ we have in mind arise as hyper-K\"ahler
resolutions of hyper-K\"ahler orbifolds of the form
${\bf C}^2/\Gamma$, where $\Gamma$ is a finite subgroup of $SU(2)$
acting linearly on ${\bf C}^2$.
These spaces are noncompact relatives of $K3$.  Indeed, $K3$ has a somewhat
analogous construction beginning with the quotient of a four-torus
by a finite group, and the singularities so obtained have the local
structure of ${\bf C}^2/\Gamma$.

Every finite subgroup $\Gamma$ of $SU(2)$
has, up to isomorphism, a finite set of irreducible
complex representations $\rho_i$.  If $\rho$ is the two-dimensional
representation associated with the embedding of $\Gamma$ in $SU(2)$,
one can decompose the tensor product $\rho\otimes\rho_i$
as the direct sum of $n_{ij}$ copies of $\rho_j$; the $n_{ij}$
are all 0 or 1.  Form a graph with the $\rho_i$ for vertices,
with two vertices $\rho_i$, $\rho_j$ connected by a line if
and only if $n_{ij}\not= 0$.  It turns out that the graph
so obtained is the extended
Dynkin diagram of a simply laced Lie group $H_\Gamma$.
The correspondence $\Gamma\leftrightarrow H_\Gamma$ is
the McKay correspondence between finite subgroups of $SU(2)$
and simply laced Lie groups.  (The subgroup ${\bf Z}_N$ of $SU(2)$
corresponds to ${\bf A}_{N-1}$, the dihedral groups correspond to the
${\bf D}_N$, and the symmetries of the three regular solids correspond
to the ${\bf E}$ series.)

The Dynkin diagram of $H_\Gamma$ appears in the geometry of $X_\Gamma$
as the intersection form of two-spheres ${\bf S}_i$ created in blowing
up the
singularities of $X_\Gamma$.  These two-spheres
give a basis of $H_2(X_\Gamma,{\bf Z})$, and the first Chern class of
a vector bundle over $X_\Gamma$ can be interpreted as a weight of $H_\Gamma$.

The hyper-K\"ahler metrics on $X_\Gamma$ are asymptotic at infinity
to ${\bf R}^4/\Gamma$.
The fundamental group at infinity is non-trivial, and admits
non-trivial homomorphisms
\eqn\lsnl{\phi:\Gamma\to G}
to the gauge group $G$.  These homomorphisms enter because requiring
that the action of an instanton should be finite does not imply
that the instanton approaches the trivial connection at infinity,
but only that it should be flat at infinity, that
is, given by a homomorphism from $\Gamma$ to $G$.

Nakajima analyzes instantons with gauge group $U(k)$ on
$X_\Gamma$.
Such an instanton determines at infinity a $k$
dimensional representation \lsnl\ of $\Gamma$, which decomposes
as a sum of $n_i$ copies of the irreducible representation
$\rho_i$, with
\eqn\psons{k=\sum_in_i\,{\rm dim}(\rho_i).}
Since the $\rho_i$
can be interpreted as points on the extended Dynkin diagram of $H_\Gamma$,
one gets a natural assignment of integers $n_i$ to the points of the
Dynkin diagram.  Those points also correspond to simple roots $e_i$
of $H_\Gamma$, so we get an assignment of integers to simple roots.
This enables us to form the sum $w_\phi=\sum_in_ie_i$, which is a positive
weight of $H_\Gamma$, so it determines a representation of $H_\Gamma$.
Even better, equation \psons\ is precisely the condition
that the representation so
determined is a highest weight of an integrable
representation of $H_\Gamma$ current algebra at level $k$.
The instanton number of these instantons is shifted from
being an integer by an amount that depends on $\phi$.

Nakajima now considers the middle dimensional cohomology $H_{\phi,c_1,n}$
of the moduli space of $U(k)$ instantons with a representation $\phi$
at infinity, arbitrary first Chern class $c_1$,
 and instanton number $n$. ($n$ is shifted from an integer by an amount
that depends on  $\phi$.)   More precisely, he considers
the cohomology of some complete hyper-K\"ahler manifolds
obtained by a particular partial compactification  and deformation of
singularities of the moduli spaces.

To describe  the results, introduce an operator
$L_0$ that has eigenvalue $n$ on $H_{\phi,c_1,n}$, and introduce the
sum
\eqn\hphi{{\cal H}_\phi=\oplus_{c_1,n}q^n H_{\phi,c_1,n},}
where the variable $q$ is formally included to keep track of the
$L_0$ action.  Nakajima's remarkable result is that ${\cal H}$
has a natural structure of irreducible representation of $H_\Gamma$
current algebra at level $k$, with the highest weight being $w_\phi$.
$L_0$ acts by multiplication by $n$, while $c_1$, interpreted as we mentioned
above as a weight of $H_\Gamma$, determines the action of the Cartan
subalgebra of $H_\Gamma$. The rest of the
action of the affine Lie algebra is defined through
some operations of twisting vector bundles along the ${\bf S}_i$;
these operations conceivably should be regarded as
analogs of vertex operators in conformal field
theory.

At any rate, the functions
\eqn\hphi{Z_\phi(q)=\sum_{c_1,n}q^{n-c/24}{\rm dim}\,H_{\phi,c_1,n}}
are therefore the characters of integrable representations of the loop
group of $H_\Gamma$.  ($c$ is the central charge of $H_\Gamma$ current
algebra at level $k$.)   So they
have modular properties and in fact transform in a unitary representation
of $SL(2,{\bf Z})$.  Because of not understanding the implications
of $S$-duality on manifolds with boundary, and for other reasons
mentioned presently, we do not know how to explain the particular
representation that arises.

To interpret the facts just sketched  in terms of $S$-duality,
one might hope that the middle dimensional cohomology of the moduli
spaces coincides with the ${\bf L}^2$ cohomology, in which case
${\rm dim}\,H_{\phi,c_1,n}$ can perhaps be
interpreted as a kind of Euler characteristic.
Then one would hope to deduce the modularity of \hphi\ from
$S$-duality.
Unfortunately, there are difficulties with this interpretation.
In particular the partition function of the twisted $N=4$
theory gives the Euler characteristic defined as a kind of curvature
integral, not the Euler characteristic of ${\bf L}^2$ cohomology of the
moduli space; therefore it is not obvious
how to compare Nakajima's results to $S$-duality.

Even if one could predict from $S$-duality the modular properties
of the ALE partition function,
that would not do justice to Nakajima's story, which involves
construction of a natural Hilbert space with group action
and not just a partition function.  It does not seem that a natural
origin of this would be in four-dimensional field theory
-- which associates Hilbert spaces with quantization on a three-manifold,
and a partition function with the path integral on a four-manifold,
but does not normally produce a Hilbert space associated
with a four-manifold.  It would seem that the full story here
should have its origins in a supersymmetric theory of dimension $\geq 5$,
which,
in quantization on $X_\Gamma\times {\bf R}$ (where ${\bf R}$ is
the ``time'' axis) would produce a Hilbert space of physical
states.  The middle dimensional cohomology of the instanton moduli
spaces would appear as a set of BPS-saturated states,
perhaps making it
possible to do justice to the results of \nakajima.
Note that the dimension of the cohomology of the instanton number
$k$ moduli space (on $K3$ or an ALE space) grows
exponentially
like ${\rm exp}(a {\sqrt k})$ for some constant $a$.
This implies that in a five dimensional
theory that would be relevant to Nakajima's results,
the multiplicity of particles of mass $m$
goes as ${\rm exp}(a {\sqrt m})$ for some constant $a$.  This
growth is huge for a field theory though it is somewhat less than what one
encounters in string theory (whose spectrum
grows as ${\rm exp}(a  m)$ because
$m^2=k$).  A natural origin for such a huge  BPS-saturated
spectrum may well require
string theory.  Anyway, the
only sensible theory that we know of in dimension $\geq 5$
is string theory, so some of Nakajima's
results may have their natural origin in $S$-duality of string theory.

\bigskip
\noindent{\it $SU(k)$}

It is also of interest to extract from Nakajima's formulas
the results for gauge group $SU(k)$.  For $SU(k)$ on compact manifolds,
we worked out precise predictions of $S$-duality in \S3.
We would like to see what happens in the ALE case.

If one wishes to consider $SU(k)$ rather than
$U(k)$ on $X_\Gamma$, one must restrict to bundles with $c_1=0$.
As the action of the Cartan subalgebra of $H_\Gamma$ is proportional
to $c_1$, this entails restriction to states annihilated by the maximal torus
$T$.  This suggests that the ``partition functions,'' in the above
sense, of the $SU(k)$ theory would be the characters of
the coset conformal field theory $H_\Gamma/T$, times the contribution
of extra oscillators coming from $T$ (which  give a non-zero modular weight).
This appears to be
so, though a verification requires a careful study of the
(apparently non-standard) way that Nakajima's partial compactification
affects the $U(1)$ factor.

Let us specialize to the
the ALE space $X_\Gamma$ with $\Gamma ={\bf Z}_N$, so $H_\Gamma=SU(N)$.
A representation $\phi$ of the fundamental group at infinity determines
a highest weight $w_\phi$ of $SU(N)$ and also a representation
$\Lambda$ of the loop group of $SU(N)$ at level $k$.
Nakajima's result for $U(k)$
means that (with the partition function $Z$ interpreted in his sense)
$$Z_{H_{\bf \Gamma}}(\tau ,U(k),\phi )=\chi^{\Lambda }(q)$$
where $\chi^\Lambda$ is the character  of the representation $\Lambda$.
$\chi^\Lambda$ can be written
\ref\kacbook{V.G. Kac,
{\it Infinite dimensional Lie algebras}, Birkhauser, Boston, 1983.}
$$\chi^{\Lambda}=\sum_{\lambda}c^{\Lambda}_\lambda \Theta_{\lambda,k}$$
where the $c^{\Lambda}_\lambda $, known as ``string functions,''
are characters of the $SU(N)/T$ coset model times the partition
function of oscillators of $T$, and
the $\Theta_{\lambda,k}$ are certain theta functions.
The $\lambda$'s are as follows: $\lambda$ is an $SU(N)$ weight
that is congruent to $w_\phi$ modulo the root lattice $L$; also one
identifies $\lambda\cong \lambda'$ if $\lambda-\lambda'\in kL$.
Thus
$\lambda$ takes $(N-1)^k$ values.  Since $L$ can be identified
with $H_2(X_\Gamma)$, $\lambda$ can be identified with the ${\bf Z}_k$-valued
magnetic flux.
The ${\bf Z}_k$
magnetic flux and $c_1$ (which is the magnetic flux for the center
$U(1)$ of $U(k)$) are correlated in this way simply because
$$U(k)={SU(k)\times U(1)\over Z_k}$$
where $Z_k$ is a diagonal subgroup of the center of $SU(k)$ and a $Z_k$
subgroup of $U(1)$.

How does $c^\Lambda_\lambda$ transforms under
$\tau \rightarrow {-1/\tau}$?  As is well known in the
study of coset models, the answer is the product of a matrix acting
on the $\Lambda$ index and a matrix acting on $\lambda$:
\eqn\strin{c^\Lambda_\lambda(-1/\tau)=({\tau \over i})^{-(N-1)/2}
{\rm const.}
\sum_{\Lambda ',\lambda'}
 b(\Lambda ,\Lambda ',\lambda ,\lambda ')c^{\Lambda'}_{\lambda'}(\tau)}
where
\eqn\nice{b(\Lambda,\Lambda',\lambda,\lambda')={\rm exp}(2\pi i
\lambda \cdot \lambda' /k)\sum_{w\in W} \epsilon (w)\ {\rm exp}
(2\pi i (\Lambda +\rho)w(\Lambda'+\rho)/k+N-1)}
where $W$ is the Weyl group of $SU(N)$ and $\rho$ is half the
sum of positive roots of $SU(N)$.  The factor in \nice\
that acts on the the magnetic flux
$\lambda$  is consistent with what one expects in
\stran .
The factor acting on $\Lambda$, which is determined by the boundary
conditions, is more mysterious;  we do not understand why
the modular transformations of the $SU(N)$ current algebra appear here.
The formula can actually be motivated to a certain extent by considering
$S$-duality for (purely bosonic) $U(1)$ gauge theory on $X_\Gamma$; in that
case one can see that $S$-duality induces a Fourier transform on the boundary
conditions, a result similar to the large
$k$ limit of the $\Lambda$-dependent factor in \nice.

Note also that \strin\ tells
us that the modular weight is determined by
$w=-(N-1)$, which is in some agreement with our earlier formulas
as $N-1$ is the Euler characteristic of the ${\bf L}^2$ cohomology
of $X_\Gamma$.  It is interesting that this value is independent
of the gauge group.
The value of $c$, however, is fractional now, and
deviates from the formula $c=(k-1)\chi$ that we found for compact
four-manifolds without boundary.

 \newsec{Computation By Physical Methods}

\nref\switten{E. Witten, ``Supersymmetry Yang-Mills Theory On A
Four-Manifold,'' IASSNS-HEP-94/5.}
\subsec{Reduction To $N=1$}

In this section, by imitating an analogous computation for Donaldson
theory \switten, we will analyze the partition function of the
twisted $N=4$ theory (with gauge group $SU(2)$ or $SO(3)$)
on an arbitrary K\"ahler manifold $X$ with
$H^{2,0}(X)\not= 0$. An answer emerges that satisfies an impressive
set of constraints and gives support therefore to the overall picture
of $S$-duality.

The key input, mentioned at the end of the introduction to
\S2, is that on a K\"ahler
manifold the theory actually has four topological symmetries, one
from each of the four underlying supersymmetries.  Therefore, a perturbation
that breaks the $N=4$ supersymmetry down to $N=1$ would still leave
us with one topological symmetry, enough to control the theory
by reducing computations to classical configurations and quantum
vacuum states.  We can attempt to find an $N=1$-invariant perturbation whose
addition to the theory brings about some simplification.

The $N=4$ theory can be viewed as an $N=1$ theory with three chiral
superfields, say $T$, $U$, and $V$, in the adjoint representation of
the gauge group.
The superpotential is
\eqn\wis{ W= \Tr \,T[U,V]. }
For the time being the gauge group is an arbitrary compact simple
Lie group $G$.
While preserving $N=1$ supersymmetry, we could add a perturbation to
the superpotential, say the mass term
$\Delta W =- {1\over 2}m\Tr \,(T^2+U^2+V^2)$ (or any other nondegenerate
quadratic expression, which would have essentially the same effect).
As in \switten, the  resulting
perturbation of the Lagrangian
is the sum of a term of the form $\{Q,\dots\}$ plus a BRST-invariant operator
${\cal O}$ of positive ghost number.  In Donaldson theory, ${\cal O}$
has non-vanishing
matrix elements and affects the theory in an important, but calculable,
way.  A simplification occurs for $N=4$: the ``vacuum'' has zero ghost
number (as there is no anomaly), so all matrix elements of ${\cal O}$ vanish;
hence the partition function is invariant under the perturbation.

Now let us find the vacuum states of the theory.  First we do this classically.
The superpotential including the perturbation is
\eqn\what{\widehat W = - {1\over 2}m\Tr\,(T^2+U^2+V^2)+\Tr\, T[U,V].}
The equations for a critical point of $W$ are
\eqn\sutwo{\eqalign{ [U,V] & = m T\cr
                     [V,T] & = m U \cr
                     [T,U] & = m V .\cr}}
These equations have the following meaning: apart from a factor
of $m$, which can be scaled out, they are the standard commutation
relations of the Lie algebra of $SU(2)$.  To complete the determination
of the classical vacuum states, one must divide the space of solutions
of \sutwo\ by the gauge group $G$, and set the $D$ terms to zero.
Those two steps combined are equivalent to dividing by the complexification
$G_{{\bf C}}$ of $G$.

The conclusion then is that the classical vacua are in one to one
correspondence with complex conjugacy classes of homomorphisms
of the $SU(2)$ Lie algebra to that of $G$.  For instance, for $G=SU(m)$,
there is a single irreducible representation -- since
$SU(2)$ has up to conjugacy a single irreducible $m$ dimensional
representation -- and various reducible representations.  The irreducible
embedding breaks the gauge symmetry completely, while the reducible
embeddings leave various subgroups of $G$ unbroken.
At the opposite extreme from the irreducible solution
of \sutwo\ is the trivial embedding, with $T=U=V=0$.  Here the unbroken
symmetry group is $G$ itself.

In what follows, we will mainly consider $G$ to be $SU(2)$ or $SO(3)$,
in which case the
only embeddings are the irreducible one and the trivial one.
For other groups, there are other, intermediate cases.  For $G=SU(N)$ with
prime $N$, a simplification arises:
the two extreme cases are the only
ones in which the
unbroken symmetry group is semi-simple; the vacua coming from
reducible but non-trivial embeddings all have
$U(1)$ factors in the unbroken symmetry group.
The partition function of the twisted $N=4$ theory vanishes for gauge
group $U(1)$ on any four-manifold, because of fermion zero modes
(which cannot be lifted as the $U(1)$ theory is free).  This
presumably means that in the
generalization of the computation that follows to $SU(N)$ with
prime $N$, the intermediate
vacua do not contribute.  That would not be so for other groups.
For instance, if the gauge group is $G=SU(nm)$ with $n,m>1$, one can consider
an embedding of $SU(2)$ in $G$ such that the $nm$ dimensional representation
of $G$ decomposes as $n$ copies of the $m$ dimensional representation
of $SU(2)$; for this embedding, the unbroken gauge group is $SU(n)$,
and hence one  would expect the corresponding vacuum to contribute
to the $SU(nm)$ theory.

\bigskip
\noindent{\it Quantum Vacua}

Now we consider the vacuum structure at the quantum level.
The trivial embedding gives at low energies the pure $N=1$ supersymmetric
gauge theory; the supermultiplets $T,U,$ and $V$ are all massive.
The pure $N=1$ super Yang-Mills theory is asymptotically free and is
believed to generate a mass gap (and to undergo confinement) at low
energies.  For gauge group $SU(m)$, the model
has a ${\bf Z}_{2m}$ global chiral
symmetry,\foot{This is explicitly broken by the coupling to the massive
superfields $T$, $U$,
and $V$, but the breaking is irrelevant at low energies and does
not affect the following remarks.}  which is believed to be spontaneously
broken down to the ${\bf Z}_2$ subgroup that permits fermion masses.
This gives an $m$-fold vacuum degeneracy, which is believed to account
for the full vacuum degeneracy of the theory.

The chiral symmetry and chiral symmetry breaking have
the following interpretation.
The gluino field $\lambda$ of the $N=1$
theory has $2m$ zero modes in an instanton field; the chiral symmetry
is therefore $\lambda\to e^{\pi i/m}\lambda$.  In an instanton field,
one finds an expectation value for the operator $(\lambda\lambda)^m$:
\eqn\pilp{\langle (\lambda\lambda)^m\rangle = \Lambda^{3m}e^{i\theta}.}
Here $\Lambda$ is the mass scale of the theory and $\theta$ is the
$\theta$ angle.  The $\theta$ dependence of \pilp, which is seen
most naively from the fact that the left hand side is non-zero at
the one instanton level, follows in a standard fashion
from the quantum numbers of $\lambda$
under an anomalous $U(1)$ symmetry of the low energy $N=1$ theory.
Quantum mechanically, $\lambda\lambda$ has a vacuum expectation value;
by cluster decomposition this must be an $m^{th}$ root of \pilp:
\eqn\ilp{\left\langle\lambda\lambda\right
\rangle_r=\Lambda^3e^{i\theta/m}e^{2\pi ir/m}.}
Here $r=1\dots m$ labels the choice of vacuum state and $\langle \,\,\,\,\,\,\,
\rangle_r$ is the expectation value in the $r^{th}$ vacuum.  Thus, \ilp\ shows
that the correlation functions in a particular vacuum are not invariant
under $\theta\to\theta+2\pi$; rather, that transformation cyclically permutes
the $m$ vacuum states.  In a particular vacuum, the correlators are
invariant only under $\theta\to\theta+2\pi m$.
In the context of strong-weak duality, with modular functions of
$\tau={\theta\over 2\pi}+{4\pi i\over g^2}$, this means that the contributions
of individual vacua are invariant only under
\eqn\onlyunder{\tau \to \tau+m.}
Under $\tau\to\tau +1$, the $m$ vacua are  cyclically permuted.

That completes our discussion of the trivial embedding.
The irreducible vacuum is much easier to analyze since the gauge symmetry
group is completely broken at the classical level, so there are no strong
quantum effects of any kind (if the gauge coupling is small) and the
trivial embedding leads to a unique quantum vacuum.
The intermediate vacua can be analyzed as a combination of the above
cases, with the extra phenomenon that because of the $U(1)$ factors
they have no mass gap.

Thus, for instance, for $G=SU(2)$ there are three vacuum states,
two from the trivial embedding and one from the irreducible embedding.
They all have mass gaps.  The first two are related by a broken
symmetry.
If indeed $w$ is the generator of the ${\bf Z}_4$ chiral symmetry,
then $w$ acts  by
\eqn\wtau{w:\tau\to\tau+1}
and exchanges the two vacua that come from the trivial embedding.
They are each invariant under $w^2$, which acts by $\tau\to\tau+2$
and generates the unbroken ${\bf Z}_2$.

\subsec{Partition Functions Of Some Simple Theories}

Now we need to practice with certain general points about
quantum field theory.

If the gauge group is $G=SU(2)$,
the irreducible vacuum does not quite have completely broken
gauge symmetry; $SU(2)$ is broken down to its center
${\bf Z}_2$.  On a four-manifold $X$, the theory in the irreducible
vacuum reduces at low energies
to ${\bf Z}_2$ gauge theory.  By ${\bf Z}_2$ gauge theory, we mean
a theory in which one sums over ${\bf Z}_2$-valued flat connections
-- there are $2^{b_1}$ of them\foot{$b_1$ is the first Betti number
of $X$.  We assume for simplicity that there is no torsion (or at least
no $2$-torsion) in the cohomology of $X$, so that the ordinary Betti
numbers coincide with the ${\bf Z}_2$-valued ones.} --
and in which dividing by the volume of the gauge group means dividing
by the number of elements in ${\bf Z}_2$, which is 2.  The partition
function of ${\bf Z}_2$ gauge theory on $X$ is therefore
\eqn\impono{Z=2^{b_1-1}.}
(Of course, this formula has a similar origin to the prefactor in \tuff.)
For a general finite abelian gauge group $\Gamma$, the result would be
$(\# \Gamma)^{b_1-1}$, with $\#\Gamma$ the number of elements in $\Gamma$.

Now let us consider another simple case, the partition function
of a theory that has a mass gap and in which all degrees of freedom
can be measured locally (there are no unbroken gauge invariances).
The mass gap ensures that a simple result will emerge if one scales
up the metric by $g\to tg$ and takes $t$ large.
The contribution to the partition
function will behave for large $t$ like $\exp(-L_{\mit eff})$,
where -- because of the mass gap -- $L_{\mit eff}$ has an expansion
in local operators:
\eqn\realll{L_{\mit eff}=\int_X d^4x\sqrt g\left(u + v R + w R^2+\dots\right).}
Now suppose in addition one knows for some reason that the partition
function $Z=\exp(-L_{\mit eff})$ is a topological invariant.
Topological invariance means that the operators that arise in \realll\
 integrate to give topological invariants.  The only
topological invariants of a four-manifold that can be written as such
integrals of a local operator are the Euler characteristic $\chi$
and the signature $\sigma$.  The contribution
to the partition function of a vacuum with a mass gap and in which
all degrees of freedom can be measured locally is thus of the form
\eqn\ombobo{e^{a\chi + b\sigma}}
where $a$ and $b$ are independent of the particular four-manifold.

For a vacuum with a mass gap that also has an abelian  group $\Gamma$
of local gauge invariances ($\Gamma$ is a finite group or there would
not be a mass gap), the partition function is the product of \ombobo\
with the partition function of the finite gauge theory:
\eqn\mombobo{Z=(\#\Gamma)^{b_1-1}e^{a\chi+b\sigma}.}

\subsec{Partition Function On A Hyper-K\"ahler Manifold}

Now we will begin an analysis that will lead to a proposal
for the partition function of the $SU(2)$ or $SO(3)$ theory on
a compact K\"ahler manifold $X$ with $H^{2,0}(X)\not= 0$.
As in \switten, we first neglect
the difference between the physical and topological theories.
This means that we will obtain formulas that are really valid only
for hyper-K\"ahler manifolds (of which there are very few examples)
since in the hyper-K\"ahler case the physical and topological theories
coincide.  Then we make a correction involving the twisting and the
canonical divisor of $X$ to understand the general case.

Because the various vacua of the twisted theory all have mass gaps,
we can use \mombobo\ with $\Gamma$ trivial for the trivial embeddings
and $\Gamma={\bf Z}_2$ for the irreducible embedding.
For hyper-K\"ahler manifolds, the parameters $a$ and $b$ in \mombobo\
cannot both be detected.  The reason is that
on a four-dimensional K\"ahler manifold, the canonical divisor $K$ obeys
\eqn\itobeys{K\cdot K = 2\chi+3\sigma.}
(Here $K\cdot K$ is the intersection pairing.)  In particular
$2\chi+3\sigma$ vanishes for hyper-K\"ahler manifolds, which have $K=0$.
Therefore, as long as we are on hyper-K\"ahler manifolds, we can only
see one linear combination of $\chi$ and $\sigma$ (the other will appear when
we make corrections involving the canonical divisor).
It turns out that particularly nice answers
emerge if we choose the combination
\eqn\defdel{\nu={\chi+\sigma\over 4}.}
which is always an integer for K\"ahler manifolds.
As long as the twisting can be ignored, the contribution to the partition
function from a vacuum state with a mass gap should be $e^{a\nu}$ with a
universal $a$ (times a factor involving the discrete gauge symmetry).
In \switten, as there was no dimensionless coupling constant, $a$
was simply a constant.  In our present problem, there is a dimensionless
coupling constant $\tau$ and $a$ is a function of $\tau$.

For the $SU(2)$ theory, there are three vacuum states, all with mass gaps,
so the partition function
ignoring the twisting would therefore be
\eqn\zeq{Z = 2^{b_1-1}e^{a(\tau)\nu}+e^{b(\tau)\nu}+
e^{c(\tau)\nu},}
where the first term is the contribution from the irreducible
embedding (with its unbroken ${\bf Z}_2$ gauge invariance)
and the last two terms are the contributions from
the two vacua that come from reducible embeddings.
Further, the contribution of the irreducible vacuum is periodic
in $\theta$, while the other two are related by a broken symmetry,
so
\eqn\getq{\eqalign{e^{a(\tau+1)}&=e^{a(\tau)} \cr
                   e^{b(\tau+1)}&=i^{-1}e^{c(\tau)}\cr
                   e^{c(\tau+1)}&=i^{-1}e^{b(\tau)}\cr}}
Moreover $a$ and $b$ are such that the sum in \zeq\ transforms
as a modular form for an appropriate subgroup of $SL(2,{\bf Z})$ as explained
in \S3.  Only the factors of
$i^{-1}$ on the right hand side of \getq\ may need special explanation
here.  The reason for these
factors is that, although the bulk theory is invariant under the
${\bf Z}_4$ symmetry generated by $w$, on a four-manifold $X$ the
path integral measure transforms with a factor of $i^\nu$; this
can be seen by counting $\lambda$ zero modes and is explained in the
derivation of equation (2.46) of \switten\ (in comparing to that
equation note that $\Delta\cong\nu$
modulo 4).

Comparing to the answer found for $K3$ for $SU(2)$ gauge group
in \S4.1, we see that it is naturally written in precisely this form,
with
\eqn\netq{\eqalign{e^{a(\tau)}=&{1\over 2}G(q^2)^{1/2}=
                     {1\over 2q\prod_{n=1}^\infty
                                      (1-q^{2n})^{12}} \cr
                   e^{b(\tau)}= &{1\over 2}G(q^{1/2})^{1/2}=
                              {1\over 2\,q^{1/4}\prod_{n=1}^\infty
                                   (1-q^{n/2})^{12}}.\cr
                   e^{c(\tau)} = &{1\over 2}G(-q^{1/2})^{1/2}=
                    {1\over 2\,q^{1/4}
         \prod_{n=1}^\infty (1-(-1)^nq^{n/2})^{12}} \cr}}
Since $\nu=2$ for $K3$, the signs on the right hand side of \netq\ are
undetermined, and have been selected for later convenience.\foot{It would
be possible to multiply this formula and all subsequent ones that involve
$a$, $b$, and $c$
by a factor of $(-1)^\nu$, which we are unable to determine since
the available computations are for manifolds with even $\nu$.}
Of course, this identification of $a$, $b$, and $c$ is
not really rigorous, but it is so natural that we will accept it.
Notice that -- since $e^a$, $e^b$, and $e^c$ cannot have zeroes --
the fact that
the $\eta$ function has no zeroes in the upper half plane is essential
for the formula to make sense.  Also, the factors of $i$ on the right
hand side of \getq\ come from the leading powers $q^{-1/4}$ in $e^b$
and $e^c$ ($e^a$ is instead strictly invariant under $\tau\to\tau+1$
as the leading power of $q$ is integral).

For $SU(m)$ the analog of \zeq\ would be
\eqn\zeq{Z = m^{b_1-1}e^{a(\tau)\nu}+e^{b_1(\tau)\nu}+e^{b_2(\tau)\nu}+\dots
+e^{b_m(\tau)\nu},}
where the first term comes from the irreducible embedding and
the other $m$ terms from the trivial embedding.  In addition to
modular invariance of the whole sum, the individual terms should obey
\eqn\zzeq{\eqalign{e^{a(\tau+1)} & =e^{a(\tau)}\cr
          e^{b_r(\tau+1)} & = e^{-i\pi/m}e^{b_{r+1}(\tau)}\cr}}
The proposal for the $SU(m)$ partition function on $K3$ that was written
at the end of \S4.1 is naturally written in this form with obvious
choices of $a$ and the $b_r$.  \zeq\ should be valid
for $m$ prime; otherwise, intermediate vacua contribute, as explained
above.

\subsec{Effects Of Twisting}

Now we want to consider what happens on more general K\"ahler manifolds
with a non-trivial canonical divisor.  The physical and twisted
models no longer coincide, and the twisting has the following effect.
Of the three superfields $T,U,$ and $V$, two, say $T$ and $U$,
are scalars in the twisted
model and the third, $V$, is a $(2,0)$-form.  (This is already clear
from the description of the quantum numbers in \S2.4.)  Moreover,
the superpotential $W$ now transforms as a $(2,0)$ form on
$X$ (as in equation (2.42) of \switten).

Given the quantum numbers, a mass term combining $V$ and one of the
other superfields, say $U$, can be written without any difficulty:
$W_1=-m\Tr UV$.  However, to give a mass term to all three superfields,
one needs as in \switten\ to pick a $c$-number holomorphic two-form $\omega$
on $X$.  (So, in particular, $H^{2,0}(X)$ must be non-zero.)
For instance, once such an $\omega$ is picked, one can give
$T$ a spatially dependent mass term
\eqn\wtwo{W_2=-\omega \Tr T^2.}

We now consider the effects of this perturbation.  Away from the zeroes
of $\omega$, the superpotential $W_1+W_2$ can be analyzed just as
we have done above, leading to the familiar classification of vacuum
states in terms of $SU(2)$ embeddings.
Near zeroes of $\omega$, the picture changes.  In analyzing
this, we will consider only gauge group $SU(2)$ or $SO(3)$, so in
the bulk theory we have only the trivial and irreducible embeddings
to consider.  Both exhibit new behavior near zeroes of $\omega$, but in
very different ways.  In working out the details, we assume, for simplicity
(as in \switten), that the zeroes of $\omega$ are a union of disjoint,
smooth complex curves $C_i,\,\,i=1\dots n$ of genus $g_i$, on which $\omega$
has
a simple zero.

\bigskip
\noindent{\it The Trivial Embedding}

First we consider the strongly coupled vacua that come from the trivial
embedding.  The $C_i$ behave like the world-sheets of
``superconducting cosmic strings''
that can trap charges (see \S3.3 of \switten).  Even after understanding
in bulk the dynamics of the strongly coupled four-dimensional theory,
one must analyze the behavior of the effective two-dimensional theory
near $C_i$.

The main points that one can make come from the symmetry structure (see \S2.7
of \switten).
In bulk, the theory has a mass gap with ${\bf Z}_4$ spontaneously
broken down to ${\bf Z}_2$.  This ${\bf Z}_2$ is realized near the
$C_i$ as a chiral symmetry that prevents fermion
masses in the effective two-dimensional theory, so it is natural that
it should be spontaneously broken, giving a two-fold vacuum degeneracy.
It is natural to assume that this strongly coupled vacuum state
has a mass gap and no vacuum degeneracy except what follows from the
symmetry breaking.

The vacuum structure (for states coming from the trivial embedding
of $SU(2)$) is thus as follows.  In bulk there are two vacuum
states, say $|+\rangle$ and $|-\rangle$, related by the
${\bf Z}_4\to {\bf Z}_2$ symmetry breaking.  Near any of the $C_i$
there is a further two-fold bifurcation of the vacuum;
$|+\rangle$, for instance, splits into $|++\rangle$ and $|+-\rangle$.

According to \wtau, the generator $w$ of the underlying ${\bf Z}_4$ symmetry
acts on the contributions of these vacua to the partition function
by $\tau\to\tau+1$.  The ${\bf Z}_2$ that is unbroken in bulk
is generated by $w^2$, which acts by $\tau\to\tau+2$.  Hence
the $|++\rangle$ and $|+-\rangle$ vacua are exchanged by $\tau\to\tau+2$.
This also exchanges $|-+\rangle$ with $|--\rangle$.

Now, consider a particular vacuum, say $|++\rangle$, along a cosmic
string component $C_i$.
Its contribution to the partition function is again of the form
$\exp(-L_{\mit eff})$, where $L_{\mit eff}$ has an expansion in local
operators as in \realll.  The only difference is that now they are
two-dimensional local operators since we are determining the contribution
localized near $C_i$ were the bulk description fails.  The only
topological invariant of $C_i$ that can be written as the integral
of a local operator (indeed, its only topological invariant)
is its Euler characteristic $\chi(C_i)=2-2g_i$.
Hence the contribution of $|++\rangle$ near $C_i$ is a factor
of $e^{(1-g_i)u(\tau)}$, where the function $u(\tau)$ is independent of the
details of $X$ and $C_i$.  Similarly, the $|+-\rangle$ vacuum contributes
a factor $e^{(1-g_i)v(\tau)}$, where $u$ and $v$ are exchanged by
$\tau\to \tau+2$:
\eqn\rtrans{\eqalign{u(\tau+2) & = v(\tau) \cr
                     u(\tau+4) & = u(\tau) .\cr}}
With all its possible bifurcations, which must
be chosen independently on each $C_i$, the $|+\rangle$ vacuum contributes
a factor
of
\eqn\confac{e^{b(\tau)\nu}\prod_{i=1}^s\left(e^{u(\tau)(1-g_i)}
+t_ie^{v(\tau)(1-g_i)}\right).}
The factors of $t_i=\pm 1$ have  the same origin as the factors of $i^{-1}$
on the right hand side of \getq\ -- they incorporate a global anomaly
(in the coupling to gravity) in the ${\bf Z}_2$ symmetry that permutes
$|++\rangle$ and $|+-\rangle$. \foot{This factor appears in equation (2.66)
of \switten\ and is explained in the derivation of that equation.}
Similarly the possible bifurcations of the $|-\rangle$ vacuum contribute
the transform of this under $\tau\to\tau+1$, or
\eqn\onfac{e^{b(\tau+1)\nu}\prod_{i=1}^s\left(e^{u(\tau+1)(1-g_i)}
+t_i e^{v(\tau+1)(1-g_i)}\right).}

Now, \itobeys\ is equivalent to the formula
\eqn\twoc{2\chi+3\sigma=\sum_i(g_i-1).}
This formula is the real reason that we need not in the bulk theory
consider an extra factor of the form $\exp(\gamma(\tau)(2\chi+3\sigma))$.
It could be absorbed in redefining the function $u$.

\bigskip
\noindent{\it The Non-trivial Embeddings}

Now we analyze what becomes of the non-trivial embeddings when
the mass terms involve a choice of holomorphic two-form.
What this means is that we consider the theory with the combined
superpotential
\eqn\son{W_{\mit tot}=-\omega \,\Tr\, T^2- m\,\Tr\, UV +\Tr\, T[U,V],}
and look for a classical solution with zero action in the twisted theory.
The analysis is feasible because by arguments similar to those in
\S2.4, the $(2,0)$ part of the curvature can be assumed to vanish,
and the structure is therefore holomorphic.

The condition for a critical point of $W_{\mit tot}$ is
\eqn\condcrit{\eqalign{[T,U] & = mU\cr
                       [T,V] & = -mV\cr
                       [U,V] & = 2\omega T.\cr}}
These equations have the following immediate consequence.  Regardless
of what $\omega$ does, the matrix $T$ is at each point in $X$ conjugate
to
\eqn\ondcrit{  {m\over 2}\pmatrix{ 1 & 0 \cr 0 & -1\cr},}
and in particular never vanishes.  Thus $T$ breaks the gauge
group down to $U(1)$.

For the kinetic energy of $T$, $U$, and $V$ to vanish in the twisted
theory does not require that they should be covariantly constant,
but only that they should be holomorphic.  The existence of a holomorphic
$T$ that is everywhere in
the conjugacy class indicated in \ondcrit\ means that the $SU(2)$ gauge bundle
$E$ splits as a sum $E\cong L \oplus L^{-1}$ with some holomorphic line
bundle $L$.   With such a splitting, $T$ then {\it globally} takes the form of
\ondcrit.  \condcrit\ then implies that $U$ and $V$
are of the form
\eqn\dondcrit{\eqalign{ U& =\pmatrix{ 0 & b\cr 0 & 0 \cr}\cr
                        V& =\pmatrix{ 0 & 0 \cr c & 0\cr},\cr}}
with
\eqn\gondcrit{ bc= m\omega.}
Here $b$ is a holomorphic section of the line bundle $L^2$,
and $c$ is a holomorphic section of the line bundle $K\otimes L^{-2}$.
($K$ enters because $V$ is a $(2,0)$-form.)

In order to make the exposition in the rest of this section as simple
as possible, we will assume first that the zeroes of $\omega$ consist
of a {\it single} smooth connected curve $C$, on which $\omega$ has
a simple zero.  There is then only a single $t_i$ in \confac\ and \onfac,
and according to \switten, equation (2.61), it is
\eqn\undod{t=(-1)^\nu.}
At the very end of the section we will return to
the case that the canonical divisor is a union of disjoint smooth curves $C_i$.

With the given assumption about $\omega$, \gondcrit\ is easy to analyze.
The product $bc$ vanishes with multiplicity one on the irreducible
smooth curve $C$, so either $c$ vanishes on $C$ and $b$ has no zeroes,
or vice-versa.  If $b$ has no zeroes, the line bundle $L$ is trivial, and up
to a complex gauge transformation, $b=m$ and $c=\omega$.
If $c$ has no zeroes, then $K\otimes L^{-2}$ is trivial, so
$L\cong K^{1/2}$  (that is, $L$ is a line bundle with $L^{\otimes 2}\cong K$)
and up to a complex gauge transformation, $b=\omega$
and $c=m$.

Once the holomorphic data are known
the metric on the gauge bundle is determined (presumably uniquely,
by a convexity argument, but certainly not explicitly) by the condition
(analogous to equation \thirde\ in \S2.4) for the $(1,1)$ part of
the curvature.

Thus, we have determined the
possible vacuum solutions: there is one on the trivial bundle $E_0$, which
has instanton number zero,
and one on the bundle $E_1=K^{1/2}\oplus K^{-1/2}$, whose instanton number is
\eqn\rufn{-{K\cdot K\over 4}= -{g-1\over 4},}
with $g$ the genus of $C$.  Note that this number is typically
negative.
Actually, we saw in  \S2.4 that bundles of negative instanton number
can contribute when the vanishing theorem fails, and we saw
(under a hypothesis that was
equivalent to having the canonical divisor connected
and $K\cdot K>0$) that the most negative possible value of a bundle
that would contribute was $-K\cdot K/4$.

Of course, it might happen that $X$ is not a spin manifold.  Then
$K^{1/2}$ does not exist.  In that case, only $E_0$ will contribute
to the $SU(2)$ theory.  However, $E_1$ will always contribute in the
$SO(3)$ theory.  Indeed, the $SO(3)$ bundle ${\rm ad}(E_1)$
derived from $E_1$,
which is $K\oplus K^{-1}\oplus {\cal O}$ (with ${\cal O}$ a trivial
bundle) always exists.
Its second Stieffel-Whitney class is $w_2({\rm ad}(E_1))=w_2(X)$,
where $w_2(X)$ is the second Stieffel-Whitney class of the tangent
bundle of $X$, which is the same as the reduction modulo two of $c_1(K)$.
Indeed, ${\rm ad}(E_1)$ is the bundle of self-dual two-forms, while
$E_1$, if it exists, is one of the two chiral spin bundles of $X$;
the obstruction to constructing the latter from the former is $w_2(X)$.

\bigskip
\noindent{\it The General Structure}

The general structure of the partition function is then as follows,
for the $SU(2)$ theory:
\eqn\rungo{
\eqalign{Z_{SU(2)}= &2^{b_1-1}
     e^{a(\tau)\nu}\left(F+G\delta_{w_2(X)=0}\right)\cr
           &+e^{b(\tau)\nu}\left(e^{(1-g)u(\tau)}+(-1)^{\nu}
e^{(1-g)u(\tau+2)}\right)\cr &
           +e^{c(\tau)\nu}\left(e^{(1-g)u(\tau+1)}+
(-1)^{\nu}e^{(1-g)u(\tau+3)}\right).\cr}}
(The factors of $(-1)^\nu$ come from the $t_i$ in \confac\ and \onfac\
via \undod.)  $F$ and $G$ are corrections near the canonical
divisor to contributions of the vacuum bundles $E_0$ and $E_1$; of
course $E_1$ only contributes to the $SU(2)$ theory if $w_2(X)=0$.  In
what follows we abbreviate $x_0=w_2(X)$.
The functions $a,b,c,$ and $u$ are universal (independent of $X$)
because of the mass gaps.  It might be possible to argue {\it a priori}
for a similar
universality of $F$ and $G$,\foot{It is not clear whether for the irreducible
vacua, the weakly coupled
theories that arise near the canonical divisor have mass gaps.}
but in any case $S$-duality implies such universality.

Now given \rungo, we want to determine the analogous formula
for the $SO(3)$ theory.  In the $SO(3)$ theory, one must sum
over all values of $w_2(E)$.  One can nearly interpret
\rungo\ as the contribution of bundles with $w_2(E)=0$ to the $SO(3)$
theory, but it is necessary to make a correction that involves
comparing the volumes of the $SU(2)$ and $SO(3)$ gauge groups,
as discussed in \S3.
 So to get from \rungo\ the contribution to the $SO(3)$ theory from
bundles with $w_2(E)=0$, one simply divides by $2^{b_1-1}$.

Once this is known, there is no problem getting the contribution
in the $SO(3)$ theory from bundles with an arbitrary value of
$x=w_2(E)$.  The $F$ and $G$ functions anyway only contribute
for a particular value of $x$, and the others because of the mass
gap only see the global object $x$ through its influence
on anomalies which determine the relative phases between contributions
of different vacua.  The  $x$-dependence of these anomalies
is explained   in \switten\ in the derivation of equation (2.79).
Putting together the anomalies and the factor of $2^{-b_1+1}$, the
contribution in the $SO(3)$ theory of bundles with $w_2(E)=x$ is
\eqn\urungo{
\eqalign{Z_{x}= &
     e^{a(\tau)\nu}\left(F\delta_{x=0}+G\delta_{x=x_0}\right)\cr
           &+2^{1-b_1}e^{b(\tau)\nu}\left(e^{(1-g)u(\tau)}+(-1)^{\nu+x\cdot
x_0}
e^{(1-g)u(\tau+2)}\right)\cr &
           +i^{-x^2}2^{1-b_1}e^{c(\tau)\nu}\left(e^{(1-g)u(\tau+1)}+
(-1)^{\nu+x\cdot x_0}e^{(1-g)u(\tau+3)}\right).\cr}}
Of course, the $F$ function, associated with the vacuum bundle $E_0$,
only contributes for $x=0$, and the $G$ function, associated with the
vacuum bundle $E_1$, only contributes if $x=x_0=w_2(X)$.

The unknown
functions can be determined as follows.  According to the blowing-up
formula of Yoshioka, blowing up a point in a K\"ahler manifold $X_0$
multiplies
the partition function by a factor of $\theta_0/\eta^2$ with
$\eta$ the Dedekind eta function and
\eqn\ompo{\theta_0=\sum_{n\in {\bf Z}} q^{n^2}.}
(In this assertion, $x$ is taken to be a pullback from $X_0$.)
If we consider the special case that $X_0$ is $K3$ (so the partition
function on its one-point blow-up $X$
is governed by \urungo\ with $g=0,\nu=2, x\cdot x_0=0$),\foot{Recall
from the end of  \S2.4 that the vanishing theorem holds for the
blowup of $K3$ -- so we can apply instanton results such as those
of Yoshioka.  Similarly, we can consider $K3$ with an arbitrary
number of points blown up.}
this implies that
\eqn\impod{ F(\tau)=e^{u(\tau)}+e^{u(\tau+2)}=e^{u(\tau+1)}+e^{u(\tau+3)}
={\theta_0\over\eta^2}. }
The function $\theta_0/\eta^2$ transforms in a two-dimensional representation
of $SL(2,{\bf Z})$, the other function that enters being $\theta_1/\eta^2$ with
\eqn\osn{\theta_1=\sum_{n\in {\bf Z}+{1\over 2}}q^{n^2}.}

Under $\tau\to -1/\tau$, $F$ and $G$ are mapped to $e^{u({\tau})}$ and
$e^{u(\tau+2)}$, respectively, as will be clear presently when we
check the modular transformation laws in detail.
Since $F$ is known (at least in this special case) $e^{u({\tau})}$
is thereby determined -- once and for all, since $u$ is a universal function.
By applying $\tau\to -1/\tau$ one now determines $F$ and $G$ in general.
In particular, one finds that $G\sim (\theta_1/\eta^2)^{1-g}
\sim q^{ (1-g)/4}(1+\dots)\eta^2$ where the leading exponent $q^{ (1-g)/4}$
is in happy agreement with the instanton number of the classical solution
that is responsible for the presence in the formula of the $G$ function.

In the following subsection, we will write down a precise formula
that was found by the reasoning just sketched and verify that it has all of
the right properties.  The reader who works through the verification
should be able to see that given the general structure that we have proposed,
the formula we write down is the only one that works.

\subsec{The Formula}

We continue
to assume temporarily that the canonical divisor of $X$ is connected;
its genus is
\eqn\ggenus{g-1=2\chi+3\sigma.}
$b_2{}^+$ and $b_2{}^-$
will denote the dimensions of the spaces $H^{2,+}$ and $H^{2,-}$ of
self-dual and anti-self-dual harmonic two-forms on $X$; then we have
\eqn\trufu{\eqalign{b_2&=b_2{}^++b_2{}^-,\cr
\sigma& =b_2{}^+-b_2{}^-\cr
\chi&=2-2b_1+b_2{}^++b_2{}^-,\cr}}
with $b_2$ the second Betti number of $X$.

The formula we propose is then
\eqn\imcin{\eqalign{Z_x & = \left({1\over 4}G(q^2)\right)
^{\nu/2}\left(\delta_{x,0}(-1)^\nu\left({\theta_0\over\eta^2}\right)^{1-g}
+\delta_{x,x_0}\left({\theta_1\over \eta^2}\right)^{1-g}\right)
\cr &+2^{1-b_1}\left({1\over 4}G(q^{1/2})\right)^{\nu/2}\left(
\left({\theta_0+\theta_1\over 2\eta^2}\right)^{1-g} +(-1)^{\nu+x\cdot x_0}
\left({\theta_0-\theta_1\over 2\eta^2}\right)^{1-g}\right)\cr
&+2^{1-b_1}i^{-x^2}\left({1\over 4}G(-q^{1/2})\right)^{\nu/2}\left(\left(
{\theta_0-i\theta_1\over 2\eta^2}\right)^{1-g} +(-1)^{\nu+x\cdot x_0}
\left({\theta_0+i\theta_1\over 2\eta^2}\right)^{1-g}\right)\cr}}

This formula manifestly transforms correctly under $\tau\to\tau+1$.
According to the precise version of the strong-weak duality conjecture
proposed in \S3.3, the transformation law under
$\tau\to -1/\tau$ should be
\eqn\pkns{Z_y(-1/\tau)=2^{-b_2/2}(-1)^{\nu}({\tau \over i})^{-\chi /2}
 \sum_x(-1)^{x\cdot y}Z_x(\tau).}
To evaluate the sum over $x$ on the right hand side, we need
various formulas:
\eqn\uful{\eqalign{&\sum_x(-1)^{x\cdot y}\delta_{x,0}  = 1 \cr
       &\sum_x(-1)^{x\cdot y}\delta_{x,x_0}  = (-1)^{y\cdot x_0}\cr
       &\sum_x (-1)^{x\cdot y}  =2^{b_2}\delta_{y,0} \cr
       &\sum_x (-1)^{x\cdot y +x\cdot x_0}  =2^{b_2}\delta_{y,x_0}\cr
       &\sum_x (-1)^{x\cdot y}i^{-x^2}  = 2^{b_2/2}\,i^{-\sigma/2+y^2}\cr
       &\sum_x (-1)^{x\cdot y}i^{-x^2}(-1)^{x\cdot x_0}  = 2^{b_2/2}\,
i^{\sigma/2-y^2} . \cr}}
The only formulas here that require comment are the last two.
The next-to-last sum in \uful, on completing the square,
is $\sum_xi^{-(x+y)^2+y^2}=i^{y^2}\sum_xi^{-x^2}$, and
so is equivalent to the sum $\sum_xi^{-x^2}$, which is a special
case of a sum that we met
in \S3.3.  The last sum in \uful\ can be treated the same
way; alternatively, using the Wu formula,
\eqn\wuformula{(-1)^{x\cdot x_0}=
(-1)^{x^2},}
the last sum is the complex conjugate
of the one before.

Using these sums,
we find that the right hand side of \pkns\ is (dropping the modular weight)
\eqn\undergop{\eqalign{&{\mit R.H.S.}
 = 2^{-b_2/2}\left({G(q^2)\over 4}\right)^{\nu/2}
\left(\left({\theta_0\over\eta^2}\right)^{1-g}+(-1)^{\nu+x\cdot x_0}
\left({\theta_1\over\eta^2}\right)^{1-g}\right) \cr &
+2^{b_2/2+1-b_1}
\left({G(q^{1/2})\over 4} \right)^{\nu/2} \left(\delta_{x,0}(-1)^\nu
\left({\theta_0  +\theta_1\over 2\eta^2}\right)^{1-g}  +
 \delta_{x,x_0}\left({\theta_0-\theta_1\over 2\eta^2}\right)^{1-g}\right)
\cr &+2^{1-b_1}
\left({G(-q^{1/2})\over 4}\right)^{\nu/2}\left( (-1)^\nu i^{x^2-\sigma/2}
\left({\theta_0-i\theta_1\over 2\eta^2}\right)^{1-g} + i^{-x^2+\sigma/2}
\left({\theta_0+i\theta_1\over 2\eta^2}\right)^{1-g}\right).\cr}}

To verify \pkns,
we need the modular transformations of the various functions.
Under $\tau\to -1/\tau$, we have (apart from modular weight)
\eqn\gipper{\pmatrix{ \theta_0\cr \theta_1\cr}\to {1\over \sqrt 2}\pmatrix{
1 & 1 \cr 1 & -1 \cr}\pmatrix{\theta_0 \cr\theta_1\cr}.}
This means, in particular, that $\theta_0$ is exchanged with
$(\theta_0+\theta_1)/\sqrt 2$ and $\theta_1$ with $(\theta_0-\theta_1)/\sqrt
2$,
while $(\theta_0\pm i \theta_1)$ is mapped to $i^{\pm 1/2}(\theta_0\mp
\theta_1)$.
These facts are convenient, since the functions just mentioned all appear
raised to the $(1-g)$ power in \imcin.
We also have (apart from the modular weight)
\eqn\ippper{\eqalign{G(q^2)^{\nu/2} & \to 2^{{3\over 2}(\chi+\sigma)}
 G(q^{1/2})^{\nu/2} \cr
     G(q^{1/2})^{\nu/2}&\to  2^{-{3\over 2}(\chi+\sigma)}
 G(q^{2})^{\nu/2} \cr
G(-q^{1/2})^{\nu/2} & \to  G(-q^{1/2})^{\nu/2}.\cr}}
With the aid of these transformation laws and identities \trufu\ and
\wuformula, it is straightforward to verify \pkns.

\bigskip
\noindent{\it Disconnected Canonical Divisor}

It remains only to generalize the above formulas for the case
that $\omega$ vanishes (with multiplicity one) on
a union of disjoint smooth components
$C_i$, $i=1\dots n$, of genus $g_i$.

This has two effects: the bulk vacua $|+\rangle$ and $|-\rangle$  that
come from the trivial embedding of $SU(2)$
bifurcate along the $C_i$ into $2^n$ separate ground states --
in a fashion that we have already described in \onfac\ and \confac,
before we specialized to $n=1$.
Also, the contribution of the irreducible embedding is more complicated.

It is straightforward to write down the contribution of the trivial
embedding since we now have determined all of the functions that
appear in \onfac\ and \confac. The contribution is
\eqn\heffo{\eqalign{2^{1-b_1}
\left({G(q^{1/2})\over 4}\right)^{\nu/2}&\prod_{i=1}^n
\left(
\left({\theta_0+\theta_1\over 2\eta^2}\right)^{1-g_i} +t_i(-1)^{x\cdot C_i}
\left({\theta_0-\theta_1\over 2\eta^2}\right)^{1-g_i}\right)\cr
+2^{1-b_1}
i^{-x^2}\left({G(-q^{1/2})\over 4}\right)^{{\nu\over 2}}&\prod_{i=1}^n
\left(\left(
{\theta_0-i\theta_1\over 2\eta^2}\right)^{1-g_i} +t_i(-1)^{x\cdot C_i}
\left({\theta_0+i\theta_1\over 2\eta^2}\right)^{1-g_i}\right).\cr}}
The $t_i$ obey
\eqn\snon{\prod_it_i=(-1)^\nu,}
as shown in equation (3.57) of \switten.  Also,
$\prod_i(-1)^{x\cdot C_i}=(-1)^{x\cdot x_0}$ as the union of the $C_i$
is the canonical divisor, whose Poincar\'e dual reduces modulo two to $x_0$.

To find the contributions of the irreducible embedding, we must reexamine
the factorization
$bc=m\omega$ of equation \gondcrit.  The possible solutions
are as follows. Given that $\omega$ vanishes
with multiplicity one on the union of the $C_i$, $b$ can
vanish with multiplicity one on any subset of the $C_i$, with $c$
vanishing on the others.  If then $L_i={\cal O}(C_i)$ is the line
bundle whose sections are functions with a simple zero on $C_i$,
the $SU(2)$ bundle $E$ is $E_{\vec\epsilon}=L_{\vec\epsilon}{}
^{1/2}\oplus L_{\vec\epsilon}{}^{-1/2}$, where
\eqn\hsn{L_{\vec\epsilon}=\otimes_{i=1}^nL_i{}^{\epsilon_i}, }
where $\epsilon_i=0\,{\rm or}\,1$ are chosen independently.
There are thus $2^n$ solutions in all
(a fact which under $SL(2,{\bf Z})$ transforms into the fact that
the $|+\rangle$ or $|-\rangle$ vacuum in bulk bifurcates into $2^n$
choices along the cosmic strings).  Of course, as $L_{\vec\epsilon}$
may not have  a square root, the bundles $E_{\vec\epsilon}$ may not
really exist as $SU(2)$ bundles, but the corresponding $SO(3)$ bundles
${\rm ad}(E_{\vec\epsilon})$ always exist -- with second Stieffel-Whitney
class
\eqn\ksm{w_{2}(\vec\epsilon)=\sum_i\epsilon_i[C_i].}
Here $[C_i]$ is the reduction modulo two of the Poincar\'e dual of
$C_i$; one has $w_2(X)=\sum_i[C_i]$.  The instanton  number of the bundle
$E_{\vec\epsilon}$ is
\eqn\snl{-\sum_i\epsilon_i{g_i-1\over 4}.}

The contribution of all $2^n$
solutions of $bc=m\omega$ is
\eqn\theircon{\left({G(q^2)\over 4}\right)
^{\nu/2}\sum_{\vec\epsilon}
\delta_{x,w_2(\vec\epsilon)}\prod_{i=1}^n
\left\{t_i{}^{\epsilon_i}\left(
{\theta_0\over\eta^2}\right)^{(1-\epsilon_i)(1-g_i)}
\left({\theta_1\over \eta^2}\right)^{\epsilon_i(1-g_i)}\right\}.}
This formula was found by requiring that it is zero except if
$x$ is equal to $w_2(\vec\epsilon)$ for some $\vec\epsilon$ and
that the sum of \theircon\ and \heffo\ transforms correctly under
$\tau\to-1/\tau$.
That last assertion can indeed be verified using
the identities that have already been exploited above.  Note also
that the smallest power of $q$ in the contribution of a given
solution agrees with the instanton number in \snl.

\newsec{Connections with RCFTs and Strings}

In this paper, especially in  \S4, formulas familiar
in the context of two-dimensional rational conformal field
theories (RCFTs) and strings made unexpected appearances.
In this section we discuss some aspects
of this mysterious phenomenon. We will first consider relations
to RCFTs and then speculate about the potential applications
to the question of $S$-duality in string theory.

We have seen in  \S4 that blowing up of a point
has the effect of multiplying the $N=4$ partition function
with $SU(2)$ gauge group
by essentially the level one characters of the $SU(2)$
WZW model.  The magnetic flux vector
labels the conformal blocks, and the
instanton number mod 1 in the gauge theory
corresponds to the conformal dimension mod 1 in the RCFT.
Moreover we saw
that the level $k$ characters of $SU(n)$ arise
for $N=4$ super Yang-Mills theory with gauge group $U(k)$ on ALE
spaces $A_{n-1}$.
One might think that the appearance of conformal field theory characters
simply reflects the fact that those objects are modular and that there
are not too many modular objects of low weight and level.
However, the connections go beyond the partition function.
For the ALE spaces, for instance, Nakajima does not just get a WZW
character but finds an action of
the affine Kac-Moody algebra on the Hilbert space
consisting of the cohomology of instanton moduli spaces.

Generalizing the structure found by Nakajima,
one might wonder, for instance, whether for a more general four-manifold
there is
a natural action of the Virasoro algebra on the cohomology
of instanton  moduli spaces if one consider all instanton numbers
at once.  And what is the analog of the two-dimensional operator
product expansion?  Can one find for each rational conformal
theory a (possibly non-compact) four-manifold  whose
$N=4$ twisted partition function gives the characters of that RCFT?

In two-dimensional RCFT's,
the conformal blocks $\chi_i$ form a representation
of $SL(2,{\bf Z})$, much in the same way as do our partition functions
${\bf Z}_v$ for $v\in H^2(M,{\bf Z}_n)$.  RCFT's have additional
structure:  From the underlying operator-product relations
one deduces a commutative, associative multiplication law called the Verlinde
algebra
on the space of blocks  \ref\ver{E. Verlinde, Nucl.Phys.B300 (1999) 360.}.
 Does it have an analog in our problem?
Since the Verlinde algebra is determined by the $S$ matrix, and since
the four-dimensional problem has an  $S$ matrix given in \stran,
we can deduce that the analog of the Verlinde-algebra should be the
multiplication law
$${\bf Z}_{v_1}\cdot {\bf Z}_{v_2}={\bf Z}_{v_1+v_2}.$$
This operation is indeed commutative and  associative
(it is associated with
the ordinary addition law of the ${\bf Z}_n$-valued flux)
further enhancing the analogy between four dimensional gauge
theories and rational conformal field theory.

\bigskip
\noindent{{\it Stringy Spectrum?}}

Before going on, let us recall some aspects of
the tests of $S$-duality in string theory \seni .
One basic test is to identify all the BPS-saturated
states with given ``electric'' charges  and find
what ``magnetic states'' they lead to under $SL(2,{\bf Z})$ transformation. One
can then try to see if these states exist.
This is usually done for the field theoretic modes.  However,
in string theory there is actually a full stringy spectrum of
massive BPS-saturated states.
For heterotic
strings the number of such states is given by the (left-moving)
bosonic string oscillator partition function
\ref\atha{ A. Dabholkar and J. A. Harvey, Phys. Rev. Lett.63
(1989) 478.\semi A. Dabholkar, G. Gibbons, J.A. Harvey and F. Ruiz,
Nucl.Phys.B340 (1990) 33-55.}.
To see this one notes that if one takes the supersymmetric oscillators
to be in the ground state, then regardless of the lattice momenta,
one gets a ``small'' supersymmetry representation corresponding to
BPS-saturated states \wo .  Thus all the states which come from
$N_R=0$ and arbitrary $N_L$ saturate the BPS bound and, if $S$-duality
is to hold, there must be corresponding magnetic
states.  For $N_L=0$, these states are the BPS monopoles that can be seen
in field theory.  For $N_L=1$ it
was suggested in \seni\ that they have the right quantum
numbers to be identifiable with so-called
$H$-monopoles \nref\khur{R. Khuri, Phys. Lett. {\bf B259} (1991) 261; Phys.
Lett. {\bf B294}
(1992) 325; Nucl. Phys. {\bf B387} (1992) 315.}
\nref\haret{J. Gauntlett, J. Harvey and J. Liu, Nucl. Phys. {\bf B409} (1993)
363.} \refs{\khur,\haret}.
Based on the $N=4$ structure, and the fact that
$H$-monopoles saturate the BPS bound, one expects the moduli space
of $H$-monopoles
to have a hyper-K\"ahler structure.  Moreover the ``magnetic''
states would correspond to the cohomology of this moduli space.
 Unfortunately this moduli space is little understood.
However, $K3$ was proposed as a candidate in \harre,
in part because it is four dimensional and hyper-K\"ahler and
has 24 dimensional cohomology.
The fact that $K3$ has 24 dimensional cohomology
matches the fact that the bosonic string at $N_L=1$ has
24 physical states (namely $\alpha_{-1}^i|0\rangle$, where
$i$ runs over the transverse directions in the light-cone gauge).
Moreover, the Lorentz spin of these 24 states agrees with that of
the cohomology of $K3$
if we identify the helicity operator in space-time
with $(F_L+F_R)/2$, where we define an element
of the cohomology group $H^{p,q}$ to have $F_L=p-1$ and $F_R=q-1$.
Indeed, there are 22 states with helicity 0, one with helicity
$+1$ and one with helicity $-1$.\foot{If
we compactify to six dimensions, there are two light cone helicity
operators; identifying them with $(F_L+F_R)/2$
and $(F_L-F_R)/2$,
the massless states of the bosonic string compactified to six dimensions
agree with the cohomology of $K3$.
This comment also applies to the type II superstring considered above.}.

Is there an analog of this for $N_L>1$?
The discussion in  \S4 makes clear a possible
analog: the string states at arbitrary $N_L$
can be associated with the cohomology of the symmetric product
of $N_L$ copies of $K3$!
The helicities work out correctly, just as they do for $K3$.
Note that the main thing needed for $K3$ to give the
bosonic string partition function is its Euler characteristic
(and the hyper-K\"ahler structure which enables one to use the orbifold
formula to compute the cohomology of the symmetric product).
So it is conceivable that $K3$ could be replaced here by another,
perhaps non-compact, space.

If one compactifies a Type II superstring on a six-torus,
there is again a stringy spectrum of BPS-saturated states,
this time in 1-1 parallel with fermionic oscillator states.
Is there any way to generate the oscillator states of a fermionic
string by taking the cohomology of a symmetric product?
The $N=4$ supersymmetry again suggests that the relevant moduli
space would be a hyper-K\"ahler maniflold.  In dimension four,
other than $K3$ there is a unique compact example, namely the
four-torus ${\bf T}^4$.
The cohomology of ${\bf T}^4$ is 16 dimensional, with 8 ``bosonic''
states (of even degree $0,2$, or 4) and 8 ``fermionic'' ones
(of odd degree 1 or 3).
This agrees with the fact that the fermionic string has 8 transverse
Bose oscillators and 8 Fermi ones.  If we take the helicity operator
to be $(F_L+F_R)/2$, then the bosonic states in the cohomology of ${\bf T}^4$
include six states of helicity zero and two of helicity $\pm 1$
while the Fermi states include four of helicity $1/2$ and four of helicity
$-1/2$.  This agrees with the quantum numbers of the transverse oscillators
of the string!
The cohomology of a symmetric product of ${\bf T}^4$'s can be analyzed
by the same orbifold techniques that we used for $K3$. It is a Fock
space derived from ``one-particle states'' which are the cohomology of
${\bf T}^4$, and so agrees with the fermionic string spectrum.

While we have little insight to offer at the moment, this relation of
the oscillator spectrum of bosonic and fermionic strings to
the cohomology of the two compact hyper-K\"ahler manifolds in four
dimensions is certainly provocative.

\bigskip
\noindent{{\it Acknowledgments}}
We would like to thank P. Kronheimer for explaining the background about $K3$
surfaces, H. Nakajima and K. Yoshioka
for patient explanations of their work and helpful suggestions,
D. Gross, D. Morrison and D. Zagier for
explanations of some points relevant to \S2-4,
C. Taubes for bringing to our attention various relevant work
in the literature as well
as discussions on instanton moduli spaces,  T. Miyake for advice
about some technical points, and G. Shimura
 and E. Verlinde for valuable comments.
In addition C.V. would like to thank
the Institute for Advanced Study for its hospitality and
for providing a stimulating research environment
 where this work was done.

The research of C.V. was supported in part by the Packard fellowship and NSF
grants PHY-92-18167 and PHY-89-57162.  The work of E.W. was supported
in part by NSF Grant PHY92-45317.
\listrefs
\end